\documentclass[11pt]{article}

\usepackage[margin=0.75in]{geometry}
\usepackage{amsmath,amssymb,amsthm}
\usepackage{natbib}
\usepackage{hyperref}
\usepackage{graphicx}
\usepackage{booktabs}
\usepackage{longtable}
\usepackage{enumitem}
\usepackage{multirow}
\usepackage{float}
\usepackage{authblk}
\usepackage{threeparttable}
\usepackage{caption}
\hypersetup{
  colorlinks=true,
  linkcolor=blue,
  citecolor=blue,
  urlcolor=blue
}

\title{A Network-Structured Bayesian Hierarchical Model for Sparse Mutation-Drug Response Associations: Application to Cancer Pharmacogenomics} 

\author[1*]{Hammed A. Olayinka}
\author[2]{Saheed O. Olayinka}

\affil[1]{Department of Mathematical Sciences, Worcester Polytechnic Institute, Worcester, MA, USA}
\affil[2]{Department of Computer Science and Artificial Intelligence, University of Ibadan, Ibadan, Nigeria}

\affil[*]{\texttt{haolayinka@wpi.edu}}

\date{ }

\begin{document}
\maketitle

\begin{abstract}
\begin{abstract}
We develop a network-structured Bayesian hierarchical model for sparse association mapping between genomic alterations and quantitative treatment-response phenotypes. The framework combines a Gaussian Markov random field prior that borrows strength across pathway-connected genes, a global-local horseshoe prior inducing sparsity, and a conjugate Gibbs sampler requiring no Metropolis-Hastings steps. Though broadly applicable to high-dimensional settings with known predictor networks, we validate it using cancer cell-line drug-sensitivity data. Applied to GDSC2 ($N=951$ cell lines, $G=219$ driver genes, $D=295$ drugs), the model identifies 126 gene-drug associations (0.195\% of 64{,}605 pairs), concentrated in EZH2 (45 drugs, all sensitivity-direction, mean effect $-0.911$ $\ln$IC50) and KMT2D (36 drugs, all sensitivity-direction, mean effect $-0.496$ $\ln$IC50). These markers show external support in an independent PRISM screen (1{,}518 compounds), with KMT2D achieving complete directional replication (36/36) and EZH2 partial replication (8/12). Five-fold cross-validated predictive log-likelihood confirms each prior layer's value: the full model outperforms the no-network ablation by $+3{,}109$ log-units per fold and the no-horseshoe ablation by $+14{,}039$ log-units, consistently across folds. Simulations under three scenarios show the full model achieves the highest precision and lowest false-discovery rate throughout, while the network prior improves sensitivity recovery under network-structured signal. A tissue-stratified extension identifies coherent subgroup refinements, including lung-specific EGFR-inhibitor sensitivity and skin-specific BRAF-Dabrafenib sensitivity. These results show the framework identifies sparse, interpretable, externally supported drug-sensitivity markers while enabling principled investigation of tissue-specific departures from shared effects.
\end{abstract}
\end{abstract}


\section{Introduction}
\label{sec:introduction}

High-dimensional pharmacogenomic studies often seek to identify sparse
associations between genomic alterations and quantitative
treatment-response phenotypes. In this setting, the scientific goal is
not simply to optimize statistical prediction, but to characterize which
genomic features are credible candidate markers or modifiers of
measured drug sensitivity across a large collection of candidate
feature-outcome relationships. Cancer
drug-sensitivity screens provide a particularly important example of
this problem: hundreds of cancer cell lines are screened against
hundreds or thousands of compounds while mutation status is recorded
across hundreds of genes, so the number of candidate mutation-drug
response associations can greatly exceed the number of independent
biological samples available to estimate them
\citep{garnett2012,barretina2012,iorio2016,corsello2020}.
An analogous challenge arises in other high-dimensional biomedical and
biological settings where sparse genomic signals must be recovered under
a known relational structure among predictors, including patient-level
pharmacogenomics, genome-wide association studies where prior pathway
knowledge may inform inference, organoid and patient-derived xenograft
screens, agricultural genomics studies predicting phenotypic response
from marker panels, and microbial genomics settings where gene presence
or absence predicts antibiotic resistance
\citep{kaplan2020bayesian,driehuis2020establishment,gao2015high,su2019genome,danilevicz2022plant}.
The statistical framework developed here is applicable to any of these
settings; we develop and validate it using cancer drug-sensitivity data.

Two structural features make naive regression approaches unreliable in
this setting. First, most candidate mutations are not expected to be
relevant to a given drug's mechanism of action. A model that estimates
an unconstrained, unshrunk effect for every gene--drug pair will
therefore overfit unless it incorporates a principled sparsity
mechanism. Second, genes do not operate independently. Genes that
participate in the same biological pathway, or that are co-selected
through shared mutational processes, may have correlated mutation
patterns and biologically related effects. This raises the question of
whether known pathway or network structure should be used to pool
statistical strength across genes rather than being ignored or treated
only as a nuisance source of correlation
\citep{lili2008,li2010variable,kaplan2020bayesian}.

Bayesian hierarchical models are a natural framework for addressing
these two issues simultaneously. A global--local shrinkage prior can
encode sparsity directly in the posterior, avoiding a separate
variable-selection step, while a network-structured prior can
incorporate known biological relationships among predictors
\citep{carvalho2010,rue2005gaussian,lili2008}.
In the present work, we develop such a model for sparse mutation-drug
response association mapping, with cancer drug sensitivity as the
primary application. The model combines a pathway-informed Gaussian
Markov random field (GMRF) prior, which encourages related genes to
have correlated drug-response effects, with a global--local horseshoe
shrinkage layer, which allows most mutation effects to be strongly
shrunk while permitting supported signals to escape shrinkage
\citep{rue2005gaussian,carvalho2010}. The framework also
accommodates an optional tissue-stratification layer that allows each
gene's drug-response effect to vary across cancer lineages while
pooling toward a shared pan-cancer mean
(Section~\ref{subsec:methods-model-group}), enabling joint
characterization of pan-cancer associations and their tissue-specific
heterogeneity.

Two existing lines of work motivate key components of this framework,
but neither addresses the full problem considered here.
\citet{samorodnitsky2020} fit a pan-cancer, polygenic Bayesian
hierarchical model relating somatic mutation status across 50 genes to
survival outcomes across 27 cancer types drawn from The Cancer Genome
Atlas. Their model allows each gene's effect to vary by cancer type
while pooling toward a shared mean, a hierarchical pooling structure
that successfully borrows strength across small and heterogeneous
subgroups and which the present model adopts as an optional layer
(Section~\ref{subsec:methods-model-group}). However, their model treats
genes as exchangeable: no two genes' coefficients are correlated in the
prior, regardless of whether the genes participate in the same
biological pathway. Moreover, gene importance is ultimately assessed
through a forward-selection procedure outside the Bayesian posterior,
evaluated by cross-validated predictive likelihood rather than
posterior uncertainty.

\citet{pham2018} take the opposite approach to network structure. Their
three-level model links gene expression to latent pathway factors via a
confirmatory factor analysis layer, then places a conditional
autoregressive (CAR) prior on those latent factors using a
pathway--pathway adjacency matrix constructed from shared Gene Ontology
annotation. This is, to our knowledge, one of the clearest existing
uses of an explicit biological network as a Bayesian prior structure in
this literature, and the pathway-derived adjacency matrix used in the
present model follows the same general construction logic
(Section~\ref{subsec:model-network-layer}). However, their CAR layer
describes correlation among latent expression factors for a biological
sample, not correlation among mutation-effect coefficients across genes.
Their outcome is gene expression under drug perturbation rather than
quantitative drug-sensitivity response, so the modelling target is
fundamentally different.

A third related model, \citet{tansey2022}'s Bayesian Tensor Filtering,
shares statistical strength across many cell lines and many drugs
through learned low-dimensional embeddings. This model is closely
related on the drug-response side, but its notion of structure is an
unsupervised latent embedding rather than a biologically interpretable
gene or pathway network, and mutation status is not incorporated as a
covariate.

The present model is, to our knowledge, the first to combine in a
single coherent Bayesian posterior: (i)~mutation-level genetic
structure as the predictor; (ii)~quantitative drug response -- rather
than survival or gene expression -- as the outcome; (iii)~an explicit
pathway-network prior governing how mutation effects on that outcome
are pooled across genes; and (iv)~a continuous sparsity mechanism
built into the posterior rather than applied as a post-hoc selection
step. Two additional bodies of prior work anticipate individual
components of this construction but not their combination.
\citet{lili2008} introduced a frequentist network-constrained
regularization penalty for genomic regression that uses the graph
Laplacian of a gene network as a smoothness penalty on regression
coefficients -- the same mathematical object underlying the GMRF prior
used here, but formulated as penalized likelihood rather than Bayesian
posterior inference and without a global--local sparsity layer.
\citet{stingo2011} developed a fully Bayesian variable-selection model
using a Markov random field prior over a gene--pathway network to
inform pathway and gene selection for a continuous outcome, directly
analogous in spirit to the network layer proposed here. Their model,
however, places the network-informed prior on selection indicators in a
spike-and-slab framework rather than on continuous mutation-effect
coefficients, and is not designed for the per-drug gene-effect
structure required here.

The present model can therefore be viewed as combining the
network-informed regularization philosophy of \citet{lili2008} and
\citet{stingo2011} with a continuous global--local shrinkage layer in
the tradition of \citet{mitchellbeauchamp1988},
\citet{georgemcculloch1993}, \citet{polsonscott2010}, and
\citet{carvalho2010}. By using continuous horseshoe shrinkage rather
than discrete spike-and-slab inclusion indicators, the model avoids a
combinatorial search over gene--drug inclusion patterns while still
producing posterior summaries that characterize the full uncertainty
over mutation--response associations.

We apply this framework to cancer drug-sensitivity data by modelling
quantitative drug response as a function of somatic mutation status. For
each drug, the vector of mutation effects is regularized by two coupled
prior components. A Gaussian Markov random field prior, defined on a
curated cancer-pathway network, induces dependence among effects for
pathway-connected genes and allows information to be shared across
biologically related predictors. A global--local horseshoe prior imposes
sparsity by shrinking most gene--drug effects strongly toward zero while
retaining heavy-tailed local scales for effects supported by the data.
This construction reflects the biological premise that true
pharmacogenomic signals are sparse, but that nonzero effects are more
likely to occur in pathway-organized patterns than as isolated,
independent events.

Two practical considerations shaped the model's final form. First, a
model with both a network prior and a horseshoe prior places two
shrinkage mechanisms on the same coefficient vector. In an initial
specification where both global variance components were assigned the
same Inverse-Gamma prior family, the corresponding shrinkage parameters
collapsed toward zero in a mutually reinforcing feedback loop. This
behaviour is consistent with known pathologies of weakly informative
variance-component priors in hierarchical models
\citep{gelman2006}, and is made more severe here because the network
and horseshoe layers both act on the same gene--drug coefficient
vector. We resolve this by adapting the Group Inverse-Gamma Gamma
(GIGG) prior of \citet{boss2021gigg}: we place Gamma priors on the
precisions of the global network and horseshoe shrinkage parameters
while retaining the Inverse-Gamma representation for the local
gene-specific horseshoe parameter
(Section~\ref{subsec:model-hyperpriors-primary}). This asymmetry stabilizes
both global shrinkage components and preserves conjugate Gibbs updates.
Second, the model is evaluated not only as a methodological
construction but also as a tool for identifying biologically
interpretable mutation-drug response associations. In the primary
cancer drug-sensitivity analysis, two chromatin-regulatory genes,
EZH2 and KMT2D, emerge as dominant mutation-level predictors of drug
response. These associations are supported by mechanistic evidence
from the cancer pharmacology literature, persist under ablation
comparisons, and demonstrate external robustness in PRISM/DepMap
analyses. We therefore use the cancer drug-sensitivity application
both to demonstrate the model's feasibility and to show that
network-structured Bayesian shrinkage produces sparse, interpretable,
and externally supported mutation--response associations.

The remainder of the paper is organized as follows.
Section~\ref{sec:methods} develops the model framework,
presenting Model~I, the primary network--horseshoe specification, and
Model~II, the tissue-group extension, together with the computational
details of the conjugate Gibbs sampler. Section~\ref{subsec:results-simulation}
presents simulation studies designed to evaluate operating
characteristics under known truth, including sensitivity, precision,
false-discovery control, and interval
coverage across distinct data-generating scenarios. 
Section~\ref{subsec:application} applies the proposed framework to cancer
pharmacogenomic data, including the primary GDSC2 analysis, ablation
comparisons, external robustness assessment using PRISM/DepMap data,
and the tissue-stratified group-layer analysis. 
Section~\ref{sec:discussion} discusses the findings in relation to the
biological literature, model behavior, limitations, and broader
methodological implications. Section~\ref{sec:conclusion} concludes the paper.

\section{Model Framework}
\label{sec:methods}

\subsection{Model I: Shared-Effect Network--Horseshoe Model}
\label{subsec:methods-model-primary}

Let $i=1,\dots,N$ index biological samples or experimental units
(e.g., cancer cell lines), $d=1,\dots,D$ index drugs, and
$g=1,\dots,G$ index genes. Let $m_{ig}\in\{0,1\}$ denote the mutation
status of gene $g$ in sample $i$. The response $y_{id}$, such as a
log-transformed drug-sensitivity measure, is modeled as
\begin{equation}
y_{id} \mid \mu_{id}, \sigma_d^2
\;\sim\;
\text{Normal}(\mu_{id},\;\sigma_d^2),
\qquad
\mu_{id}
=
\alpha_d
+
\sum_{g=1}^G m_{ig}\,\widetilde{\beta}_{gd},
\label{eq:methods-likelihood}
\end{equation}
\noindent where $\alpha_d$ is a drug-specific intercept and
$\widetilde{\beta}_{gd}$ is the shared effect of a somatic mutation in
gene $g$ on response to drug $d$.

\subsubsection{Network-Structured Prior}
\label{subsec:model-network-layer}

Let $W$ denote a fixed, undirected gene--gene adjacency matrix defined
over the subset of $G'$ genes with available network information. For
each drug $d$, the corresponding sub-vector
$\boldsymbol\beta_{\cdot d}^{(\text{net})}$ receives a Gaussian Markov
random field (GMRF) prior \citep{besag1974,besagyorkmollie1991}:
\begin{equation}
\boldsymbol\beta_{\cdot d}^{(\text{net})} \mid \kappa_d^2
\;\sim\;
\text{Normal}\!\Big(\mathbf{0},\;
\kappa_d^2 (D_W - \delta W)^{-1}\Big),
\label{eq:methods-gmrf}
\end{equation}
where $D_W=\mathrm{diag}(W\mathbf{1})$ is the degree matrix and
$\delta\in(0,1)$ is a fixed smoothing parameter chosen so that
$D_W-\delta W$ is positive-definite. Fixing $\delta$, rather than
estimating it via the adaptive Metropolis step used by \citet{pham2018}
for an analogous CAR scaling parameter, is the first of two deliberate
simplifications made to keep every step of posterior computation a
closed-form conjugate draw.

\subsubsection{Global--Local Horseshoe Prior}
\label{subsec:model-horseshoe-layer}

Because most candidate mutations are not relevant to a given drug's
mechanism, $\widetilde{\beta}_{gd}$ additionally receives a
global--local horseshoe prior
\citep{polsonscott2010,carvalho2010,makalic2016}, a
continuous-shrinkage alternative to discrete spike-and-slab variable
selection \citep{mitchellbeauchamp1988,georgemcculloch1993} that avoids
a combinatorial search over inclusion indicators:
\begin{equation}
\widetilde{\beta}_{gd}
\mid
\lambda_{gd}^2,\zeta_d^2
\;\sim\;
\text{Normal}(0,\lambda_{gd}^2\zeta_d^2),
\qquad
\lambda_{gd}^2 \mid \nu_{gd}
\;\sim\;
\text{Inverse-Gamma}\!\left(\frac{1}{2},\frac{1}{\nu_{gd}}\right),
\label{eq:methods-horseshoe}
\end{equation}
with
\[
\nu_{gd}
\sim
\text{Inverse-Gamma}\!\left(\frac{1}{2},1\right).
\]
Here $\lambda_{gd}$ is the local gene--drug-specific shrinkage parameter
and $\zeta_d$ is the global drug-specific shrinkage parameter. The GMRF
and horseshoe prior factors act on the same shared coefficient
$\widetilde{\beta}_{gd}$: the network prior induces dependence across
pathway-connected genes, while the horseshoe layer controls sparse
global--local shrinkage.

\subsubsection{Hyperpriors and Variance-Component Parameterization}
\label{subsec:model-hyperpriors-primary}

For Model I, we assign weakly informative hyperpriors to the
drug-specific network and sparsity variance components through their
precision parameterization:
\begin{align}
\tau_{\kappa,d}=1/\kappa_d^2
&\sim \mathrm{Gamma}(1,1),
\label{eq:prior-kappa} \\
\tau_{\zeta,d}=1/\zeta_d^2
&\sim \mathrm{Gamma}(1,1),
\label{eq:prior-zeta} \\
\nu_{gd}
&\sim \mathrm{Inverse\text{-}Gamma}\left(\frac{1}{2},1\right),
\label{eq:prior-nu} \\
\alpha_d
&\sim \mathrm{Normal}(0,100),
\label{eq:prior-alpha} \\
\sigma_d^2
&\sim \mathrm{Inverse\text{-}Gamma}(0.01,0.01).
\label{eq:prior-sigma2}
\end{align}
Gamma distributions are parameterized by shape and rate, and
Inverse-Gamma distributions are also parameterized by shape and rate.

The Gamma priors on the precision parameters
$\tau_{\kappa,d}$ and $\tau_{\zeta,d}$ are used instead of highly
diffuse Inverse-Gamma priors directly on the corresponding variance
components. This precision-based parameterization follows the
stabilization principle underlying the Group Inverse-Gamma Gamma prior
construction \citep{boss2021gigg} and avoids undesirable boundary
behavior that can arise from near-improper Inverse-Gamma priors on
variance components \citep{gelman2006}. In particular, placing highly
diffuse Inverse-Gamma priors directly on $\kappa_d^2$ and $\zeta_d^2$
can produce unstable variance updates and excessive shrinkage toward
zero for some drug-specific effects. Parameterizing these components
through their precisions yields stable conjugate updates while retaining
adaptive drug-specific control over network smoothing and global
sparsity. Full conditional derivations are provided in Supplementary
Section~S2.

\subsubsection{Joint Posterior for Model I}
\label{subsec:model-joint-primary}

The joint posterior for
$\Theta_{\text{I}} = \{\widetilde{\beta}_{gd}, \alpha_d, \sigma_d^2,
\kappa_d^2, \zeta_d^2, \lambda_{gd}^2, \nu_{gd}\}.$

\begin{align}
\pi(\Theta_{\text{I}} \mid \mathbf{y}) \;\propto\;&
\prod_{i,d}
  \text{N}(y_{id};\;\mu_{id},\;\sigma_d^2)
\cdot
\prod_d
  \text{N}(\boldsymbol\beta_{\cdot d}^{(\text{net})};\;
  \mathbf{0},\;\kappa_d^2(D_W-\delta W)^{-1})
\nonumber\\
&\times\;
\prod_{g,d}
 \text{N}(\widetilde{\beta}_{gd};\;0,\;\lambda_{gd}^2\zeta_d^2)
  \cdot
  \text{IG}(\lambda_{gd}^2;\;\tfrac{1}{2},\;\nu_{gd}^{-1})
  \cdot
  \text{IG}(\nu_{gd};\;\tfrac{1}{2},\;1)
\nonumber\\
&\times\;
\prod_d
  \text{Gamma}(\tau_{\kappa,d};\;1,\;1)
  \cdot
  \text{Gamma}(\tau_{\zeta,d};\;1,\;1)
  \cdot
  \text{N}(\alpha_d;\;0,\;100)
  \cdot
  \text{IG}(\sigma_d^2;\;0.01,\;0.01).
\label{eq:joint-primary}
\end{align}

Full derivations of all conditional posterior distributions needed for the Gibbs sampler steps are in Supplementary
Section~S2.

\subsection{Model II: Tissue-Group Extension}
\label{subsec:methods-model-group}

As an extension, Model II allows mutation effects to vary across
predefined groups, such as tissue of origin. Let
$c_i\in\{1,\dots,K\}$ denote the group membership of sample $i$. The
likelihood becomes
\begin{equation}
y_{id} \mid \cdot
\;\sim\;
\text{Normal}(\mu_{id}^{(c_i)},\;\sigma_d^2),
\qquad
\mu_{id}^{(c_i)}
=
\alpha_{c_i d}
+
\sum_{g=1}^G m_{ig}\,\beta_{gd}^{(c_i)},
\label{eq:methods-likelihood-group}
\end{equation}
where $\alpha_{kd}$ is a group-and-drug-specific intercept and
$\beta_{gd}^{(k)}$ is the group-specific gene effect. Each
group-specific effect is pooled toward a shared mean
$\widetilde{\beta}_{gd}$:

\begin{equation}
\beta_{gd}^{(k)} \mid \tilde\beta_{gd}, \eta_{gd}^2
\;\sim\;
\text{Normal}(\tilde\beta_{gd},\;\eta_{gd}^2),
\qquad k = 1,\dots,K,
\label{eq:methods-group-prior}
\end{equation}
where $\eta_{gd}^2$ is the group-heterogeneity variance. The shared
mean $\tilde\beta_{gd}$ then receives the same GMRF network prior
(\ref{eq:methods-gmrf}) and horseshoe prior
(\ref{eq:methods-horseshoe}) as in Model I.

\subsubsection{Preprocessing: Minimum Group-Size Requirement}
\label{subsec:methods-group-preprocessing}

The group-specific update for group $k$ and outcome $d$ requires
sampling a $G$-dimensional coefficient vector using the precision matrix
\begin{equation}
A_{kd}
=
\frac{1}{\sigma_d^2}M_{kd}' M_{kd}
+
\mathrm{diag}(1/\eta_{gd}^2)
+
\varepsilon I,
\label{eq:Akd}
\end{equation}
where $M_{kd}$ is the $n_{kd}\times G$ design submatrix for the
$n_{kd}$ observations in group $k$ with observed outcome $d$. The matrix
$M_{kd}' M_{kd}$ has rank at most $\min(n_{kd},G)$, so when
$n_{kd}<G$, the likelihood contribution alone identifies only a
low-dimensional subspace of the $G$ coefficient dimensions. The remaining
directions are stabilized by the prior precision
$\mathrm{diag}(1/\eta_{gd}^2)$ and by the ridge term $\varepsilon I$.

To avoid fitting group-specific effects for groups with very limited
within-group information, we impose a minimum group-size requirement
$n_{\min}$. Groups with $n_k<n_{\min}$ are merged into a single
\textup{``Other''} category prior to fitting the group-layer model. This
preprocessing step ensures that each retained group contributes enough
observations to support stable group-specific coefficient updates while
preserving all observations in the likelihood.

In the analyses in Section~\ref{subsec:application}, we set $n_{\min}=50$, a conservative threshold
chosen to ensure adequate within-group replication relative to the
$G=219$ predictors in the real-data application. More generally, users
applying the group-layer model to other datasets should choose
$n_{\min}$ based on the number of predictors, the number of outcomes, and
the degree of within-group replication. Groups below the chosen threshold
should either be merged into an \textup{``Other''} category or the model
should be fitted without the group layer.

\subsubsection{Hyperpriors for Model II}
\label{subsec:methods-hyperpriors-group}

All hyperpriors for $\tilde\beta_{gd}$, $\kappa_d^2$, $\zeta_d^2$,
$\lambda_{gd}^2$, $\sigma_d^2$, and $\alpha_{kd}$ are identical to
Model I (\ref{eq:prior-kappa}--\ref{eq:prior-sigma2}). 
The group-heterogeneity variance $\eta_{gd}^2$ requires a proper prior.
Using Inverse-Gamma$(0.01, 0.01)$ -- analogous to the near-improper
prior on $\sigma_d^2$ -- caused numerical overflow in $A_{kd}$
(\ref{eq:Akd}) because $\eta_{gd}^2$ could take arbitrarily
large values early in the chain, driving
$\mathrm{diag}(1/\eta_{gd}^2) \to 0$ and leaving $A_{kd}$
near-singular. We therefore assign the proper prior

\begin{equation}
\eta_{gd}^2
\;\sim\;
\text{Inverse-Gamma}(a_\eta,b_\eta),
\label{eq:prior-eta2}
\end{equation}

\noindent With $a_\eta=2$ and $b_\eta=0.5$, this prior has mode
$b_\eta/(a_\eta+1)=0.167$ and mean $b_\eta/(a_\eta-1)=0.5$., placing the group-level standard deviation $\sqrt{\eta_{gd}^2}$
in the range $[0.2,\, 0.7]$ a priori -- a scale consistent with
documented tissue-level heterogeneity in cancer drug response.. The conjugate conditional posterior is

\begin{equation}
\eta_{gd}^2 \mid \cdot \;\sim\;
\text{Inverse-Gamma}\!\left(
a_\eta + \frac{K}{2},\;\;
b_\eta + \frac{1}{2}\sum_{k=1}^K
\bigl(\beta_{gd}^{(k)} - \tilde\beta_{gd}\bigr)^2
\right).
\label{eq:update-eta2}
\end{equation}

Additionally, a ridge term $\varepsilon I$ with $\varepsilon = 10^{-4}$
is added to $A_{kd}$ before Cholesky factorization
(\ref{eq:Akd}). This is equivalent to a weak additional
$\text{Normal}(0,\,\varepsilon^{-1})$ prior on each group-specific
coefficient and guarantees positive-definiteness of $A_{kd}$ regardless
of the realized value of $\eta_{gd}^2$.

The hyperpriors for Model II therefore differ from Model I in exactly
two respects: (i) $\eta_{gd}^2 \sim \text{IG}(a_\eta,b_\eta)$ replaces the
absence of this parameter in Model I; and (ii) tissue-specific
intercepts $\alpha_{kd} \sim \text{Normal}(0,100)$ replace the
drug-specific intercept $\alpha_d$ of Model I.

\subsubsection{Joint Posterior for Model II}
\label{subsec:methods-joint-group}

The joint posterior for
$\Theta_{\text{II}} = \{\beta_{gd}^{(k)}, \tilde\beta_{gd},
\alpha_{kd}, \sigma_d^2, \kappa_d^2, \zeta_d^2,
\lambda_{gd}^2, \nu_{gd}, \eta_{gd}^2\}$ is

\begin{align}
p(\Theta_{\text{II}} \mid \mathbf{y}) \;\propto\;&
\prod_{i,d}
  \text{N}\!\left(y_{id};\;
  \alpha_{c_id} + \textstyle\sum_g m_{ig}\beta_{gd}^{(c_i)},\;
  \sigma_d^2\right)
\cdot
\prod_{k,g,d}
  \text{N}(\beta_{gd}^{(k)};\;\tilde\beta_{gd},\;\eta_{gd}^2)
\nonumber\\
&\times\;
\prod_{g,d}
  \text{IG}(\eta_{gd}^2;\;a_\eta,\;b_\eta)
\cdot
\prod_d
  \text{N}(\tilde{\boldsymbol\beta}_{\cdot d}^{(\text{net})};\;
  \mathbf{0},\;\kappa_d^2(D_W-\delta W)^{-1})
\nonumber\\
&\times\;
\prod_{g,d}
  \text{N}(\tilde\beta_{gd};\;0,\;\lambda_{gd}^2\zeta_d^2)
  \cdot
  \text{IG}(\lambda_{gd}^2;\;\tfrac{1}{2},\;\nu_{gd}^{-1})
  \cdot
  \text{IG}(\nu_{gd};\;\tfrac{1}{2},\;1)
\nonumber\\
&\times\;
\prod_d
  \text{Gamma}(\tau_{\kappa,d};\;1,\;1)
  \cdot
  \text{Gamma}(\tau_{\zeta,d};\;1,\;1)
  \cdot
  \prod_d\text{IG}(\sigma_d^2;\;0.01,\;0.01)
\nonumber\\
&\times\;
\prod_{k,d}
  \text{N}(\alpha_{kd};\;0,\;100).
\label{eq:joint-group}
\end{align}

Full derivations of all Gibbs sampler steps for Model II, including the
group-specific updates for $\alpha_{kd}$, $\beta_{gd}^{(k)}$,
$\eta_{gd}^2$, and $\widetilde{\beta}_{gd}$, are provided in
Supplementary Section~S2.2. The corresponding Model I derivations are
provided in Supplementary Section~S2.1.

\subsection{Posterior Computation}
\label{subsec:methods-computation}

Both models are fitted by Gibbs sampling
\citep{gemangeman1984,gelfandsmith1990}. The hierarchical specification
was chosen so that the full conditional distributions for the regression
coefficients, intercepts, residual variances, network-scale parameters,
and global--local shrinkage components are available in closed form. In
particular, the required updates are Normal, Inverse-Gamma, or Gamma
draws, yielding a fully conjugate sampler with no Metropolis--Hastings
step. This is in contrast to related network-structured Bayesian models
that require adaptive Metropolis updates or more general-purpose
posterior samplers for at least one parameter
\citep{samorodnitsky2020,pham2018}. Full conditional posterior
distributions and derivations are provided in Supplementary
Section~S2.

For continuous drug-response outcomes, the Gaussian likelihood in
\eqref{eq:methods-likelihood} leads directly to conjugate
Gaussian updates for the regression coefficients and intercepts. The
current sampler is therefore designed for continuous response summaries,
such as log-transformed IC50, AUC, or viability scores. It cannot be
applied directly to a binary responder/non-responder outcome without
modifying the likelihood, because a Bernoulli likelihood would break the
Gaussian conjugacy used in the coefficient updates.

If the response were instead a binary responder indicator, one natural
extension would be to use a probit formulation with the Albert--Chib
data-augmentation scheme \citep{albertchib1993}. Specifically, one could
introduce a latent continuous response $z_{id}$ such that
\[
y_{id} = \mathbf{1}\{z_{id}>0\}, \qquad
z_{id} \mid \cdot \sim \mathrm{Normal}(\mu_{id},1),
\]
where $\mu_{id}$ is defined using the same linear predictor as in
\eqref{eq:methods-likelihood}. Conditional on the binary
outcome, the latent variable $z_{id}$ is sampled from a Normal
distribution truncated above or below zero. Conditional on the augmented
latent responses $\{z_{id}\}$, the remaining model again has a Gaussian
working likelihood, so the same conjugate updates for the regression
effects, network prior, shrinkage parameters, and group-layer extension
can be retained with minor modifications. The residual variance is fixed
to one in the probit model for identifiability. We do not pursue this
binary-response extension here, because the present application focuses
on continuous drug-sensitivity summaries.

Two computational simplifications are used throughout. First,
parameter-independent cross-product terms are precomputed within each
drug-specific update whenever possible. Second, a single Cholesky
factorization is reused both to solve for the posterior mean and to
generate the Gaussian random draw. For Model~I, these simplifications
substantially reduce the cost of each drug-specific coefficient update.
Model~II retains the same conjugate structure but is computationally more
expensive because, for each drug and MCMC iteration, it additionally
updates $K$ group-specific $G$-dimensional coefficient vectors, requiring
up to $K$ Cholesky factorizations of $G\times G$ precision matrices per
drug.

\section{Simulation Study}
\label{subsec:results-simulation}

Because ground truth is unknown in any real data, we
evaluate the model's operating characteristics under three controlled
data-generating scenarios where the true gene--drug associations are
known. The simulation design is fully described in Supplementary Section~S4;
we summarize it here.

\subsection{Simulation Design}
\label{subsec:simulation-design}

\textbf{Simulation design.} For each replicate, $N=400$ cell lines,
$G=150$ genes, $D=50$ drugs, and $G'=100$ network-mapped genes are
used. Mutation indicators $m_{ig} \sim \text{Bernoulli}(p_g)$ with
$p_g \sim \text{Uniform}(0.01,0.25)$. Drug-specific intercepts
$\alpha_d \sim \text{Normal}(0,\sigma_\alpha^2)$. Responses follow
$y_{id} = \alpha_d + \sum_g m_{ig}\beta_{gd} + \varepsilon_{id}$,
$\varepsilon_{id} \sim \text{Normal}(0,1)$. Three scenarios vary how
true effects $\beta_{gd}$ are generated.

\textbf{Scenario 1 (Sparse independent):} A sparse set $\mathcal{S}_d$
of genes is selected independently of the network per drug; effects
$\beta_{gd}=s_{gd}b_{gd}$, $b_{gd} \sim
\text{Normal}(\mu_\beta,\tau_\beta^2)$, $s_{gd}\in\{-1,+1\}$. No
pathway structure in the truth. Tests whether the horseshoe controls
false discoveries when the network prior is irrelevant.

\textbf{Scenario 2 (Network-structured):} For each drug $d$, an active
network module $\mathcal{P}_d$ is generated by seeding a random
network-mapped gene and iteratively adding adjacent genes until the
desired module size is reached. To induce correlated effects within the
active module, network-correlated perturbations are drawn as
$
\boldsymbol{u}_{\cdot d}
\sim
\text{Normal}
\left(
\mathbf{0},
\tau_{\text{net}}^2
(D_W-\delta W+\varepsilon I)^{-1}
\right),
$
where $W$ is the gene adjacency matrix and $D_W$ is the corresponding
degree matrix. True effects are then set to
$
\beta_{gd}=s_d\theta+u_{gd}, \; g\in\mathcal{P}_d,
$
and $\beta_{gd}=0$ for $g\notin\mathcal{P}_d$, with $s_d\in\{-1,+1\}$
determining the common direction of the drug-specific module effect.
This scenario tests whether the GMRF prior improves recovery and
false-discovery control when true mutation--response effects are
localized within connected pathway modules.

\textbf{Scenario 3 (Dense weak noise with sparse strong signals):}
Sparse strong effects ($\beta_{gd}=s_{gd}b_{gd}^{(S)}$,
$b_{gd}^{(S)} \sim \text{Normal}(\mu_S,\tau_S^2)$) are mixed with
many weak effects ($\beta_{gd}=b_{gd}^{(W)}$,
$b_{gd}^{(W)} \sim \text{Normal}(0,\tau_W^2)$, $\tau_W \ll \tau_S$).
Only the strong set is marked as truly associated. This tests whether the
horseshoe distinguishes sparse strong signals from weak nuisance
effects.

A gene--drug pair is classified as selected when its 95\% posterior
credible interval excludes zero. We report sensitivity, precision, FDR,
coefficient RMSE (overall and decomposed by true-association status),
and 95\% CI coverage. The simulation sparsity rate is approximately
3.3\% (250 of 7{,}500 pairs), intentionally higher than the real-data
rate (0.20\%); an extremely sparse simulation would yield near-zero
sensitivity for all models.

\subsection{Simulation Results}
\label{subsec:simulation-results}

\begin{figure}[H]
\centering
\includegraphics[width=0.7\textwidth]{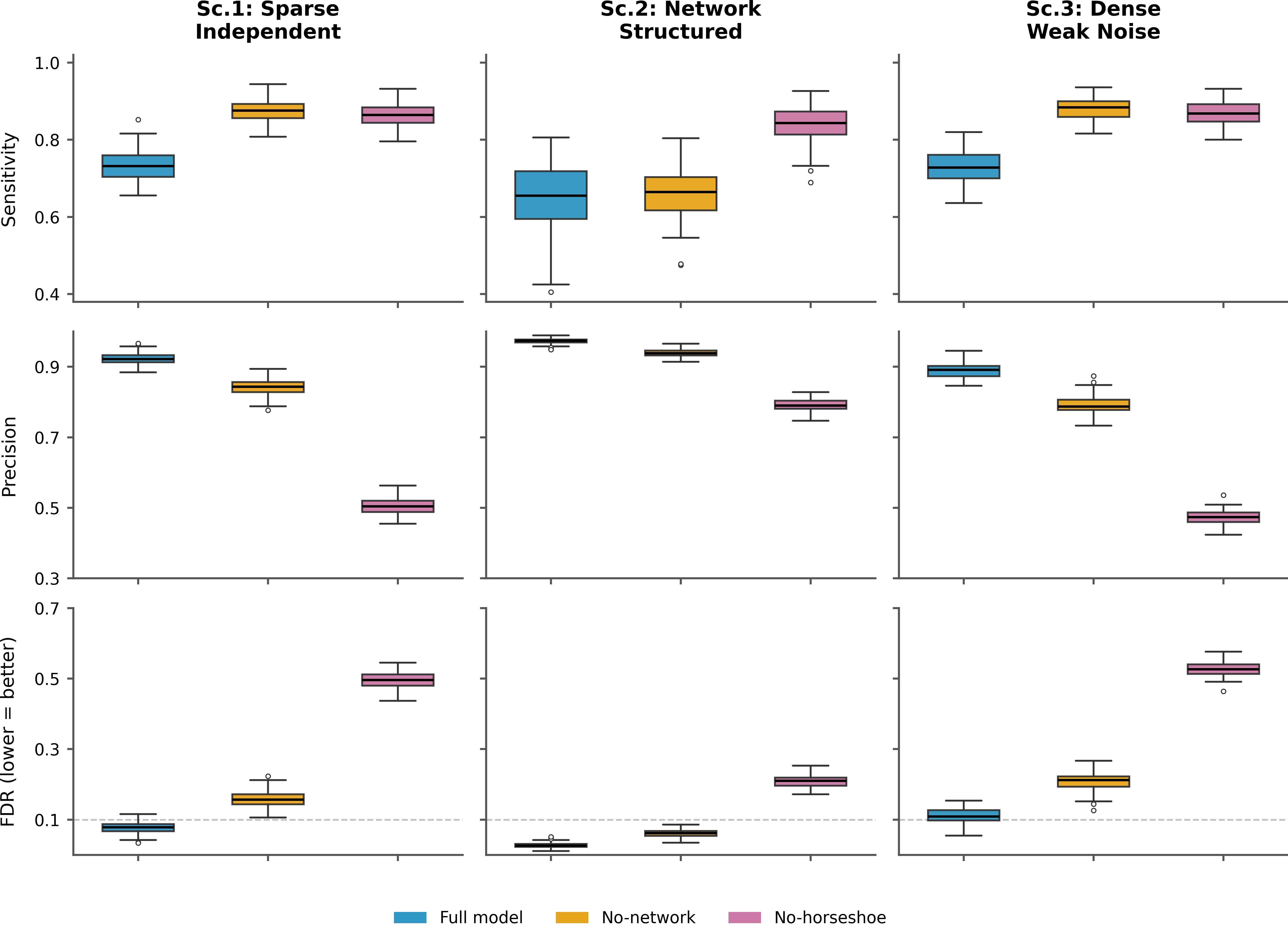}
\caption{Simulation study results across three scenarios
($n=100$ replicates each). \emph{Rows:} sensitivity, precision, and FDR
(lower is better for FDR; dashed reference line at FDR = 0.10).
\emph{Columns:} Scenario 1 (sparse independent), Scenario 2
(network-structured), and Scenario 3 (dense weak noise with sparse strong
signals). Colors denote the full model, no-network ablation, and
no-horseshoe ablation.} 
\label{fig:sim-boxplots}
\end{figure}

Four findings emerge from the simulation studies results. First, the full model achieves the highest precision and
lowest FDR in all three scenarios (Table~\ref{tab:simulation},
Figure~\ref{fig:sim-boxplots}). The full model selects fewer gene--drug
pairs, but a larger fraction of its selected pairs are true
associations. The no-network model is more sensitive but this comes at
the cost of substantially higher FDR. The no-horseshoe model has FDR
near or above 0.50 in Scenarios~1 and~3, indicating substantial
over-selection when the global--local sparsity layer is removed.

\begin{table}[H]
\caption{Simulation results at the 95\% credible-interval operating point.}
\label{tab:simulation}
\centering
\small
\setlength{\tabcolsep}{3.5pt}
\begin{threeparttable}
\begin{tabular}{llccccc}
\toprule
Scenario & Model & Sensitivity & Precision & FDR &
  RMSE$_\beta^{(\text{all})}$ & Coverage \\
\midrule
\multirow{3}{*}{1: Sparse Indep.}
 & Full
   & 0.734 (0.039) & \textbf{0.923 (0.017)} & \textbf{0.077 (0.017)}
   & 0.122 (0.005) & 0.977 \\
 & No-net
   & \textbf{0.874 (0.029)} & 0.843 (0.023) & 0.157 (0.023)
   & \textbf{0.112 (0.015)} & 0.992 \\
 & No-HS
   & 0.864 (0.031) & 0.505 (0.021) & 0.495 (0.021)
   & 0.180 (0.010) & 0.957 \\
\midrule
\multirow{3}{*}{2: Network Struct.}
 & Full
   & 0.654 (0.084) & \textbf{0.972 (0.008)} & \textbf{0.028 (0.008)}
   & \textbf{0.127 (0.009)} & 0.953 \\
 & No-net
   & 0.659 (0.060) & 0.939 (0.010) & 0.061 (0.010)
   & 0.130 (0.006) & 0.981 \\
 & No-HS
   & \textbf{0.841 (0.046)} & 0.791 (0.016) & 0.209 (0.016)
   & 0.174 (0.011) & 0.958 \\
\midrule
\multirow{3}{*}{3: Dense Weak}
 & Full
   & 0.732 (0.039) & \textbf{0.890 (0.021)} & \textbf{0.110 (0.021)}
   & 0.125 (0.005) & 0.967 \\
 & No-net
   & \textbf{0.879 (0.027)} & 0.791 (0.025) & 0.209 (0.025)
   & \textbf{0.112 (0.006)} & 0.988 \\
 & No-HS
   & 0.868 (0.028) & 0.473 (0.020) & 0.527 (0.020)
   & 0.181 (0.012) & 0.956 \\
\bottomrule
\end{tabular}
\begin{tablenotes}[flushleft]
\footnotesize
\item \textbf{Note.} Results are based on $n=100$ simulation replicates. Values are mean (SD), except coverage, which is reported as the mean empirical 95\% credible-interval coverage. Scenario labels denote sparse independent effects, network-structured effects, and dense weak noise with sparse strong signals. No-net = No-network model; No-HS = No-horseshoe model. RMSE denotes RMSE$_\beta^{(\mathrm{all})}$ over all $G \times D = 7{,}500$ gene--drug pairs. Bold entries mark the best-performing model per metric within each scenario. Full model results at $n \in \{1,10,20,50, 100\}$ are in Supplementary Table~2.
\end{tablenotes}
\end{threeparttable}
\end{table}

\begin{figure}[H]
\centering
\includegraphics[width=\textwidth]{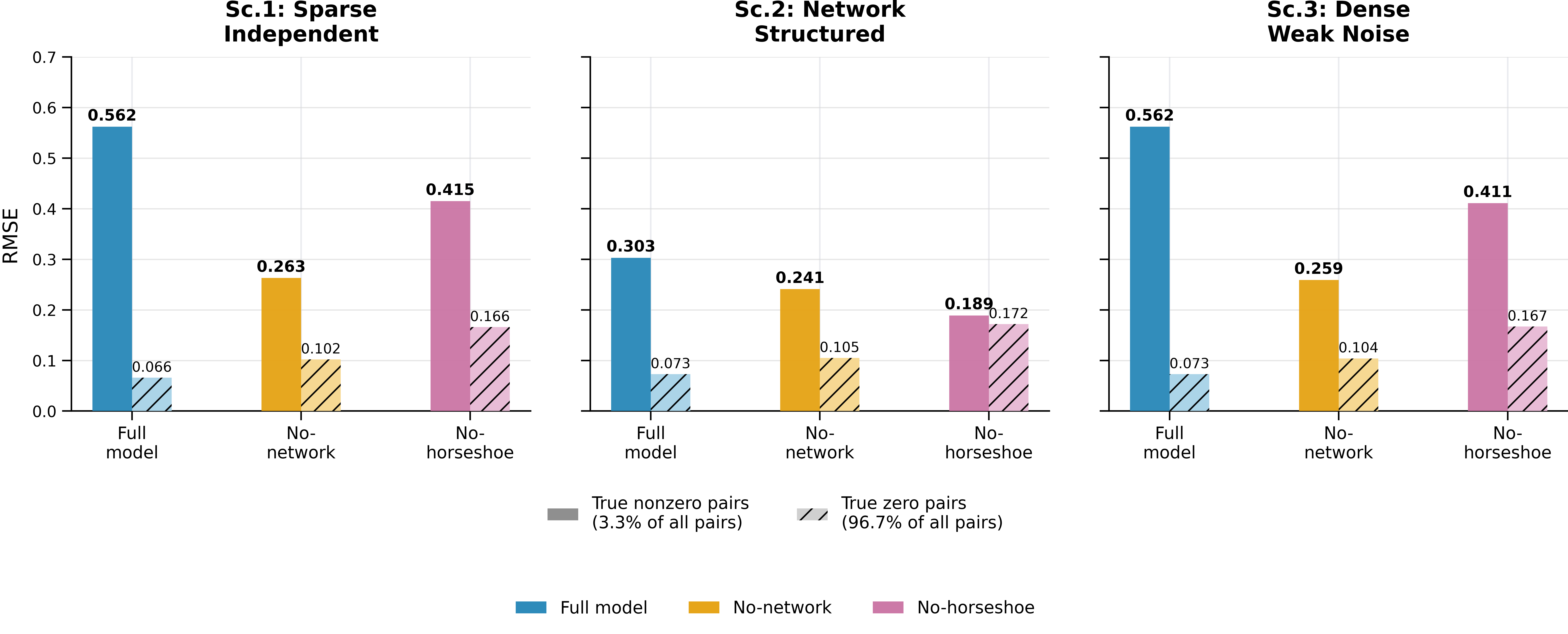}
\caption{RMSE decomposed by true-association status ($n{=}100$
replicates, mean values). \emph{Solid bars:} RMSE restricted to the
250 truly associated pairs (3.3\%). \emph{Hatched bars:} RMSE
restricted to the 7{,}250 truly null pairs (96.7\%).}
\label{fig:rmse-decomposed}
\end{figure}

Second, in Scenarios~1 and~3 (non-network DGPs), the
no-network model achieves lower RMSE$_\beta^{(\text{all})}$ than the
full model. In Scenario~2 (network-structured DGP), the ordering
reverses: the full model achieves the lowest aggregate RMSE (0.127 vs.
0.130 for the no-network model), confirming that  the GMRF prior improves overall coefficient estimation when the true effects are
localized within the same pathway structure as the prior. This result
is unique to Scenario~2 and is consistent with the matched-FDR
sensitivity advantage. In all three scenarios, aggregate RMSE alone does not fully characterize
model performance. When RMSE is decomposed by true association status
(Table~\ref{tab:rmse-decomp}, Figure~\ref{fig:rmse-decomposed}), the
tradeoff becomes clear: the no-network model has lower RMSE on true 

\begin{table}[H]
\caption{Threshold-Based and RMSE-Decomposed Simulation Results}
\label{tab:rmse-decomp}
\centering
\small
\begin{threeparttable}
\begin{tabular}{llccc}
\toprule
Scenario & Model &
  Sens.\ at FDR$\le$10\% &
  RMSE$_\beta^{(\text{signal})}$ &
  RMSE$_\beta^{(\text{null})}$ \\
\midrule
\multirow{3}{*}{1: Sparse Independent}
 & Full         & 0.747 & 0.562 & \textbf{0.066} \\
 & No-network   & \textbf{0.859} & \textbf{0.263} & 0.102 \\
 & No-horseshoe & 0.764 & 0.415 & 0.166 \\
\midrule
\multirow{3}{*}{2: Network Structured}
 & Full         & \textbf{0.772} & 0.303 & \textbf{0.073} \\
 & No-network   & 0.700 & 0.241 & 0.105 \\
 & No-horseshoe & 0.769 & \textbf{0.189} & 0.172 \\
\midrule
\multirow{3}{*}{3: Dense Weak Noise}
 & Full         & 0.727 & 0.562 & \textbf{0.073} \\
 & No-network   & \textbf{0.849} & \textbf{0.259} & 0.104 \\
 & No-horseshoe & 0.748 & 0.411 & 0.167 \\
\bottomrule
\end{tabular}
\begin{tablenotes}[flushleft]
\footnotesize
\item \textbf{Note.} Sens.\ at FDR$\le$10\% $=$ sensitivity of the largest
selected set satisfying empirical FDR $\le 10\%$ using posterior sign
probability $q_{gd}$; RMSE$_\beta^{(\text{signal})}$ $=$ RMSE on the
250 truly associated pairs; RMSE$_\beta^{(\text{null})}$ $=$ RMSE on
the 7{,}250 truly null pairs. Bold $=$ best per metric per scenario.
In Scenario~2, the full model wins matched-FDR sensitivity (0.772 vs.\
0.700 for no-network), confirming that the GMRF prior provides a
genuine advantage when the DGP matches the prior structure.
\end{tablenotes}
\end{threeparttable}
\end{table}

\noindent signals in Scenarios~1 and~3 (less shrinkage bias), while the full
model has the lowest RMSE on truly null pairs in every scenario
(stronger shrinkage of irrelevant associations). Since false positives
are the main practical concern for gene--drug association
prioritization, the full model's lower null RMSE and lower FDR are
more aligned with the intended use of the model.

\begin{figure}[H]
\centering
\includegraphics[width=\textwidth]{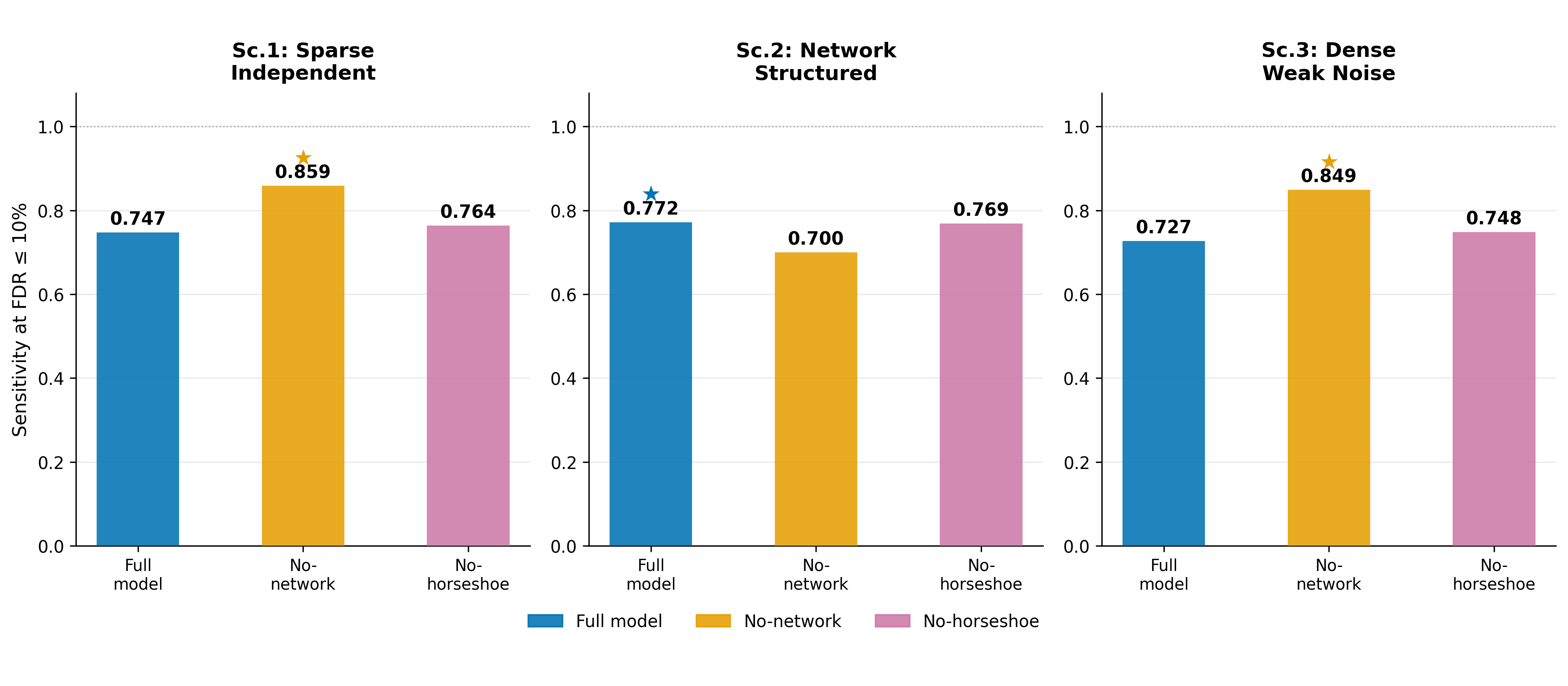}
\caption{Sensitivity at a matched FDR of 10\% for each model across
three simulation scenarios ($n{=}100$ replicates, mean values). Stars
mark the best-performing model in each scenario.}
\label{fig:auc-fdr}
\end{figure}

Third, the threshold-based evaluation
(Table~\ref{tab:rmse-decomp}, Figure~\ref{fig:auc-fdr}) reveals a
scenario-dependent pattern that directly validates the GMRF network
prior. In Scenario~2 (network-structured DGP), where true effects are
generated using the same pathway network structure as the model's
prior, the full model achieves the highest matched-FDR sensitivity
(0.772), surpassing the no-network model (0.700) and the no-horseshoe
model (0.769). This confirms that the GMRF prior provides a genuine advantage 
in recovering true associations when pathway structure is real: by borrowing

\noindent information across network-connected genes, the
full model identifies associations that the no-network model misses. 
In Scenarios~1 and 3 (non-network DGPs), where true effects are
generated independently of network structure, the no-network model
has higher matched-FDR sensitivity (0.849--0.859 vs.\ 0.727--0.747
for the full model). This is the expected bias-variance tradeoff: when
no pathway structure exists in the signal, the GMRF prior introduces
mild smoothing toward network configurations, reducing sensitivity
relative to an unconstrained model. The result across all three
scenarios -- the full model wins when the prior matches the DGP and
loses when it does not -- is precisely the behavior a correctly
specified network prior should exhibit. At the primary 95\%
credible-interval operating point, the full model maintains superior
FDR control in all three scenarios (2.8--11\% vs.\ 6.1--21\% for the
no-network model and 20.9--53\% for the no-horseshoe model;
Table~\ref{tab:simulation}).

Fourth, the 95\% credible-interval coverage is well-calibrated
for the full model (Table~\ref{tab:simulation}), ranging from 0.957
to 0.977 across scenarios. The no-network model over-covers
substantially (0.981--0.992), consistent with inflated posterior
variance when the sparsity-inducing layer is removed. The no-horseshoe
model has coverage near 0.957--0.958 but substantially worse FDR,
indicating that nominal interval coverage alone is insufficient for
reliable association selection.

Overall, the simulation study shows that the full model behaves as a
conservative structured-shrinkage procedure: it consistently provides
the most reliable selected association set (highest precision, lowest
FDR, strongest null shrinkage), and the GMRF prior provides a genuine
recovery advantage precisely when pathway structure is real. The
no-network model is more aggressive and ranks true signals higher
when signal has no network structure, but at the cost of substantially
more false discoveries. The no-horseshoe model performs worst on FDR
in every scenario, confirming that the global--local shrinkage layer
is the primary mechanism responsible for false-discovery control.

\section{Application to Cancer Pharmacogenomics}
\label{subsec:application}

\subsection{Data Sources and Network Construction}
\label{subsec:methods-data}

We use the Genomics of Drug Sensitivity in Cancer (GDSC2) dataset
\citep{yang2013,garnett2012,iorio2016}, which provides fitted
dose-response summaries (median inhibitory concentration, IC50, on the
natural log scale) for a panel of cancer cell lines screened against a
panel of drugs, together with somatic mutation calls from Cell Model
Passports \citep{vandermeer2019} restricted to a curated set of cancer
driver genes. GDSC2 was preferred to the legacy GDSC1 release (improved
screening protocol; GDSC's own documentation recommends GDSC2 values
where both are available) and to genome-wide mutation calling (the
curated driver-gene panel removes the need to separately filter
passenger mutations); a comparable resource, the Cancer Cell Line
Encyclopedia \citep{barretina2012}, was considered but not used for the
primary analysis, and is noted in Section~\ref{sec:discussion} as a
candidate for external validation. After restricting mutation calls to
coding, driver-flagged variants and intersecting the resulting gene set
with the GDSC2 cell-line panel, the final analysis dataset comprises
$N = 951$ cell lines, $D = 295$ drugs, and $G = 219$ genes, with
$237{,}566$ (cell line, drug) response observations (no missing values
among them).

The pathway network used in the model's prior structure
(Section~\ref{subsec:model-network-layer}) was constructed from KEGG
(Kyoto Encyclopedia of Genes and Genomes) \citep{kanehisagoto2000}
signaling, regulatory, and cellular-process pathway gene sets. We
excluded broad metabolic maps and disease- or infection-specific maps
before constructing the graph because these categories can create dense,
less specific gene connections that may weaken the interpretability of
network smoothing. This filtering choice is biologically conservative
rather than exhaustive: metabolic reprogramming, immune signalling, and
inflammatory pathways can influence cancer drug response, but the goal
of the GMRF prior here was to encode a curated set of cancer-relevant
regulatory and signalling relationships rather than a complete causal
map of all drug-response biology. This filtering reduced 320 candidate
KEGG pathways to 171 after pathway-category exclusion, and to 141 after
restricting to pathways containing at least one gene in the 219-gene
panel. Let $W$ denote the resulting weighted gene--gene
adjacency matrix. Two genes were linked in $W$ if and only if they shared membership in at
least one retained pathway, with edge weights defined by the Jaccard
index of their pathway-membership sets. After restricting to non-isolated
panel genes, the final KEGG-derived graph used by the GMRF prior
contained 134 genes. The remaining 85 genes were retained in the
likelihood and global--local shrinkage layer but received no
network-correlated prior structure.

For this network, we constructed $D_W=\mathrm{diag}(W\mathbf{1})$ and
verified that $D_W-\delta W$ was positive-definite for
$\delta\in\{0.5,0.7,0.8,0.9\}$. The primary analysis used
$\delta=0.8$, which was also used in the simulation study.

\subsection{Primary GDSC2 Analysis}
\label{sec:results}

\subsubsection{Posterior Computation and MCMC Diagnostics}
\label{subsec:results-convergence}

We fit Model~I from Section~\ref{subsec:methods-model-primary} to the full GDSC2
dataset ($N=951$ cell lines, $G=219$ genes, and $D=295$ drugs) using two
Markov chains of 5{,}000 iterations each, with 1{,}000 burn-in iterations
and thinning by 2, yielding 2{,}000 retained posterior samples per chain.
We assessed MCMC behavior using the Gelman--Rubin potential scale
reduction factor $\hat R$ \citep{gelmanrubin1992}, effective sample size
(ESS), and the Geweke diagnostic \citep{geweke1992}, computed across all
$219\times295=64{,}605$ gene--drug coefficients.

The diagnostic summaries indicated satisfactory overall mixing. Only 27
coefficients (0.04\%) had $\hat R>1.1$, and only 14 coefficients
(0.02\%) had ESS below 800, with a median ESS of 3{,}822. The Geweke
diagnostic produced two-sided $p$-values below 0.05 for 4.47\% of
coefficients, which are not concentrated in any particular gene or drug. These results
suggest that posterior summaries for the primary real-data analysis are
not driven by widespread chain instability.

\subsubsection{A Sparse, Concentrated Set of Mutation-Drug Associations}
\label{subsec:results-drivers}

For each of the $64{,}605$ gene-drug pairs, we assessed posterior
inclusion via whether $\beta_{gd}$'s 95\% credible interval excludes
zero. 126 pairs (0.195\%) met this criterion, spanning 20 genes and
105 of the 295 drugs. As presented in Table~\ref{tab:top-genes}, the associations were highly concentrated in two
genes: \textbf{EZH2}, mutated in only 22 of 951 cell lines (2.3\%),
was associated with 45 drugs, and \textbf{KMT2D}, mutated in 103 cell
lines (10.8\%), with 36 drugs -- together accounting for 64\% of all
flagged associations. Every one of EZH2's 45 associations and KMT2D's
36 were in the same direction (negative; mutation associated with lower
$\ln$IC50, i.e.\ greater drug sensitivity). 

\begin{table}[H]
\caption{Top recurrent gene--drug response associations under the full model.}
\label{tab:top-genes}
\centering
\small
\setlength{\tabcolsep}{20pt}
\begin{threeparttable}
\begin{tabular}{lrrrrr}
\toprule
Gene & Mut.\ freq. & Drugs & \% of pairs &
  $\overline{\hat\beta}$ & $\overline{\text{SD}}$ \\
\midrule
EZH2  & 2.3\%  & 45 & 35.7\% & $-0.911$ & $0.357$ \\
KMT2D & 10.8\% & 36 & 28.6\% & $-0.496$ & $0.210$ \\
TET2  & 2.0\%  &  8 &  6.3\% & $-1.002$ & $0.410$ \\
ASXL1 & 3.3\%  &  7 &  5.6\% & $-1.230$ & $0.384$ \\
PBRM1 & 3.0\%  &  5 &  4.0\% & $\phantom{-}0.644$ & $0.263$ \\
BRAF  & 8.6\%  &  4 &  3.2\% & $-0.134$ & $0.057$ \\
TP53  & 67.3\% &  1 &  0.8\% & $\phantom{-}1.351$ & $0.144$ \\
\bottomrule
\end{tabular}
\begin{tablenotes}[flushleft]
\footnotesize
\item \textbf{Note.} Genes are ranked by the number of associated drugs among 295 screened drugs. A gene--drug pair is counted as associated when the 95\% posterior credible interval for $\beta_{gd}$ excludes zero. $\overline{\hat \beta}$ and  $\overline{\text{SD}}$ are the posterior mean and standard deviation of $\beta_{gd}$ averaged across each gene's associated drugs. In total, 126 of 64,605 gene--drug pairs met the inclusion criterion, spanning 20 genes; the table reports the seven most frequent genes. Full per-drug posterior summaries for all flagged gene--drug pairs are provided in the Supplementary Section~S6, Table~3.
\end{tablenotes}
\end{threeparttable}
\end{table}

\begin{figure}[H]
\centering
\includegraphics[width=0.9\textwidth]{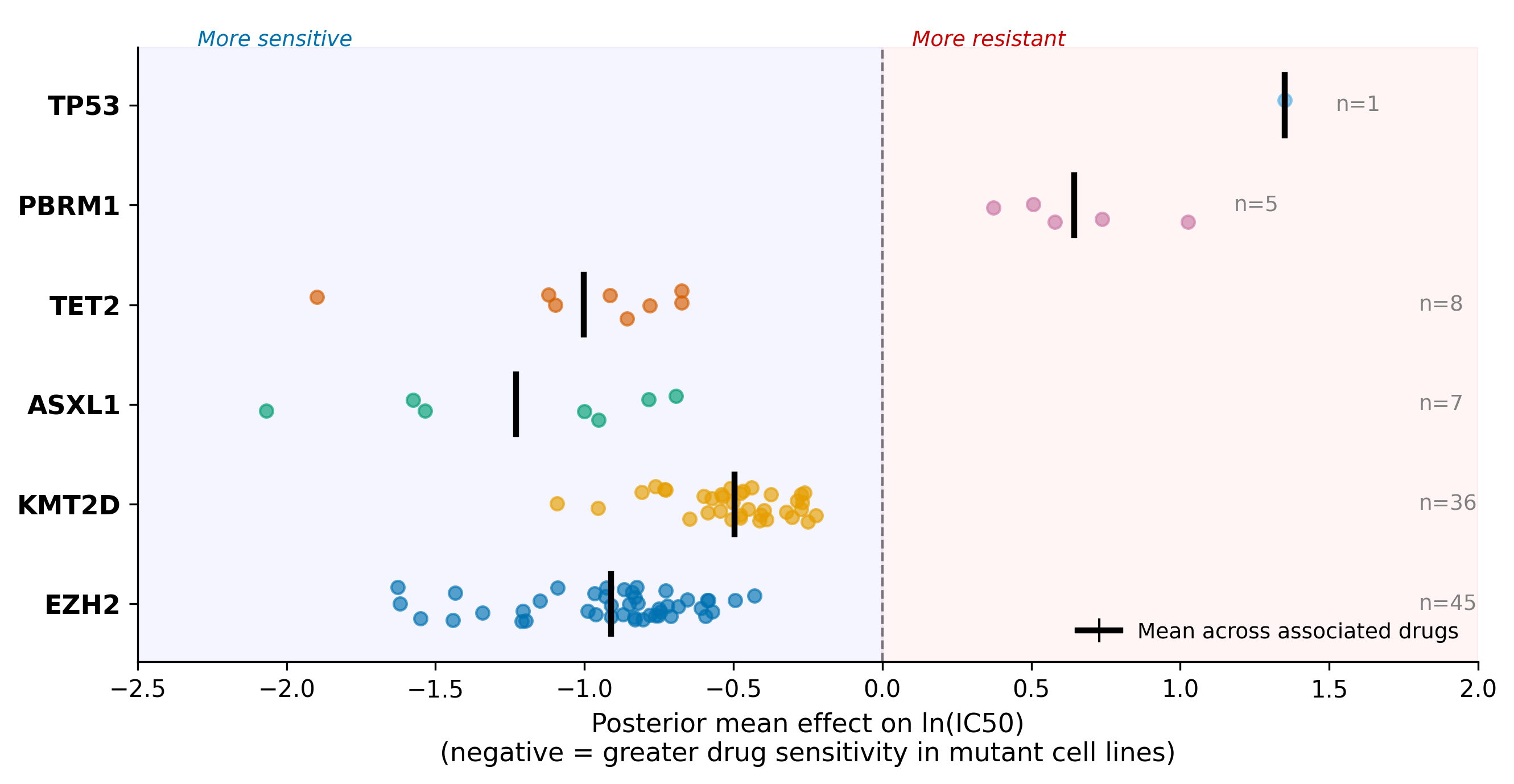}
\caption{Posterior mean effect distributions for the seven most-associated
driver genes (full model, $\delta{=}0.8$). Each point represents one
associated drug whose 95\% posterior credible interval excludes zero;
the bold vertical tick marks the mean posterior effect across all
associated drugs for that gene. EZH2 ($n{=}45$) and KMT2D ($n{=}36$)
cluster uniformly in the sensitivity (negative-effect) region. PBRM1
($n{=}5$) and TP53 ($n{=}1$, Nutlin-3a) are the only genes with
positive mean effects; TP53's single hit is mechanistically explained
in the text. Blue shaded region = greater sensitivity; red = greater
resistance.}
\label{fig:real-effects}
\end{figure}

The recurrent genes in Table~\ref{tab:top-genes} also suggest a broader
chromatin- and epigenetic-regulator pattern beyond the two dominant
genes. TET2 and ASXL1 were the third and fourth most recurrent genes,
with 8 and 7 associated drugs, respectively, and both showed average
effects in the sensitivity direction. This is biologically plausible:
TET2 is a DNA methylation regulator frequently altered in lymphoid and
myeloid malignancies, and ASXL1 is a chromatin-associated regulator of
epigenetic marks \citep{lio2019dysregulation,inoue2018asxl1}. Thus,
although EZH2 and KMT2D account for most flagged associations, the
broader pattern among the most recurrent genes is consistent with the
hypothesis that alterations in chromatin or epigenetic regulators may
mark increased in-vitro drug sensitivity. This interpretation is
hypothesis-generating, since we did not perform a formal enrichment test
for epigenetic regulators as a gene class. 

\textbf{TP53}, the most frequently mutated gene in the panel (67.3\%
of cell lines) and the dominant predictor in \citet{samorodnitsky2020},
was associated with only 1 of 295 drugs: Nutlin-3a, an MDM2 inhibitor
(posterior mean $\hat\beta = 1.351$, SD $0.144$, 95\% CI $[1.063,
1.612]$), with a \emph{positive} sign indicating TP53 mutation is
associated with greater resistance (Figure~\ref{fig:real-effects}). This direction is mechanistically
expected: Nutlin-3a works by blocking MDM2-mediated degradation of
wild-type p53, which has no substrate to act on in TP53-mutant cells.
We return to TP53's broader near-absence in
Section~\ref{sec:discussion}.

\subsection{Biological Interpretation and Pharmacological Support}
\label{subsec:results-pharmacology}

\subsubsection{Pharmacological Support for the EZH2 and KMT2D Associations}
\label{subsec:results-EK}

We cross-referenced the drugs associated with EZH2 and KMT2D against
their documented mechanisms of action. EZH2 is the catalytic subunit
of Polycomb Repressive Complex 2 (PRC2), responsible for histone H3
lysine 27 trimethylation and transcriptional silencing of target genes;
somatic EZH2 mutations are recurrently observed across solid and
hematological cancers \citep{laugesen2016}. The 45 drugs associated
with EZH2 mutation included GSK343, a direct, selective small-molecule
EZH2 inhibitor, as well as several chromatin- and
transcription-regulatory compounds, including BET inhibitors
(I-BET-762, JQ1, I-BRD9), HDAC inhibitors (Vorinostat, Entinostat,
PCI-34051), and DOT1L inhibitors (EPZ004777 and EPZ5676)
\citep{daigle2013}. \citet{knutson2014} report that lymphoma cell
lines carrying EZH2 catalytic-domain point mutations show selective
sensitivity to EZH2 inhibition both in vitro and in vivo; the direction
of our model's estimated effect is consistent with this finding for all
45 associated drugs.

The 36 drugs associated with KMT2D mutation included several PARP
inhibitors (Olaparib, Veliparib, Rucaparib, Niraparib, Talazoparib)
and DNA-damaging or DNA-repair-related agents (Cisplatin, Oxaliplatin,
Camptothecin, Irinotecan, Cytarabine, Gemcitabine, Fludarabine).
KMT2D's documented role in chromatin-mediated DNA damage response
provides a plausible mechanistic rationale: the close paralog
KMT2C/MLL3 has been shown directly to regulate DNA-repair gene
expression, with KMT2C-depleted cells becoming dependent on PARP1/2
for survival \citep{rampias2019kmt2c}. This is structurally analogous
to -- though mechanistically distinct from -- the well-established
sensitization of BRCA1/2-deficient tumors to PARP inhibition
\citep{bryant2005,farmer2005}.

\subsubsection{No Coordinated Signal from a Correlated Co-Mutation Cluster}
\label{subsec:results-msi}

A preliminary diagnostic check identified a cluster of six genes --
RPL22, ACVR2A, BAX, BMPR2, KMT2B, and STAT5B -- whose mutation
status is highly correlated across cell lines, consistent with their
being co-targets of a shared microsatellite-instability (MSI)
mechanism: ACVR2A and BAX in particular are coding-microsatellite
frameshift targets with directly measured mutation rates in DNA
mismatch-repair-deficient cells \citep{chung2008msi}. Of the $295
\times 6 = 1{,}770$ possible gene-drug associations among these six
genes, only one met the inclusion criterion (RPL22, one drug), and no
drug showed more than one cluster gene as associated simultaneously.
The model does not detect a coordinated MSI-driven drug-response signal
distinct from individual gene effects at this sample size and gene
count.

\subsection{Model Ablation and Predictive Assessment}
\label{subsec:results-ablation}

To assess whether the network and horseshoe priors are each
contributing to, rather than merely complicating, the result above, we
refit the model twice more at a shorter chain length (1{,}500
iterations, 300 burn-in, 2 chains), once with the network prior
removed (\textsc{no-network}: $\tilde\beta_{gd}$ retains only the
horseshoe prior) and once with the horseshoe prior removed
(\textsc{no-horseshoe}: $\tilde\beta_{gd}$ retains only the network
prior, plus a flat $\text{Normal}(0,100)$ fallback for the 84 genes
outside the network).

Both reduced models identify far more associations than the full model
(Table~\ref{tab:ablation}). KRAS illustrates the mechanism: in the
full model, KRAS shows a small, consistent positive effect (posterior
mean $\approx 0.023$), correctly identified as real but too diffuse to
meet the inclusion criterion (1 of 295 drugs); with the horseshoe
removed, the same effect crosses the threshold for 161 drugs -- the  largest count of any gene in that model -- which we interpret as
inflated rather than genuinely informative. By contrast, EZH2 and
KMT2D survive both ablations (EZH2: 95/12 drugs without horseshoe/network;
KMT2D: 111/30), corroborating the finding of
Section~\ref{subsec:results-drivers}. 

\begin{table}[H]
\caption{GDSC2 model ablation comparison of flagged gene--drug associations.}
\label{tab:ablation}
\centering
\small
\setlength{\tabcolsep}{20pt}
\begin{threeparttable}
\begin{tabular}{lrr}
\toprule
Model & Associated pairs (of 64{,}605) & Ratio to full model \\
\midrule
Full model    & 126 (0.195\%)    & ---          \\
No-network    & 867 (1.34\%)     & $6.9\times$  \\
No-horseshoe  & 2{,}011 (3.11\%) & $16.0\times$ \\
\bottomrule
\end{tabular}
\begin{tablenotes}[flushleft]
\footnotesize
\item \textbf{Note.} A gene--drug pair is counted as associated when the
95\% posterior credible interval for $\beta_{gd}$ excludes zero. Percentages
are computed relative to all $64{,}605$ tested gene--drug pairs. The ratio
column reports the number of associated pairs relative to the full model.
\end{tablenotes}
\end{threeparttable}
\end{table}

\subsubsection{Cross-Validated Predictive Likelihood}
\label{subsec:results-cv}

To move from qualitative inclusion-behavior comparison to a formal
out-of-sample predictive comparison following
\citet{samorodnitsky2020}'s Section~2.4, we computed the $k$-fold
($k=5$) cross-validated posterior predictive log-likelihood for all
three model configurations, splitting cell lines into 5 equal folds
($\approx 190$ held-out per fold). 

\begin{table}[H]
\caption{Five-fold cross-validated predictive performance of the full model and ablation variants.}
\label{tab:kfold}
\centering
\small
\setlength{\tabcolsep}{13pt}
\begin{threeparttable}
\begin{tabular}{lrrrrr}
\toprule
& \multicolumn{5}{c}{Held-out log-likelihood by fold} \\
\cmidrule{2-6}
Model & Fold 1 & Fold 2 & Fold 3 & Fold 4 & Fold 5 \\
\midrule
Full model
 & $-82{,}356$ & $-81{,}418$ & $-85{,}267$ & $-84{,}822$ & $-80{,}376$ \\
No-network
 & $-84{,}914$ & $-84{,}963$ & $-89{,}062$ & $-87{,}035$ & $-83{,}812$ \\
No-horseshoe
 & $-95{,}247$ & $-98{,}238$ & $-99{,}997$ & $-95{,}802$ & $-95{,}150$ \\
\midrule
& \multicolumn{2}{c}{Mean (all folds)}
& \multicolumn{2}{c}{Full model advantage} & \\
\cmidrule{2-3}\cmidrule{4-5}
Model & & & vs.\ No-network. & vs.\ No-horseshoe. & \\
\midrule
Full model
 & \multicolumn{2}{c}{$-82{,}848$} & --- & --- & \\
No-network
 & \multicolumn{2}{c}{$-85{,}957$} & $+3{,}109$ & --- & \\
No-horseshoe
 & \multicolumn{2}{c}{$-96{,}887$} & --- & $+14{,}039$ & \\
\bottomrule
\end{tabular}
\begin{tablenotes}[flushleft]
\footnotesize
\item \textbf{Note.} Values are held-out log-likelihoods; higher values, or less negative values, indicate better predictive performance. The full model advantage is the mean held-out log-likelihood of the full model minus that of the corresponding ablation model across the five folds. Positive values indicate better held-out predictive performance by the full model.
\end{tablenotes}
\end{threeparttable}
\end{table}

As presented in Table~\ref{tab:kfold}, the full model achieves the highest held-out predictive
log-likelihood in every one of the five folds, without exception.
Its mean per-fold advantage over the no-network variant is $+3{,}109$
log-units (range $+2{,}213$ to $+3{,}795$), and over the no-horseshoe
variant is $+14{,}039$ log-units (range $+10{,}980$ to $+16{,}820$).
These are not marginal: the no-horseshoe model's predictive performance
is 14.5\% worse and the no-network model's is 3.6\% worse on average.
Each prior layer independently contributes to out-of-sample predictive
accuracy, not merely to in-sample posterior behavior.

\subsection{Robustness to Network-Smoothing Parameter $\delta$}
\label{subsec:results-delta}

The GMRF network prior contains one fixed parameter, $\delta \in
(0,1)$, controlling smoothing strength between adjacent network genes.
We refit the full model and both ablation variants at $\delta \in
\{0.5, 0.7, 0.9\}$ and repeated the posterior inclusion scan.

\begin{table}[H]
\caption{Delta-sensitivity analysis under the full model.}
\label{tab:delta-sensitivity}
\centering
\small
\setlength{\tabcolsep}{14pt}
\begin{threeparttable}
\begin{tabular}{crrrrrrr}
\toprule
$\delta$ & EZH2 & KMT2D & Total & Genes & Drugs &
  $\bar{\hat\beta}_{\text{EZH2}}$ & $\bar{\hat\beta}_{\text{KMT2D}}$ \\
\midrule
0.5 & 44 & 36 & 125 & 20 & 105 & $-0.922$ & $-0.495$ \\
0.7 & 44 & 36 & 125 & 20 & 105 & $-0.922$ & $-0.496$ \\
\textbf{0.8} & \textbf{45} & \textbf{36} & \textbf{126} & \textbf{20}
  & \textbf{105} & $\mathbf{-0.911}$ & $\mathbf{-0.496}$ \\
0.9 & 45 & 36 & 125 & 20 & 105 & $-0.911$ & $-0.496$ \\
\bottomrule
\end{tabular}
\begin{tablenotes}[flushleft]
\footnotesize
\item \textbf{Note.} EZH2 and KMT2D drug counts, total
flagged pairs, distinct driver genes, distinct associated drugs, and
mean posterior effect at each value of $\delta$. Bold row = primary
analysis. $\bar{\hat\beta}$ is negative throughout, indicating greater
drug sensitivity in mutant cell lines.
\end{tablenotes}
\end{threeparttable}
\end{table}

\begin{figure}[H]
\centering
\includegraphics[width=0.85\textwidth]{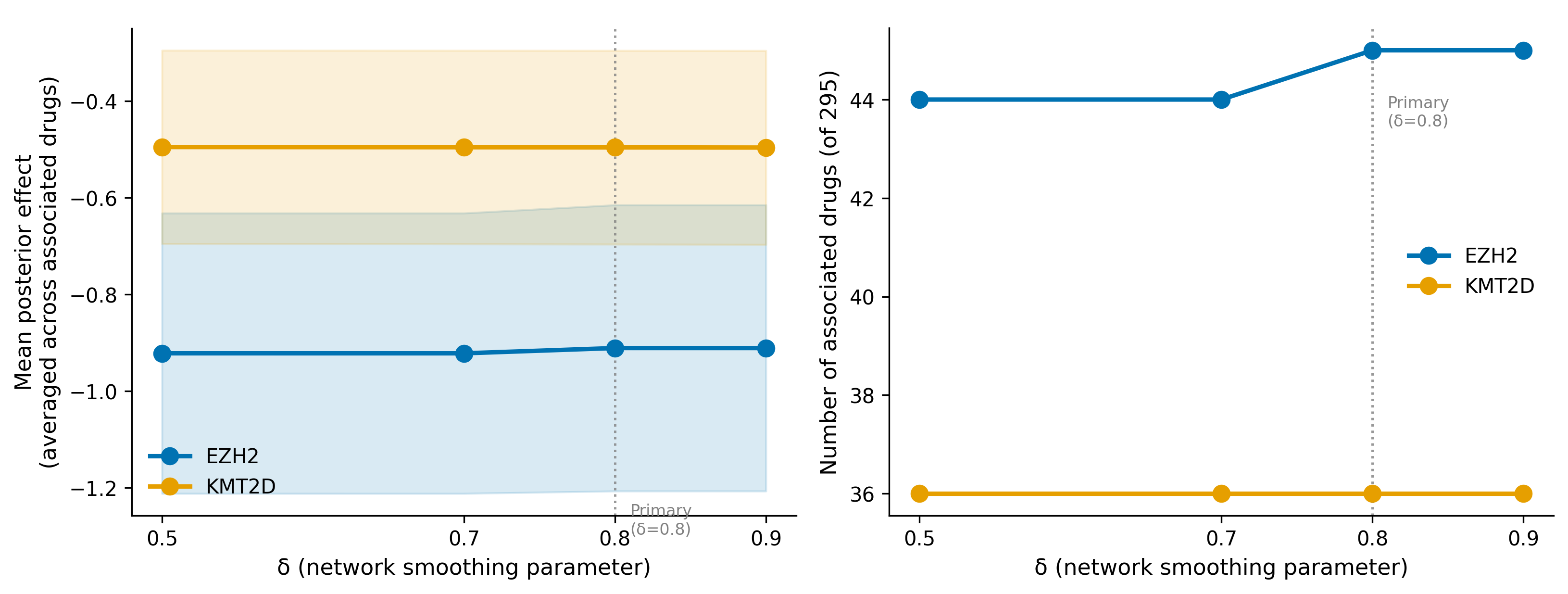}
\caption{Stability of EZH2 and KMT2D associations across four values
of $\delta \in \{0.5, 0.7, 0.8, 0.9\}$. \emph{Left:} Mean posterior
effect size ($\pm$1 SD across associated drugs) as a function of
$\delta$. EZH2's mean effect ranges from $-0.922$ to $-0.911$;
KMT2D's from $-0.496$ to $-0.495$. \emph{Right:} Number of associated
drugs as a function of $\delta$. KMT2D's count is constant at 36; EZH2
varies between 44 and 45. The dotted vertical line marks the primary
analysis value $\delta{=}0.8$.}
\label{fig:delta-sensitivity}
\end{figure}

\textit{KMT2D's 36 associated drugs are identical across all four
$\delta$ values.} EZH2 varies by at most one drug; mean effect sizes
vary by less than 0.015; the ablation ordering full $\prec$ no-network
$\prec$ no-horseshoe is preserved at every $\delta$. These results 
(Table~\ref{tab:delta-sensitivity}, Figure~\ref{fig:delta-sensitivity}) 
confirm that the paper's primary findings are not an
artifact of the specific choice $\delta = 0.8$ within the retained
KEGG-derived network. This analysis does not, however, assess
sensitivity to alternative KEGG pathway-category inclusion rules; that
question concerns the boundary of the network itself rather than the
strength of smoothing over a fixed network. 

\subsection{External Validation in PRISM/DepMap}
\label{subsec:results-prism}

To assess whether the EZH2 and KMT2D associations identified in GDSC2
replicate in an independent pharmacogenomic screen using a different
assay technology, we applied the full model to data from the PRISM
Repurposing Public 24Q2 dataset \citep{corsello2020} paired with DepMap
26Q1 somatic mutation calls. PRISM measures log$_2$ fold-change in
pooled cell-line viability at a fixed dose of 2.5\,$\mu$M, providing
both a larger compound panel and a methodologically independent readout.
After intersecting cell lines and restricting to our 219-gene panel,
the PRISM analysis dataset comprised $N {=} 899$ cell lines, $G {=} 218$
genes (one absent from DepMap 26Q1),
and $D {=} 1{,}518$ compounds, yielding $330{,}924$ gene-compound pairs. 

\begin{table}[H]
\caption{External PRISM validation of recurrent GDSC2 mutation-drug response associations.}
\label{tab:prism-comparison}
\centering
\small
\setlength{\tabcolsep}{10pt}
\begin{threeparttable}
\begin{tabular}{lrrrll}
\toprule
Gene & GDSC2 & PRISM & Direction &
  Mean $\hat\beta$ (GDSC2) & Mean $\hat\beta$ (PRISM) \\
\midrule
EZH2  & 45 & 12 & 8 sens.\ / 4 resist. & $-0.911$ & $-0.421$ \\
KMT2D & 36 & 36 & 36 sens.\ / 0 resist. & $-0.496$ & $-0.302$ \\
\bottomrule
\end{tabular}
\begin{tablenotes}[flushleft]
\footnotesize
\item \textbf{Note.} GDSC2 and PRISM columns report the number of associated compounds for each gene. ``Direction'' denotes the number of associated PRISM compounds with negative posterior mean effects, interpreted as greater sensitivity, versus positive posterior mean effects, interpreted as resistance. Mean $\hat{\beta}$ values are averaged across associated compounds within each dataset. Effect-size attenuation from GDSC2 to PRISM is expected because PRISM uses a fixed single-dose viability readout rather than the fitted dose--response IC50 summary used in GDSC2.
\end{tablenotes}
\end{threeparttable}
\end{table}

\textit{KMT2D replicates with complete directional consistency.}
As shown in Table~\ref{tab:prism-comparison}, all 36 compounds flagged for KMT2D in PRISM show negative posterior
mean effects, identical in direction to all 36 GDSC2 associations.
The drug count is identical despite a 5-fold larger compound panel
in PRISM, suggesting KMT2D's sensitivity association is concentrated
in a specific mechanistic class of compounds. Effect size attenuates
from $\bar{\hat\beta}{=}{-}0.496$ (GDSC2) to $-0.302$ (PRISM),
consistent with fixed single-dose attenuation.

\textit{EZH2 shows partial replication.} As presented in Table~\ref{tab:prism-comparison}, 
twelve of 1{,}518 PRISM compounds are flagged for EZH2 (vs.\ 45 of 295 in GDSC2), of which
8 (67\%) show the same negative direction. The top two
negative-direction compounds are directly mechanistically coherent:
CPI-169 is a selective EZH2 inhibitor \citep{balasubramanian2014};
CYC065 (fadraciclib) is a CDK2/9 inhibitor, and CDK2 directly
phosphorylates EZH2 \citep{fadhraciclib2020}. The four
resistance-direction compounds have no established EZH2-related
mechanism and likely reflect noise at the 2.5\,$\mu$M fixed dose.

\textit{Assay differences explain the attenuation.} PRISM's fixed
dose captures a single point on the dose-response curve; compounds with
shallow slopes or high IC50 values will show attenuated LFC signals
even when the biological association is real. The reduction in EZH2's
flagged count (45 to 12) and four resistance-direction compounds are
consistent with increased measurement noise rather than genuine reversal
of the GDSC2 finding.

\textit{Observations.} KMT2D's association replicates exactly across
two independent screens with fundamentally different assay technologies.
EZH2's association replicates in directional majority and in the two
mechanistically strongest hits. Together, these results provide
independent, cross-platform support for the primary GDSC2 findings.

\textit{Ablation comparison.} The model ordering observed in GDSC2
replicates in PRISM: the full model flags 1{,}077 of 330{,}924
gene--compound pairs (0.325\%), compared to 2{,}544 for the no-network
model (0.769\%; $2.4\times$) and 8{,}549 for the no-horseshoe model
(2.583\%; $7.9\times$). The ordering full $\prec$ no-network $\prec$
no-horseshoe by inclusion count is preserved, confirming that each
prior layer's sparsity-inducing role operates consistently across both
screens.

\subsection{Targeted Tissue-Group Analysis}
\label{subsec:results-group-layer}

We next used the group-layer model in Section~\ref{subsec:methods-model-group} as a targeted follow-up analysis to
assess whether recurrent mutation-drug associations showed
tissue-specific departures from the shared effect. Because storing
posterior draws for all $K\times G\times D$ group-specific coefficients
is memory intensive, we refit the full group-layer model while retaining
posterior draws of $\beta_{gd}^{(k)}$ for 25 target genes selected from
the primary 

\begin{table}[H]
\centering
\small
\setlength{\tabcolsep}{3.2pt}
\renewcommand{\arraystretch}{1.15}
\begin{threeparttable}

\caption{Selected tissue-group-specific mutation-drug effects.}
\label{tab:group-layer-selected}

\begin{tabular*}{\textwidth}{@{\extracolsep{\fill}}p{2.25cm}p{1.15cm}p{2.15cm}p{2.85cm}p{2.85cm}p{1.35cm}@{}}
\toprule
Group $(n_k)$ & Gene & Drug &
$\hat{\beta}^{(k)}$ \newline (95\% CI) &
$\hat{\Delta}$ \newline (95\% CI) &
Direction \\
\midrule
Skin (57) & BRAF & Dabrafenib
& $-4.143$ [$-5.274,\,-3.016$]
& $-4.139$ [$-5.265,\,-3.008$]
& Sensitive \\

Lung (187) & EGFR & Sapitinib
& $-4.071$ [$-5.406,\,-2.690$]
& $-4.068$ [$-5.415,\,-2.660$]
& Sensitive \\

Lung (187) & EGFR & Osimertinib
& $-3.918$ [$-4.814,\,-3.045$]
& $-3.916$ [$-4.810,\,-3.050$]
& Sensitive \\

Lung (187) & EGFR & Gefitinib
& $-3.489$ [$-4.375,\,-2.590$]
& $-3.487$ [$-4.380,\,-2.587$]
& Sensitive \\

Lung (187) & EGFR & AZD3759
& $-3.446$ [$-4.219,\,-2.668$]
& $-3.443$ [$-4.215,\,-2.664$]
& Sensitive. \\

Lung (187) & EGFR & Erlotinib
& $-3.287$ [$-4.237,\,-2.367$]
& $-3.284$ [$-4.251,\,-2.354$]
& Sensitive \\

Lung (187) & EGFR & Afatinib
& $-3.281$ [$-4.364,\,-2.217$]
& $-3.280$ [$-4.362,\,-2.207$]
& Sensitive \\

Haem./Lymph. (164) & TP53 & Nutlin-3a (-)
& $2.535$ [$2.069,\,3.000$]
& $2.531$ [$2.066,\,2.996$]
& Resistant \\

Haem./Lymph. (164) & ASXL1 & Nilotinib
& $-3.662$ [$-4.510,\,-2.784$]
& $-3.419$ [$-4.546,\,-1.682$]
& Sensitive \\

Haem./Lymph. (164) & NRAS & PD0325901
& $-2.049$ [$-2.862,\,-1.242$]
& $-2.046$ [$-2.859,\,-1.227$]
& Sensitive \\

Breast (51) & PTEN & AZD8186
& $-1.823$ [$-3.036,\,-0.754$]
& $-1.814$ [$-3.031,\,-0.736$]
& Sensitive \\

Other (437) & STAG2 & Talazoparib
& $-3.860$ [$-4.991,\,-2.692$]
& $-3.852$ [$-4.985,\,-2.679$]
& Sensitive \\

Breast (51) & STAG2 & Foretinib
& $3.584$ [$0.645,\,6.582$]
& $3.583$ [$0.641,\,6.567$]
& Resistant \\
\bottomrule
\end{tabular*}
\begin{tablenotes}[flushleft]
\footnotesize
\item \textbf{Note.} $\hat{\beta}^{(k)}$ is the posterior mean of the
group-specific mutation-drug effect for the indicated group. $\hat{\Delta}$
is the posterior mean of the tissue/group-specific deviation
$\Delta_{kgd}=\beta_{gd}^{(k)}-\widetilde{\beta}_{gd}$ from the shared
effect. CI denotes posterior credible interval. Negative effects indicate
greater sensitivity and positive effects indicate resistance under the
response orientation used in the main analysis. Haem./Lymph. denotes the
haematopoietic and lymphoid group. The \textup{``Other''} group contains
tissue categories merged because their sample sizes were below the
pre-specified minimum group-size threshold. Rows shown are selected
examples in which both the group-specific effect and the deviation from
the shared effect had 95\% credible intervals excluding zero.
\end{tablenotes}
\end{threeparttable}
\end{table}

\noindent full-model findings and the exploratory group-layer screen.
The analysis used a minimum tissue-group size threshold of
$n_{\min}=50$, yielding $K=6$ final groups after merging smaller tissue
categories into \textup{``Other''}. For each retained target
gene--drug--group combination, we computed posterior summaries for the
group-specific effect $\beta_{gd}^{(k)}$ and for the tissue-specific
deviation from the shared effect,
\[
\Delta_{kgd}
=
\beta_{gd}^{(k)}-\widetilde{\beta}_{gd}.
\]

Across the $6\times 25\times 295=44{,}250$ scanned
group--gene--drug combinations, 1{,}216 group-specific effects had 95\%
credible intervals excluding zero, and 1{,}103 tissue-specific
deviations had 95\% credible intervals excluding zero. These results indicate that
the targeted group-layer analysis identified a selective subset of
group-specific mutation-drug effects and deviations from the shared
effect.

The strongest tissue-specific refinements were biologically coherent
(Table~\ref{tab:group-layer-selected}). EGFR showed pronounced
lung-specific sensitivity effects across multiple EGFR inhibitors,
including Sapitinib, Osimertinib, Gefitinib, AZD3759, Erlotinib, and
Afatinib. For all six compounds, both the group-specific effect and the
deviation from the shared effect had 95\% credible intervals excluding
zero, indicating that the EGFR-associated sensitivity signal was
substantially amplified in the lung group. Similarly, BRAF showed a
strong skin-specific sensitivity effect for Dabrafenib
($\widehat{\beta}_{gd}^{(k)}=-4.143$, 95\% CI:
$[-5.274,-3.016]$; $\widehat{\Delta}_{kgd}=-4.139$, 95\% CI:
$[-5.265,-3.008]$). Additional strong tissue-specific effects included haematopoietic 
and lymphoid ASXL1--Nilotinib sensitivity and STAG2-associated effects for Talazoparib and Foretinib. 

The gene-level summary further showed that tissue-specific effects were
not restricted to a single gene. The largest numbers of 95\% credible
group-specific effects were observed for KRAS, RB1, STAG2, TP53, and
BRAF, with corresponding 95\% tissue-deviation counts of 272, 200, 157,
109, and 93, respectively. In contrast, EZH2 and KMT2D, although
important in the primary shared-effect analysis, showed comparatively
limited evidence of tissue-specific deviation in the targeted
group-layer scan. Thus, the group-layer model refined the primary
analysis by identifying tissue contexts in which selected
mutation-drug effects were substantially amplified, rather than simply
recapitulating the most recurrent shared-effect genes.

\begin{figure}[H]
\centering
\includegraphics[width=0.85\textwidth]{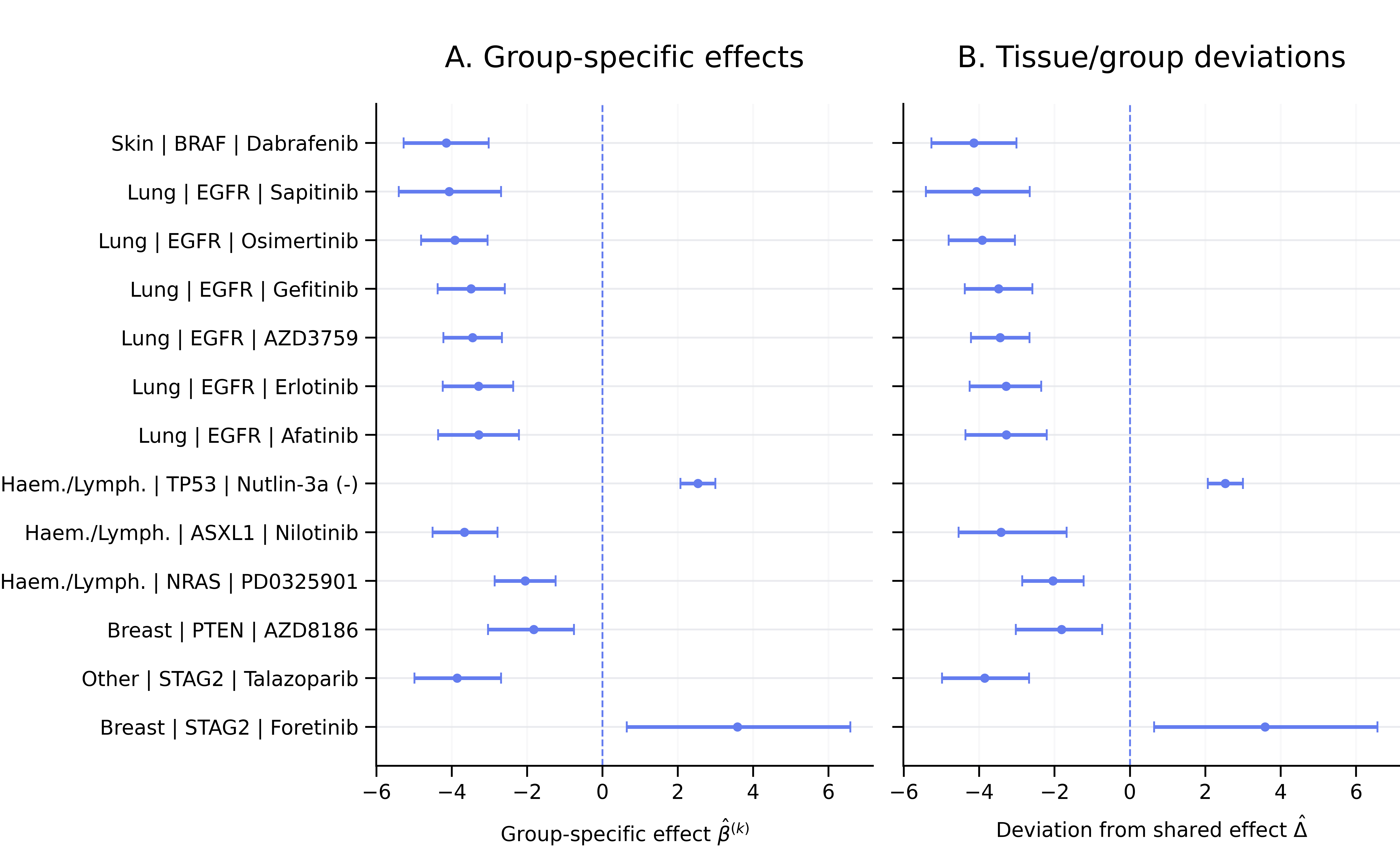}
\caption{Selected tissue/group-specific effects from the targeted group-layer analysis.
Panel A shows posterior means and 95\% credible intervals for the
group-specific effects $\beta_{gd}^{(k)}$. Panel B shows posterior means
and 95\% credible intervals for the deviations from the shared effect,
$\Delta_{kgd}=\beta_{gd}^{(k)}-\widetilde{\beta}_{gd}$. For several
selected examples, the shared effect $\widetilde{\beta}_{gd}$ is close to
zero, so the group-specific effect and the deviation from the shared
effect are visually similar; the two panels nevertheless represent
distinct posterior summaries. The dashed vertical line denotes zero.
Negative values indicate greater sensitivity, whereas positive values
indicate resistance under the response orientation used in the main
analysis.}
\label{fig:group-layer-selected-effects}
\end{figure}

The paired intervals in Figure~\ref{fig:group-layer-selected-effects} visually confirms that the
selected tissue/group-specific effects were not only nonzero within their
respective groups, but also represented departures from the shared
pan-sample effect. The most coherent patterns were the lung-specific EGFR
inhibitor effects and the skin-specific BRAF-Dabrafenib effect,
supporting the interpretation that the group-layer model identified
tissue contexts in which selected mutation-drug effects were amplified.

\section{Discussion}
\label{sec:discussion}

The central finding of this paper is specific: somatic mutations in
EZH2 and KMT2D are reproducible, independently corroborated candidate
mutation-level markers of in-vitro drug sensitivity, identified here
across a large cancer cell-line pharmacogenomics panel. 
Of the 64{,}605 gene-drug pairs examined, 126 (0.195\%) meet
the 95\% posterior credible-interval criterion, with EZH2 and KMT2D
together accounting for 64\% of all flagged associations. Every
association for both genes is in the sensitivity direction (negative
$\hat\beta_{gd}$), the effect sizes are large relative to their
posterior uncertainty (mean $|\hat\beta|=0.911$ for EZH2 and $0.496$
for KMT2D), and the findings survive each specification check applied:
removal of the network prior, removal of the horseshoe prior, variation
in the network smoothing parameter $\delta$, five-fold
cross-validation, and replication in an independent screen. 
These genes also have plausible biological relevance: EZH2 is acatalytic 
component of Polycomb repressive complex 2 and an established
epigenetic drug target, while KMT2D is a frequently altered chromatin
regulator whose loss can reprogram enhancer and transcriptional states
in cancer \citep{kim2016targeting,dhar2021cancer}. Moreover, TET2 and
ASXL1 also appear among the most recurrent genes, suggesting that the
dominant EZH2 and KMT2D findings may be part of a broader
chromatin- and epigenetic-regulator-associated sensitivity pattern
\citep{lio2019dysregulation,inoue2018asxl1}. We discuss below what each
model check contributes to interpreting the main result, and what the
findings collectively imply for the method and for cancer
pharmacogenomics.

Although EZH2 and KMT2D are emphasized because they account for the
largest share of flagged associations, the recurrence of TET2 and ASXL1
among the top genes is also biologically informative. Both genes are
involved in epigenetic regulation, and both showed average sensitizing
effects across their associated drugs. This pattern does not establish
formal enrichment of epigenetic regulators, but it supports a
hypothesis-generating interpretation that mutation-level alterations in
chromatin and epigenetic control may define a subset of cell lines with
broadly altered in-vitro drug sensitivity.

\paragraph{Cross-validated predictive accuracy and the value of each
prior layer.}
The most direct evidence that the model's complexity is warranted comes
from the five-fold cross-validated predictive log-likelihood
(Section~\ref{subsec:results-cv}). The full model outperforms the
no-network variant by a mean of $+3{,}109$ log-units per fold (range
$+2{,}213$ to $+3{,}795$) and the no-horseshoe variant by $+14{,}039$
log-units (range $+10{,}980$ to $+16{,}820$), with the full model
winning in every fold for both comparisons. These differences indicate
that the network and sparsity layers improve out-of-sample prediction,
not merely in-sample posterior summaries. This is important because
large pharmacogenomic screens are high-dimensional by design: they seek
to relate molecular features to drug-response phenotypes across many
cell lines and many compounds \citep{barretina2012,
garnett2012,iorio2016}. In this setting, an
association model must control false discoveries while retaining signals
that generalize to unseen samples. The cross-validation results suggest
that each prior layer earns its complexity: the GMRF prior contributes
biologically informed borrowing across pathway-connected genes, while
the horseshoe prior provides the sparsity needed to avoid overfitting in
a large gene--drug search space.

\paragraph{TP53: a positive-control result and a model limitation.}
TP53, the most frequently mutated gene in the panel (67.3\% of samples)
and the dominant predictor in the pan-cancer survival analysis of
\citet{samorodnitsky2020}, is associated with only one drug in the full
model: Nutlin-3a, in the resistance direction
($\hat\beta=1.351$). This result is mechanistically coherent. Nutlin-3a
is an MDM2 antagonist that inhibits the MDM2--p53 interaction, thereby
stabilizing and activating p53; its activity is therefore expected to
depend on an intact p53 pathway \citep{vassilev2004inVivo,
gu2008mdm2}. Consistent with this mechanism, prior experimental work has
shown that Nutlin-3 is cytotoxic in wild-type p53 leukemia cells but has
little effect in p53-mutant or p53-null cells \citep{gu2008mdm2}. Thus,
the resistance-direction TP53--Nutlin-3a association in our analysis
serves as a useful positive-control signal.

The near-absence of TP53 associations across the remaining 294 drugs is
also informative. TP53 mutation status is common in the dataset, which
reduces contrast between mutated and non-mutated groups, and TP53
mutations are functionally heterogeneous. Different TP53 variants can
produce loss-of-function, dominant-negative, or gain-of-function
effects, with distinct consequences for DNA damage response, apoptosis,
metabolism, and drug sensitivity \citep{monti2020tp53,
alvaradoortiz2021mutant}. A binary mutation indicator cannot distinguish
these classes. The contrast between TP53's limited role here and its
importance in the survival analysis of \citet{samorodnitsky2020} is
therefore not a discrepancy: that study examined patient survival,
whereas the present analysis examines quantitative in-vitro drug
response. These outcomes capture different aspects of TP53 biology.

\paragraph{The MSI co-mutation cluster.}
The six-gene microsatellite-instability cluster (RPL22, ACVR2A, BAX,
BMPR2, KMT2B, STAT5B) was checked specifically for coordinated
drug-response signal and produced only one flagged association (RPL22,
one drug). This negative result should be distinguished from the
estimation difficulty caused by co-mutation. These genes are highly
correlated in mutation status across samples, which increases posterior
uncertainty on each gene's individual effect. The model is therefore
correctly reporting that it cannot separate individual contributions of
co-mutated genes and that no coordinated MSI-driven drug-response signal
is detectable at this sample size. This is an honest negative finding,
not a model failure. More generally, this result illustrates why
high-dimensional pharmacogenomic association analyses should not treat
correlated genomic features as independent sources of evidence without
explicit regularization or uncertainty propagation.

\paragraph{External validation and the meaning of partial replication.}
The PRISM replication analysis (Section~\ref{subsec:results-prism})
provides independent cross-platform support for the primary findings.
KMT2D replicates with complete directional consistency across a compound
panel larger than GDSC2: all 36 PRISM-flagged compounds show the
sensitivity direction, matching the direction of all 36 GDSC2
associations. This level of directional consistency across two distinct
screening platforms is strong evidence that the KMT2D signal is not an
artifact of a single dataset. EZH2 shows partial replication: 8 of 12
flagged PRISM compounds are in the sensitivity direction, with the most
mechanistically interpretable hits including an EZH2 inhibitor and a
CDK2/9 inhibitor. This pattern is consistent with EZH2's established
role as a chromatin regulator and therapeutic target, but the
PRISM-only resistance-direction compounds should be interpreted
cautiously because they do not have an obvious EZH2-directed mechanism
\citep{kim2016targeting}.

Effect-size attenuation from GDSC2 to PRISM (mean $\hat\beta$ from
$-0.911$ to $-0.421$ for EZH2; from $-0.496$ to $-0.302$ for KMT2D) is
also expected. GDSC2 estimates drug response using dose-response
screening and fitted sensitivity summaries, whereas the PRISM
repurposing screen is a high-throughput pooled viability platform that
supports much broader compound coverage \citep{iorio2016,
corsello2020}. Differences in assay format, compound
concentration, screening duration, and response summarization can
attenuate or reshape effect sizes even when the underlying biological
direction is preserved. The important validation result is therefore not
identity of effect magnitude, but reproducibility of direction and the
persistence of sparse, biologically interpretable signals across
platforms.

\paragraph{Simulation study.}
The simulation study (Section~\ref{subsec:results-simulation}) provides
controlled evidence about the conditions under which each model
component adds value. Across all three scenarios, the full model
achieves the highest precision and lowest false-discovery rate at the
primary 95\% credible-interval operating point, the operating
characteristic most relevant for an application in which every predicted
association may require laboratory follow-up. The matched-FDR
sensitivity comparison further clarifies the role of the network layer.
The full model wins sensitivity in Scenario~2, where nonzero effects are
generated in connected pathway modules, whereas the no-network model can
be more sensitive when the data-generating process is independent of
network structure. This behavior is expected: graph- or
network-informed regularization is beneficial when the prior network
captures meaningful dependence among predictors, but it need not improve
power when true signals are independent of that structure
\citep{lili2008,li2010variable}. The simulation results therefore
support a conditional interpretation of the network prior: it is useful
to the extent that the supplied pathway graph captures biologically
relevant clustering of mutation effects.

\paragraph{Tissue-level heterogeneity and the limits of pan-sample
association.}
The group-layer extension
(Section~\ref{subsec:results-group-layer}) provides a targeted
assessment of whether mutation-drug associations detected in the pooled
analysis are amplified, attenuated, or reversed in specific tissue
contexts. This is important because cancer pharmacogenomic studies have
shown that tissue lineage can mediate drug response and influence how
genomic alterations translate into sensitivity or resistance
\citep{iorio2016}. Rather than treating the pan-sample effect
as the only scientifically relevant estimand, Model~II decomposes each
selected association into a shared effect and a tissue/group-specific
deviation,
\[
\Delta_{kgd}=\beta_{gd}^{(k)}-\widetilde{\beta}_{gd}.
\]
This decomposition allows the model to distinguish broadly shared
associations from effects that are concentrated in particular tissue
contexts.

The strongest tissue-specific refinements in the targeted group-layer
scan were biologically coherent. The clearest examples were
lung-specific EGFR sensitivity effects across multiple EGFR inhibitors,
including Sapitinib, Osimertinib, Gefitinib, AZD3759, Erlotinib, and
Afatinib. This pattern is consistent with the established role of EGFR
alterations as predictive markers for EGFR tyrosine kinase inhibitors in
non-small-cell lung cancer \citep{holleman2019first}. Similarly, BRAF
showed a strong skin-specific sensitivity effect for Dabrafenib, in
agreement with the established therapeutic relevance of BRAF inhibition
in BRAF-mutant melanoma \citep{kainthla2014dabrafenib}. TP53 showed
recurrent tissue-specific resistance effects to Nutlin-3a, particularly
in the haematopoietic and lymphoid group, again consistent with the
dependence of MDM2 antagonism on a functional p53 pathway
\citep{vassilev2004inVivo,gu2008mdm2}. Additional strong deviations
were observed for STAG2, ASXL1, NRAS, PTEN, and related target genes.
These findings illustrate the value of the group-layer model as a
refinement of the primary pan-sample analysis: it identifies tissue
contexts in which selected mutation-drug effects are substantially
stronger than the shared effect.

The targeted group-layer analysis also clarifies which primary
pan-sample signals do \emph{not} appear to be strongly tissue-specific.
Although EZH2 and KMT2D were important in the primary shared-effect
analysis, they showed comparatively limited evidence of tissue-specific
departure in the targeted group-layer scan. In particular, EZH2 had only
two 95\% credible group-specific effects and no 95\% credible deviations
from the shared effect, while KMT2D had eight 95\% credible
group-specific effects but only one 95\% credible deviation. Thus, the
group-layer model does not simply reproduce the most recurrent
pan-sample associations. Instead, it separates broadly shared
associations from effects that are concentrated in particular tissue
contexts.

More generally, these results emphasize a limitation of purely
pan-sample association models in cancer pharmacogenomics. A pooled model
can efficiently detect mutation-drug associations that are stable across
samples, but it may average over heterogeneous subgroup-specific effects.
The group-layer extension provides a principled Bayesian mechanism for
investigating this heterogeneity while retaining partial pooling toward a
shared effect. The same framework could be applied in other stratified
settings, including patient subgroups defined by ancestry, sex,
treatment history, molecular subtype, or clinical context, whenever the
scientific question concerns both an overall association and its
variation across biologically meaningful subgroups.

Several limitations should be noted. First, the mutation indicator is
binary, conflating gain-of-function, dominant-negative, and
loss-of-function variants whose pharmacological consequences may differ
substantially. This limitation is especially relevant for genes such as
TP53, where different classes of alteration have distinct biological and
therapeutic implications \citep{monti2020tp53,alvaradoortiz2021mutant}.
Second, the pathway network used in the GMRF prior is necessarily
incomplete and reflects both current database coverage and the
pathway-category filtering choices used to construct the graph, rather
than a complete causal map of pathway interactions. The robustness of
the primary findings to variation in $\delta$ addresses sensitivity to
the strength of smoothing within the retained network, but it does not
address uncertainty about the network's boundaries, missing biological
edges, or the exclusion of broad metabolic and immune/inflammatory
pathway categories. Future work could compare alternative
pathway-filtering schemes, including networks that retain metabolic or
immune-related pathways, to assess whether the same shared-effect
findings persist under broader biological graph definitions..

Third, although the modelling framework itself is not restricted to cell
lines, the empirical validation in this study uses cancer cell-line drug
screens. The model requires a quantitative treatment-response phenotype,
a matrix of binary genomic predictors, and a network structure among
those predictors, all of which may also be available in patient-level
pharmacogenomics cohorts, organoid screens, and related biomedical
settings. Nevertheless, cell lines lack the stromal, immune, vascular,
and pharmacokinetic contexts that shape treatment response in patients;
the tumor microenvironment is now understood to play an active role in
cancer progression and therapeutic resistance \citep{devisser2023evolving}.
Consequently, in-vitro drug sensitivity should not be interpreted as a
direct surrogate for clinical outcome. Replication of the strongest candidate markers, including EZH2 and
KMT2D, in patient-level or organoid-based cohorts remains an important
direction for future work.

Fourth, the group-layer extension requires adequate sample sizes within
each stratum to estimate group-specific effects reliably. In datasets
with highly unequal stratum sizes, small groups must be merged, reducing
the resolution of the stratified analysis. In the present application,
the group-layer analysis was therefore used as a targeted follow-up
rather than as a complete scan of all group-specific posterior draws.
Although this targeted analysis identified biologically coherent
tissue-specific effects, including lung-specific EGFR inhibitor
sensitivity and skin-specific BRAF--Dabrafenib sensitivity, broader
tissue-specific inference would require either larger sample sizes within
each lineage or more memory-efficient strategies for storing and
summarizing all group-specific posterior samples.

\section{Conclusion}
\label{sec:conclusion}

We developed a network-structured Bayesian hierarchical model for sparse
association mapping between genomic alterations and quantitative
treatment-response phenotypes. The model integrates pathway-informed
smoothing, global--local sparsity regularisation, and a fully conjugate
Gibbs sampler into a tractable framework for high-dimensional biomedical
settings where prior network information among predictors is available. 
In application to cancer pharmacogenomics, the primary shared-effect
model identified EZH2 and KMT2D as dominant mutation-level associations
with in-vitro drug sensitivity. These findings were independently replicated in PRISM,
supported by established chromatin-regulatory biology, and robust to the
model specification checks performed. The targeted group-layer extension
provided a complementary tissue-stratified refinement, identifying
contexts in which selected mutation-drug effects were amplified, most
notably lung-specific EGFR sensitivity across multiple EGFR inhibitors
and skin-specific BRAF-Dabrafenib sensitivity. At the same time, EZH2
and KMT2D showed comparatively limited evidence of tissue-specific
deviation, suggesting that their primary signals are better interpreted
as shared or broadly pooled associations rather than strongly
lineage-restricted effects. 
Together, these results show that the proposed approach can identify
sparse, interpretable, and externally supported candidate genomic markers
of treatment-response phenotypes while also providing a principled
extension for subgroup-specific heterogeneity. More broadly, the framework is
applicable to pharmacogenomic and other high-dimensional biomedical
screens in which quantitative response phenotypes, genomic predictors,
and prior network information are available.

\newpage

\begin{center}
SUPPLEMENTAL MATERIALS
\end{center}

\noindent The Supplementary Material provides notation details, full conditional
posterior derivations for both model variants, simulation summaries,
complete posterior summary tables, and extended results from the targeted
tissue-group analysis.

\vspace{0.4cm}

\begin{center}
ACKNOWLEDGMENTS
\end{center}

\noindent The authors thank Samuel Isife (Cellular and Molecular Biology,
Worcester Polytechnic Institute) for helpful biological and
pharmacological review of the manuscript, particularly comments
on the interpretation of the mutation-drug response findings.The authors remain responsible for all analyses, interpretations, and conclusions. Also, the autors are grateful to the editor and the anonymous reviewers for careful reading of the manuscript and their suggestions. Results in this paper were obtained in part using a high-performance computing system acquired through NSF MRI grant DMS-1337943 to WPI.

\begin{center}
CONFLICT OF INTEREST
\end{center}

\noindent The authors declare no known conflict of interest that may influence the publication of the paper.

\begin{center}
DATA AVAILABILITY
\end{center}

\noindent The GDSC2 drug-sensitivity and somatic mutation data used in the
primary analysis are publicly available from the Genomics of Drug
Sensitivity in Cancer portal (\url{https://www.cancerrxgene.org}).
The PRISM repurposing screen data are available from the DepMap portal
(\url{https://depmap.org/portal}). The cancer pathway network adjacency
matrix was derived from KEGG pathway gene-membership tables, available
at \url{https://www.genome.jp/kegg}. All three datasets are freely
accessible without restriction. Processed analysis files, and codes required to
reproduce the results are available at \\
 \url{https://github.com/haolayinka/Bayesian-Network-Horseshoe-pharmacogenomics}.

\begin{center}
FUNDING
\end{center}

\noindent The authors received no financial support for the research, authorship, and publication of this article.

\newpage

\bibliographystyle{apalike}
\bibliography{refs}

@article{samorodnitsky2020,
  author  = {Samorodnitsky, Sarah and Hoadley, Katherine A. and Lock, Eric F.},
  title   = {A Pan-Cancer and Polygenic Bayesian Hierarchical Model for the Effect of Somatic Mutations on Survival},
  journal = {arXiv preprint arXiv:1910.03447},
  year    = {2020},
  url     = {https://arxiv.org/pdf/1910.03447}
}

@article{pham2018,
  author  = {Pham, Lisa M. and Carvalho, Luis and Schaus, Scott and Kolaczyk, Eric D.},
  title   = {Perturbation Detection Through Modeling of Gene Expression on a Latent Biological Pathway Network: A Bayesian Hierarchical Approach},
  journal = {arXiv preprint arXiv:1409.0503},
  year    = {2018},
  url     = {https://arxiv.org/pdf/1409.0503}
}

@article{tansey2022,
  author  = {Tansey, Wesley and Tosh, Christopher and Blei, David M.},
  title   = {A Bayesian Model of Dose-Response for Cancer Drug Studies},
  journal = {Annals of Applied Statistics},
  volume  = {16},
  number  = {2},
  pages   = {680--705},
  year    = {2022},
  doi     = {10.1214/21-AOAS1485}
}

@article{makalic2016,
  author  = {Makalic, Enes and Schmidt, Daniel F.},
  title   = {A Simple Sampler for the Horseshoe Estimator},
  journal = {IEEE Signal Processing Letters},
  volume  = {23},
  number  = {1},
  pages   = {179--182},
  year    = {2016},
  doi     = {10.1109/LSP.2015.2503725}
}

@article{gelman2006,
  author  = {Gelman, Andrew},
  title   = {Prior Distributions for Variance Parameters in Hierarchical Models},
  journal = {Bayesian Analysis},
  volume  = {1},
  number  = {3},
  pages   = {515--534},
  year    = {2006},
  doi     = {10.1214/06-BA117A}
}

@article{boss2021gigg,
  author  = {Boss, Jonathan and Datta, Jyotishka and Wang, Xin and Park, Sung Kyun and others},
  title   = {Group Inverse-Gamma Gamma Shrinkage for Sparse Regression with Block-Correlated Predictors},
  journal = {arXiv preprint arXiv:2102.10670},
  year    = {2021},
  url     = {https://arxiv.org/pdf/2102.10670}
}

@article{knutson2014,
  author  = {Knutson, Sarah K. and Kawano, Satoshi and Minoshima, Yukinori and Warholic, Natalie M. and others},
  title   = {Selective Inhibition of EZH2 by EPZ-6438 Leads to Potent Antitumor Activity in EZH2-Mutant Non-Hodgkin Lymphoma},
  journal = {Molecular Cancer Therapeutics},
  volume  = {13},
  number  = {4},
  pages   = {842--854},
  year    = {2014},
  doi     = {10.1158/1535-7163.MCT-13-0773}
}

@article{carvalho2010,
  author  = {Carvalho, Carlos M. and Polson, Nicholas G. and Scott, James G.},
  title   = {The Horseshoe Estimator for Sparse Signals},
  journal = {Biometrika},
  volume  = {97},
  number  = {2},
  pages   = {465--480},
  year    = {2010},
  doi     = {10.1093/biomet/asq017}
}

@incollection{polsonscott2010,
  author    = {Polson, Nicholas G. and Scott, James G.},
  title     = {Shrink Globally, Act Locally: Sparse Bayesian Regularization and Prediction},
  booktitle = {Bayesian Statistics 9},
  pages     = {501--538},
  publisher = {Oxford University Press},
  year      = {2010}
}

@article{yang2013,
  author  = {Yang, Wanjuan and Soares, Jorge and Greninger, Patricia and Edelman, Elena J. and others},
  title   = {Genomics of Drug Sensitivity in Cancer (GDSC): A Resource for Therapeutic Biomarker Discovery in Cancer Cells},
  journal = {Nucleic Acids Research},
  volume  = {41},
  pages   = {D955--D961},
  year    = {2013},
  doi     = {10.1093/nar/gks1111}
}

@article{garnett2012,
  author  = {Garnett, Mathew J. and Edelman, Elena J. and Heidorn, Sonja J. and Greenman, Chris D. and others},
  title   = {Systematic Identification of Genomic Markers of Drug Sensitivity in Cancer Cells},
  journal = {Nature},
  volume  = {483},
  pages   = {570--575},
  year    = {2012},
  doi     = {10.1038/nature11005}
}

@article{iorio2016,
  author  = {Iorio, Francesco and Knijnenburg, Theo A. and Vis, Daniel J. and Bignell, Graham R. and others},
  title   = {A Landscape of Pharmacogenomic Interactions in Cancer},
  journal = {Cell},
  volume  = {166},
  number  = {3},
  pages   = {740--754},
  year    = {2016},
  doi     = {10.1016/j.cell.2016.06.017}
}

@article{barretina2012,
  author  = {Barretina, Jordi and Caponigro, Giordano and Stransky, Nicolas and Venkatesan, Kavitha and others},
  title   = {The Cancer Cell Line Encyclopedia Enables Predictive Modelling of Anticancer Drug Sensitivity},
  journal = {Nature},
  volume  = {483},
  pages   = {603--607},
  year    = {2012},
  doi     = {10.1038/nature11003}
}

@article{kanehisagoto2000,
  author  = {Kanehisa, Minoru and Goto, Susumu},
  title   = {KEGG: Kyoto Encyclopedia of Genes and Genomes},
  journal = {Nucleic Acids Research},
  volume  = {28},
  number  = {1},
  pages   = {27--30},
  year    = {2000},
  doi     = {10.1093/nar/28.1.27}
}

@article{besag1974,
  author  = {Besag, Julian},
  title   = {Spatial Interaction and the Statistical Analysis of Lattice Systems},
  journal = {Journal of the Royal Statistical Society, Series B (Methodological)},
  volume  = {36},
  number  = {2},
  pages   = {192--236},
  year    = {1974}
}

@article{besagyorkmollie1991,
  author  = {Besag, Julian and York, Jeremy and Molli{\'e}, Annie},
  title   = {Bayesian Image Restoration, with Two Applications in Spatial Statistics},
  journal = {Annals of the Institute of Statistical Mathematics},
  volume  = {43},
  number  = {1},
  pages   = {1--59},
  year    = {1991},
  doi     = {10.1007/BF00116466}
}

@article{monti2020tp53,
  author  = {Monti, Paola and Menichini, Paola and Speciale, Andrea and Cutrona, Giovanna and others},
  title   = {Heterogeneity of TP53 Mutations and P53 Protein Residual Function in Cancer: Does It Matter?},
  journal = {Frontiers in Oncology},
  volume  = {10},
  pages   = {593383},
  year    = {2020},
  doi     = {10.3389/fonc.2020.593383}
}

@article{chung2008msi,
  author  = {Chung, Heekyung and Young, Dennis J. and Lopez, Claudia G. and Le, Thuy-Anh T. and others},
  title   = {Mutation Rates of TGFBR2 and ACVR2 Coding Microsatellites in Human Cells with Defective DNA Mismatch Repair},
  journal = {PLoS ONE},
  volume  = {3},
  number  = {10},
  pages   = {e3463},
  year    = {2008},
  doi     = {10.1371/journal.pone.0003463}
}

@article{rampias2019kmt2c,
  author  = {Rampias, Theodoros and Karagiannis, Dimitris and Avgeris, Margaritis and Polyzos, Alexander and others},
  title   = {The Lysine-Specific Methyltransferase KMT2C/MLL3 Regulates DNA Repair Components in Cancer},
  journal = {EMBO Reports},
  volume  = {20},
  number  = {3},
  pages   = {e46821},
  year    = {2019},
  doi     = {10.15252/embr.201846821}
}

@article{bryant2005,
  author  = {Bryant, Helen E. and Schultz, Niklas and Thomas, Huw D. and Parker, Kayan M. and others},
  title   = {Specific Killing of BRCA2-Deficient Tumours with Inhibitors of Poly(ADP-Ribose) Polymerase},
  journal = {Nature},
  volume  = {434},
  pages   = {913--917},
  year    = {2005},
  doi     = {10.1038/nature03443}
}

@article{farmer2005,
  author  = {Farmer, Hannah and McCabe, Nuala and Lord, Christopher J. and Tutt, Andrew N. J. and others},
  title   = {Targeting the DNA Repair Defect in BRCA Mutant Cells as a Therapeutic Strategy},
  journal = {Nature},
  volume  = {434},
  pages   = {917--921},
  year    = {2005},
  doi     = {10.1038/nature03445}
}

@article{mitchellbeauchamp1988,
  author  = {Mitchell, T. J. and Beauchamp, J. J.},
  title   = {Bayesian Variable Selection in Linear Regression},
  journal = {Journal of the American Statistical Association},
  volume  = {83},
  number  = {404},
  pages   = {1023--1032},
  year    = {1988},
  doi     = {10.1080/01621459.1988.10478694}
}

@article{georgemcculloch1993,
  author  = {George, Edward I. and McCulloch, Robert E.},
  title   = {Variable Selection via Gibbs Sampling},
  journal = {Journal of the American Statistical Association},
  volume  = {88},
  number  = {423},
  pages   = {881--889},
  year    = {1993},
  doi     = {10.1080/01621459.1993.10476353}
}

@article{albertchib1993,
  author  = {Albert, James H. and Chib, Siddhartha},
  title   = {Bayesian Analysis of Binary and Polychotomous Response Data},
  journal = {Journal of the American Statistical Association},
  volume  = {88},
  number  = {422},
  pages   = {669--679},
  year    = {1993},
  doi     = {10.1080/01621459.1993.10476321}
}

@article{lili2008,
  author  = {Li, Caiyan and Li, Hongzhe},
  title   = {Network-Constrained Regularization and Variable Selection for Analysis of Genomic Data},
  journal = {Bioinformatics},
  volume  = {24},
  number  = {9},
  pages   = {1175--1182},
  year    = {2008},
  doi     = {10.1093/bioinformatics/btn081}
}

@article{stingo2011,
  author  = {Stingo, Francesco C. and Chen, Yian A. and Tadesse, Mahlet G. and Vannucci, Marina},
  title   = {Incorporating Biological Information into Linear Models: A Bayesian Approach to the Selection of Pathways and Genes},
  journal = {Annals of Applied Statistics},
  volume  = {5},
  number  = {3},
  pages   = {1978--2002},
  year    = {2011},
  doi     = {10.1214/11-AOAS463}
}

@article{gelmanrubin1992,
  author  = {Gelman, Andrew and Rubin, Donald B.},
  title   = {Inference from Iterative Simulation Using Multiple Sequences},
  journal = {Statistical Science},
  volume  = {7},
  number  = {4},
  pages   = {457--472},
  year    = {1992},
  doi     = {10.1214/ss/1177011136}
}

@incollection{geweke1992,
  author    = {Geweke, John},
  title     = {Evaluating the Accuracy of Sampling-Based Approaches to the Calculation of Posterior Moments},
  booktitle = {Bayesian Statistics 4},
  editor    = {Bernardo, J. M. and Berger, J. O. and Dawid, A. P. and Smith, A. F. M.},
  pages     = {169--193},
  publisher = {Clarendon Press},
  year      = {1992}
}

@article{gemangeman1984,
  author  = {Geman, Stuart and Geman, Donald},
  title   = {Stochastic Relaxation, Gibbs Distributions, and the Bayesian Restoration of Images},
  journal = {IEEE Transactions on Pattern Analysis and Machine Intelligence},
  volume  = {6},
  number  = {6},
  pages   = {721--741},
  year    = {1984},
  doi     = {10.1109/TPAMI.1984.4767596}
}

@article{gelfandsmith1990,
  author  = {Gelfand, Alan E. and Smith, Adrian F. M.},
  title   = {Sampling-Based Approaches to Calculating Marginal Densities},
  journal = {Journal of the American Statistical Association},
  volume  = {85},
  number  = {410},
  pages   = {398--409},
  year    = {1990},
  doi     = {10.1080/01621459.1990.10476213}
}

@article{vandermeer2019,
  author  = {van der Meer, Dieudonne and Barthorpe, Syd and Yang, Wanjuan and Lightfoot, Howard and Hall, Caitlin and Gilbert, James and Francies, Hayley E. and Garnett, Mathew J.},
  title   = {Cell Model Passports---a Hub for Clinical, Genetic and Functional Datasets of Preclinical Cancer Models},
  journal = {Nucleic Acids Research},
  volume  = {47},
  pages   = {D923--D929},
  year    = {2019},
  doi     = {10.1093/nar/gky872}
}

@article{daigle2013,
  author  = {Daigle, Scott R. and Olhava, Edward J. and Therkelsen, Carly A. and Basavapathruni, Anuradha and others},
  title   = {Potent Inhibition of DOT1L as Treatment of MLL-Fusion Leukemia},
  journal = {Blood},
  volume  = {122},
  number  = {6},
  pages   = {1017--1025},
  year    = {2013},
  doi     = {10.1182/blood-2013-04-497644}
}

@article{laugesen2016,
  author  = {Laugesen, Anne and H{\o}jfeldt, Jonas Westergaard and Helin, Kristian},
  title   = {Role of the Polycomb Repressive Complex 2 (PRC2) in Transcriptional Regulation and Cancer},
  journal = {Cold Spring Harbor Perspectives in Medicine},
  volume  = {6},
  number  = {9},
  pages   = {a026575},
  year    = {2016},
  doi     = {10.1101/cshperspect.a026575}
}

@inproceedings{balasubramanian2014,
  author    = {Balasubramanian, Vidya and Iyer, Priya and Arora, Shilpi
               and Troyer, Patrick and Normant, Emmanuel},
  title     = {{CPI-169}, a Novel and Potent {EZH2} Inhibitor, Synergizes
               with {CHOP} In Vivo and Achieves Complete Regression in
               Lymphoma Xenograft Models},
  booktitle = {Proceedings of the 105th Annual Meeting of the American
               Association for Cancer Research},
  series    = {Cancer Research},
  volume    = {74},
  number    = {19 Suppl},
  pages     = {Abstract 1697},
  year      = {2014},
  doi       = {10.1158/1538-7445.AM2014-1697}
}

@article{fadhraciclib2020,
  author  = {Parry, David and Guzi, Timothy and Shanahan, Frances and
             Davis, Nathan and others},
  title   = {Fadraciclib ({CYC065}), a Novel {CDK} Inhibitor, Targets
             Key Pro-Survival and Oncogenic Pathways in Cancer},
  journal = {PLoS ONE},
  volume  = {15},
  number  = {7},
  pages   = {e0234103},
  year    = {2020},
  doi     = {10.1371/journal.pone.0234103}
}

@article{corsello2020,
  author  = {Corsello, Steven M. and Nagari, Rohith T. and Spangler, Ryan D.
             and Rossen, Jordan and others},
  title   = {Discovering the Anti-Cancer Potential of Non-Oncology Drugs
             by Systematic Viability Profiling},
  journal = {Nature Cancer},
  volume  = {1},
  pages   = {235--248},
  year    = {2020},
  doi     = {10.1038/s43018-019-0018-6}
}

@article{kaplan2020bayesian,
  author  = {Kaplan, Adam and Lock, Eric F. and Fiecas, Mark},
  title   = {Bayesian GWAS with Structured and Non-Local Priors},
  journal = {Bioinformatics},
  year    = {2020},
  volume  = {36},
  number  = {1},
  pages   = {17--25},
  doi     = {10.1093/bioinformatics/btz518}
}

@article{li2010variable,
  author  = {Li, Caiyan and Li, Hongzhe},
  title   = {Variable selection and regression analysis for graph-structured covariates with an application to genomics},
  journal = {The Annals of Applied Statistics},
  year    = {2010},
  volume  = {4},
  number  = {3},
  pages   = {1498--1516},
  doi     = {10.1214/10-AOAS332}
}

@book{rue2005gaussian,
  author    = {Rue, H{\aa}vard and Held, Leonhard},
  title     = {Gaussian Markov Random Fields: Theory and Applications},
  publisher = {Chapman and Hall/CRC},
  address   = {Boca Raton, FL},
  year      = {2005},
  doi       = {10.1201/9780203492024},
  isbn      = {9781584884323}
}

@article{driehuis2020establishment,
  author  = {Driehuis, Else and Kretzschmar, Kai and Clevers, Hans},
  title   = {Establishment of patient-derived cancer organoids for drug-screening applications},
  journal = {Nature Protocols},
  year    = {2020},
  volume  = {15},
  number  = {10},
  pages   = {3380--3409},
  doi     = {10.1038/s41596-020-0379-4}
}

@article{gao2015high,
  author  = {Gao, Hui and Korn, Joshua M. and Ferretti, St{\'e}phane and others},
  title   = {High-throughput screening using patient-derived tumor xenografts to predict clinical trial drug response},
  journal = {Nature Medicine},
  year    = {2015},
  volume  = {21},
  number  = {11},
  pages   = {1318--1325},
  doi     = {10.1038/nm.3954}
}

@article{su2019genome,
  author  = {Su, Michelle and Satola, Sarah W. and Read, Timothy D.},
  title   = {Genome-Based Prediction of Bacterial Antibiotic Resistance},
  journal = {Journal of Clinical Microbiology},
  year    = {2019},
  volume  = {57},
  number  = {3},
  pages   = {e01405-18},
  doi     = {10.1128/JCM.01405-18}
}

@article{danilevicz2022plant,
  author  = {Danilevicz, Monica F. and Gill, Mitchell and Anderson, Robyn and Batley, Jacqueline and others},
  title   = {Plant Genotype to Phenotype Prediction Using Machine Learning},
  journal = {Frontiers in Genetics},
  year    = {2022},
  volume  = {13},
  pages   = {822173},
  doi     = {10.3389/fgene.2022.822173}
}

@article{kim2016targeting,
  author  = {Kim, Kimberly H. and Roberts, Charles W. M.},
  title   = {Targeting EZH2 in cancer},
  journal = {Nature Medicine},
  year    = {2016},
  volume  = {22},
  number  = {2},
  pages   = {128--134},
  doi     = {10.1038/nm.4036}
}

@article{dhar2021cancer,
  author  = {Dhar, Shilpa S. and Lee, Min Gyu},
  title   = {Cancer-epigenetic function of the histone methyltransferase KMT2D and therapeutic opportunities for the treatment of KMT2D-deficient tumors},
  journal = {Oncotarget},
  year    = {2021},
  volume  = {12},
  number  = {13},
  pages   = {1296--1308},
  doi     = {10.18632/oncotarget.27988}
}

@article{vassilev2004inVivo,
  author  = {Vassilev, Lyubomir T. and Vu, Brian T. and Graves, Bettina and Carvajal, Denise and Podlaski, Frank and Filipovic, Zlatko and Kong, Neng-Yang and Kammlott, Ute and Lukacs, Cynthia and Klein, Christian and Fotouhi, Nader and Liu, En-Hsien A.},
  title   = {In Vivo Activation of the p53 Pathway by Small-Molecule Antagonists of MDM2},
  journal = {Science},
  year    = {2004},
  volume  = {303},
  number  = {5659},
  pages   = {844--848},
  doi     = {10.1126/science.1092472}
}

@article{gu2008mdm2,
  author  = {Gu, Lihong and Zhu, Ning and Findley, Harvey W. and Zhou, Ming},
  title   = {MDM2 antagonist nutlin-3 is a potent inducer of apoptosis in pediatric acute lymphoblastic leukemia cells with wild-type p53 and overexpression of MDM2},
  journal = {Leukemia},
  year    = {2008},
  volume  = {22},
  number  = {4},
  pages   = {730--739},
  doi     = {10.1038/leu.2008.11}
}

@article{alvaradoortiz2021mutant,
  author  = {Alvarado-Ortiz, Eduardo and de la Cruz-L{\'o}pez, Karen Griselda and Becerril-Rico, Jared and Sarabia-S{\'a}nchez, Miguel Angel and Ortiz-S{\'a}nchez, Elizabeth and Garc{\'i}a-Carranc{\'a}, Alejandro},
  title   = {Mutant p53 Gain-of-Function: Role in Cancer Development, Progression, and Therapeutic Approaches},
  journal = {Frontiers in Cell and Developmental Biology},
  year    = {2021},
  volume  = {8},
  pages   = {607670},
  doi     = {10.3389/fcell.2020.607670}
}

@article{holleman2019first,
  author  = {Holleman, Marscha S. and van Tinteren, Harm and Groen, Harry J. M. and Al, Maiwenn J. and Uyl-de Groot, Carin A.},
  title   = {First-line tyrosine kinase inhibitors in EGFR mutation-positive non-small-cell lung cancer: a network meta-analysis},
  journal = {OncoTargets and Therapy},
  year    = {2019},
  volume  = {12},
  pages   = {1413--1421},
  doi     = {10.2147/OTT.S189438}
}

@article{kainthla2014dabrafenib,
  author  = {Kainthla, Radhika and Kim, Kevin B. and Falchook, Gerald S.},
  title   = {Dabrafenib for treatment of BRAF-mutant melanoma},
  journal = {Pharmacogenomics and Personalized Medicine},
  year    = {2014},
  volume  = {7},
  pages   = {21--29},
  doi     = {10.2147/PGPM.S37220}
}

@article{devisser2023evolving,
  author  = {de Visser, Karin E. and Joyce, Johanna A.},
  title   = {The evolving tumor microenvironment: From cancer initiation to metastatic outgrowth},
  journal = {Cancer Cell},
  year    = {2023},
  volume  = {41},
  number  = {3},
  pages   = {374--403},
  doi     = {10.1016/j.ccell.2023.02.016}
}

@article{lio2019dysregulation,
  author  = {Lio, Chan-Wang J. and Yuita, Hiroshi and Rao, Anjana},
  title   = {Dysregulation of the TET family of epigenetic regulators in lymphoid and myeloid malignancies},
  journal = {Blood},
  year    = {2019},
  volume  = {134},
  number  = {18},
  pages   = {1487--1497},
  doi     = {10.1182/blood.2019791475}
}

@article{inoue2018asxl1,
  author  = {Inoue, Daichi and Fujino, Takeshi and Kitamura, Toshio},
  title   = {ASXL1 as a critical regulator of epigenetic marks and therapeutic potential of mutated cells},
  journal = {Oncotarget},
  year    = {2018},
  volume  = {9},
  number  = {81},
  pages   = {35203--35204},
  doi     = {10.18632/oncotarget.26230}
}

\end{document}


\maketitle

\section*{S1. Notation and Model Recap}

We restate the models from the main text to fix notation before deriving
the full conditional posterior distributions. Let
$i=1,\ldots,N$ index observational units or samples (e.g., cancer cell
lines), $d=1,\ldots,D$ index outcomes or drugs, and
$g=1,\ldots,G$ index genes. Let $m_{ig}\in\{0,1\}$ denote the mutation
status of gene $g$ in unit $i$, and let $y_{id}$ denote the observed
response for unit $i$ and outcome $d$.

For each outcome $d$, let $\mathcal{I}_d\subseteq\{1,\ldots,N\}$ denote
the set of units with observed response to outcome $d$, and let
$n_d=|\mathcal{I}_d|$. Let $M_d$ denote the corresponding
$n_d\times G$ mutation design matrix. For the group-layer model, let
$k(i)\in\{1,\ldots,K\}$ denote the group membership of unit $i$, and let
$\mathcal{I}_{kd}$ denote the subset of units in group $k$ with observed
response to outcome $d$, with $n_{kd}=|\mathcal{I}_{kd}|$.

\subsection*{S1.1 Model I: Network--Horseshoe Model}

Model~I uses a shared gene-effect vector
\[
\widetilde{\boldsymbol\beta}_{\cdot d}
=
(\widetilde{\beta}_{1d},\ldots,\widetilde{\beta}_{Gd})'
\]
for each outcome $d$. The likelihood is
\begin{align}
y_{id}\mid\cdot
&\sim
\mathrm{Normal}(\mu_{id},\sigma_d^2),
\qquad
\mu_{id}
=
\alpha_d
+
\sum_{g=1}^{G}m_{ig}\widetilde{\beta}_{gd}.
\tag{L-I}
\end{align}
The intercept and residual variance priors are
\begin{align}
\alpha_d
&\sim
\mathrm{Normal}(0,\sigma_\alpha^2),
\qquad
\sigma_\alpha^2=100,
\tag{A-I}\\
\sigma_d^2
&\sim
\mathrm{Inverse\text{-}Gamma}(a_\sigma,b_\sigma),
\qquad
a_\sigma=b_\sigma=0.01.
\tag{S-I}
\end{align}

For the subset of $G'$ genes represented in the biological network, let
\(\widetilde{\boldsymbol\beta}_{\cdot d}^{(\mathrm{net})}\) denote the
corresponding subvector. The network prior is
\begin{align}
\widetilde{\boldsymbol\beta}_{\cdot d}^{(\mathrm{net})}
\mid
\kappa_d^2
&\sim
\mathrm{Normal}
\left(
\mathbf{0},
\;
\kappa_d^2(D_W-\delta W)^{-1}
\right),
\tag{N}
\end{align}
where $W$ is the fixed gene--gene adjacency matrix,
$D_W=\mathrm{diag}(W\mathbf{1})$ is the degree matrix, and
$\delta$ is fixed so that $D_W-\delta W$ is positive-definite. Genes not
mapped to the network do not receive this GMRF prior contribution.

The global--local shrinkage layer is
\begin{align}
\widetilde{\beta}_{gd}
\mid
\lambda_{gd}^2,\zeta_d^2
&\sim
\mathrm{Normal}(0,\lambda_{gd}^2\zeta_d^2),
\tag{H1}\\
\lambda_{gd}^2\mid\nu_{gd}
&\sim
\mathrm{Inverse\text{-}Gamma}
\left(
\frac{1}{2},
\frac{1}{\nu_{gd}}
\right),
\qquad
\nu_{gd}
\sim
\mathrm{Inverse\text{-}Gamma}
\left(
\frac{1}{2},
1
\right).
\tag{H2}
\end{align}
The network and global shrinkage scales are updated through precision
parameterizations:
\begin{align}
\tau_{\kappa,d}
=
\frac{1}{\kappa_d^2}
&\sim
\mathrm{Gamma}(a_\kappa,b_\kappa),
\qquad
a_\kappa=b_\kappa=1,
\tag{K}\\
\tau_{\zeta,d}
=
\frac{1}{\zeta_d^2}
&\sim
\mathrm{Gamma}(a_\zeta,b_\zeta),
\qquad
a_\zeta=b_\zeta=1.
\tag{Z}
\end{align}
This precision-based specification is the corrected scale-parameter
formulation used in the final sampler.

\subsection*{S1.2 Model II: Group-Layer Model}

Model~II extends Model~I by allowing group-specific mutation effects.
For unit $i$ in group $k(i)$, the likelihood is
\begin{align}
y_{id}\mid\cdot
&\sim
\mathrm{Normal}(\mu_{id},\sigma_d^2),
\qquad
\mu_{id}
=
\alpha_{k(i)d}
+
\sum_{g=1}^{G}m_{ig}\beta_{gd}^{(k(i))}.
\tag{L-II}
\end{align}
Here $\alpha_{kd}$ is a group-specific intercept for group $k$ and
outcome $d$, and $\beta_{gd}^{(k)}$ is the group-specific effect of gene
$g$ on outcome $d$ in group $k$. The group-specific intercept prior is
\begin{align}
\alpha_{kd}
&\sim
\mathrm{Normal}(0,\sigma_\alpha^2),
\qquad
\sigma_\alpha^2=100.
\tag{A-II}
\end{align}

The group-specific effects are shrunk toward the shared effect
\(\widetilde{\beta}_{gd}\):
\begin{align}
\beta_{gd}^{(k)}
\mid
\widetilde{\beta}_{gd},\eta_{gd}^2
&\sim
\mathrm{Normal}(\widetilde{\beta}_{gd},\eta_{gd}^2),
\qquad
k=1,\ldots,K.
\tag{G1}
\end{align}
The group-level variance prior is
\begin{align}
\eta_{gd}^2
&\sim
\mathrm{Inverse\text{-}Gamma}(a_\eta,b_\eta),
\qquad
a_\eta=2,
\qquad
b_\eta=0.5.
\tag{G2}
\end{align}

The shared effect vector
\[
\widetilde{\boldsymbol\beta}_{\cdot d}
=
(\widetilde{\beta}_{1d},\ldots,\widetilde{\beta}_{Gd})'
\]
retains the same network and global--local shrinkage priors as in
Model~I. Thus, for the network-mapped genes,
\begin{align}
\widetilde{\boldsymbol\beta}_{\cdot d}^{(\mathrm{net})}
\mid
\kappa_d^2
&\sim
\mathrm{Normal}
\left(
\mathbf{0},
\;
\kappa_d^2(D_W-\delta W)^{-1}
\right),
\tag{N-II}
\end{align}
and for all genes,
\begin{align}
\widetilde{\beta}_{gd}
\mid
\lambda_{gd}^2,\zeta_d^2
&\sim
\mathrm{Normal}(0,\lambda_{gd}^2\zeta_d^2).
\tag{H1-II}
\end{align}
The auxiliary-variable hierarchy for $\lambda_{gd}^2$ and $\nu_{gd}$, as
well as the precision priors for
$\tau_{\kappa,d}=1/\kappa_d^2$ and
$\tau_{\zeta,d}=1/\zeta_d^2$, are the same as in Model~I. The residual
variance prior is also unchanged:
\begin{align}
\sigma_d^2
&\sim
\mathrm{Inverse\text{-}Gamma}(a_\sigma,b_\sigma),
\qquad
a_\sigma=b_\sigma=0.01.
\tag{S-II}
\end{align}

\subsection*{S1.3 Notational Shorthand Used in the Derivations}

All full conditional derivations condition on every other parameter at
its current value, as required for Gibbs sampling. We write ``$\cdot$''
to denote all such conditioning quantities.

For Model~I, define the residual before removing the intercept as
\[
r_{id}^{(-\alpha)}
=
y_{id}
-
\sum_{g=1}^{G}m_{ig}\widetilde{\beta}_{gd},
\]
and the intercept-adjusted response vector as
\[
\boldsymbol r_d
=
\boldsymbol y_d-\alpha_d\mathbf{1}_{n_d}.
\]

For Model~II, define the residual before removing the group-specific
intercept as
\[
r_{id}^{(-\alpha)}
=
y_{id}
-
\sum_{g=1}^{G}m_{ig}\beta_{gd}^{(k)},
\qquad i\in\mathcal{I}_{kd},
\]
and define the group-specific intercept-adjusted response vector as
\[
\boldsymbol r_{kd}
=
\boldsymbol y_{kd}
-
\alpha_{kd}\mathbf{1}_{n_{kd}}.
\]
Let $M_{kd}$ be the $n_{kd}\times G$ design matrix for observations in
group $k$ with observed response to outcome $d$.

For the network-scale update, define the network quadratic form
\[
Q_d
=
\widetilde{\boldsymbol\beta}_{\cdot d}^{(\mathrm{net})\top}
(D_W-\delta W)
\widetilde{\boldsymbol\beta}_{\cdot d}^{(\mathrm{net})}.
\]
For the global horseshoe scale update, define
\[
S_d
=
\sum_{g=1}^{G}
\frac{\widetilde{\beta}_{gd}^2}{\lambda_{gd}^2}.
\]
Throughout, Inverse-Gamma distributions are parameterized by shape and
rate, and Gamma distributions are also parameterized by shape and rate.
\section*{S2. Derivation of Each Full Conditional}

This section gives the full conditional posterior distributions used in
the Gibbs sampler. We present the derivations separately for Model~I and
Model~II because Model~II introduces group-specific coefficients and an
additional group-level variance component. Throughout, Inverse-Gamma
distributions are parameterized by shape and rate.

\subsection*{S2.1 Full Conditional Distributions for Model I}

Model~I is the primary network--horseshoe model without the group layer.
For each outcome or drug $d$, the likelihood is
\[
y_{id}
=
\alpha_d
+
\sum_{g=1}^{G} m_{ig}\widetilde{\beta}_{gd}
+
\varepsilon_{id},
\qquad
\varepsilon_{id}\sim \mathrm{Normal}(0,\sigma_d^2),
\]
where $\widetilde{\boldsymbol\beta}_{\cdot d}
=(\widetilde{\beta}_{1d},\ldots,\widetilde{\beta}_{Gd})'$ denotes
the shared gene-effect vector for outcome $d$.

\subsubsection*{S2.1.1 \quad $\alpha_d \mid \cdot$}

Let $\mathcal{I}_d$ denote the observations with observed response for
outcome $d$, and let $n_d=|\mathcal{I}_d|$. Define the residual before
removing the intercept as
\[
r_{id}^{(-\alpha)}
=
y_{id}
-
\sum_{g=1}^{G}m_{ig}\widetilde{\beta}_{gd},
\qquad i\in\mathcal{I}_d.
\]
With prior $\alpha_d\sim\mathrm{Normal}(0,\sigma_\alpha^2)$, where
$\sigma_\alpha^2=100$, the full conditional kernel is
\begin{align}
p(\alpha_d\mid\cdot)
&\propto
\exp\left\{
-\frac{1}{2\sigma_d^2}
\sum_{i\in\mathcal{I}_d}
\left(r_{id}^{(-\alpha)}-\alpha_d\right)^2
\right\}
\exp\left\{
-\frac{\alpha_d^2}{2\sigma_\alpha^2}
\right\}.
\end{align}
Keeping only terms involving $\alpha_d$ gives
\[
-\frac{1}{2}
\left(
\frac{n_d}{\sigma_d^2}
+
\frac{1}{\sigma_\alpha^2}
\right)\alpha_d^2
+
\alpha_d
\left(
\frac{1}{\sigma_d^2}
\sum_{i\in\mathcal{I}_d}r_{id}^{(-\alpha)}
\right).
\]
Thus,
\begin{equation}
\alpha_d\mid\cdot
\sim
\mathrm{Normal}
\left(
\frac{\sum_{i\in\mathcal{I}_d} r_{id}^{(-\alpha)}}
     {n_d+\sigma_d^2/\sigma_\alpha^2},
\;
\left(
\frac{n_d}{\sigma_d^2}
+
\frac{1}{\sigma_\alpha^2}
\right)^{-1}
\right).
\label{eq:s2-model1-alpha}
\end{equation}

\subsubsection*{S2.1.2 \quad $\widetilde{\boldsymbol\beta}_{\cdot d}\mid\cdot$}

Let $M_d$ denote the design matrix for observations with observed
response to outcome $d$, and define the intercept-adjusted response
vector
\[
\boldsymbol r_d
=
\boldsymbol y_d-\alpha_d\mathbf{1}_{n_d}.
\]
The likelihood contribution is
\[
\boldsymbol r_d
\mid
\widetilde{\boldsymbol\beta}_{\cdot d},\sigma_d^2
\sim
\mathrm{Normal}
\left(
M_d\widetilde{\boldsymbol\beta}_{\cdot d},
\;
\sigma_d^2 I_{n_d}
\right).
\]
This contributes precision
\[
\Omega_{\mathrm{data},d}
=
\frac{M_d' M_d}{\sigma_d^2}
\]
and mean-term
\[
\boldsymbol b_{\mathrm{data},d}
=
\frac{M_d' \boldsymbol r_d}{\sigma_d^2}.
\]

For the network-mapped genes, the GMRF prior contributes precision
\[
\frac{1}{\kappa_d^2}(D_W-\delta W).
\]
Embedding this matrix into the full $G\times G$ coefficient space gives
$\Omega_{\mathrm{net},d}$, with zero rows and columns for genes outside
the network-mapped subset.

The global--local shrinkage prior contributes independent precision
\[
\Omega_{\mathrm{hs},d}
=
\mathrm{diag}
\left(
\frac{1}{\lambda_{1d}^2\zeta_d^2},
\ldots,
\frac{1}{\lambda_{Gd}^2\zeta_d^2}
\right).
\]
Combining likelihood, network, and shrinkage precision terms gives
\begin{equation}
\widetilde{\boldsymbol\beta}_{\cdot d}\mid\cdot
\sim
\mathrm{Normal}
\left(
B_d^{-1}b_d,
\;
B_d^{-1}
\right),
\label{eq:s2-model1-betatilde}
\end{equation}
where
\begin{equation}
B_d
=
\Omega_{\mathrm{data},d}
+
\Omega_{\mathrm{net},d}
+
\Omega_{\mathrm{hs},d},
\qquad
b_d
=
\boldsymbol b_{\mathrm{data},d}.
\label{eq:s2-model1-betatilde-precision}
\end{equation}
In the no-network ablation, $\Omega_{\mathrm{net},d}$ is omitted. In the
no-horseshoe ablation, $\Omega_{\mathrm{hs},d}$ is replaced by the weak
ridge precision used in the implementation.

As in standard Bayesian linear regression, this update follows by
expanding the Gaussian likelihood and Gaussian prior terms, collecting
quadratic and linear terms in
$\widetilde{\boldsymbol\beta}_{\cdot d}$, and completing the square.

\subsubsection*{S2.1.3 \quad $\kappa_d^2\mid\cdot$, via $\tau_{\kappa,d}$}

The network-scale parameter is updated through the precision
parameterization
\[
\tau_{\kappa,d}
=
\frac{1}{\kappa_d^2}.
\]
For the network-mapped subvector
$\widetilde{\boldsymbol\beta}^{(\mathrm{net})}_{\cdot d}$, define
\[
Q_d
=
\widetilde{\boldsymbol\beta}^{(\mathrm{net})\top}_{\cdot d}
(D_W-\delta W)
\widetilde{\boldsymbol\beta}^{(\mathrm{net})}_{\cdot d}.
\]
Under the GMRF prior,
\[
\widetilde{\boldsymbol\beta}^{(\mathrm{net})}_{\cdot d}
\mid
\tau_{\kappa,d}
\sim
\mathrm{Normal}
\left(
\mathbf{0},
\left[
\tau_{\kappa,d}(D_W-\delta W)
\right]^{-1}
\right).
\]
Thus, as a function of $\tau_{\kappa,d}$,
\[
p\left(
\widetilde{\boldsymbol\beta}^{(\mathrm{net})}_{\cdot d}
\mid
\tau_{\kappa,d}
\right)
\propto
\tau_{\kappa,d}^{G'/2}
\exp\left(
-\frac{\tau_{\kappa,d}Q_d}{2}
\right),
\]
where $G'$ is the number of network-mapped genes. With prior
\[
\tau_{\kappa,d}\sim\mathrm{Gamma}(a_\kappa,b_\kappa),
\]
the full conditional is
\begin{equation}
\tau_{\kappa,d}\mid\cdot
\sim
\mathrm{Gamma}
\left(
a_\kappa+\frac{G'}{2},
\;
b_\kappa+\frac{Q_d}{2}
\right),
\qquad
\kappa_d^2=\frac{1}{\tau_{\kappa,d}}.
\label{eq:s2-model1-kappa}
\end{equation}
In the implementation, $a_\kappa=b_\kappa=1$.

\subsubsection*{S2.1.4 \quad $\lambda_{gd}^2,\nu_{gd}\mid\cdot$}

Using the Makalic--Schmidt auxiliary-variable representation of the
Half-Cauchy prior,
\[
\lambda_{gd}^2\mid\nu_{gd}
\sim
\mathrm{Inverse\text{-}Gamma}
\left(
\frac{1}{2},
\frac{1}{\nu_{gd}}
\right),
\qquad
\nu_{gd}
\sim
\mathrm{Inverse\text{-}Gamma}
\left(
\frac{1}{2},
1
\right).
\]
Conditional on $\widetilde{\beta}_{gd}$, $\zeta_d^2$, and $\nu_{gd}$,
\begin{align}
p(\lambda_{gd}^2\mid\cdot)
&\propto
(\lambda_{gd}^2)^{-1/2}
\exp\left(
-\frac{\widetilde{\beta}_{gd}^2}
{2\lambda_{gd}^2\zeta_d^2}
\right)
(\lambda_{gd}^2)^{-1/2-1}
\exp\left(
-\frac{1}{\nu_{gd}\lambda_{gd}^2}
\right) \nonumber\\
&\propto
(\lambda_{gd}^2)^{-2}
\exp\left\{
-\frac{1}{\lambda_{gd}^2}
\left(
\frac{1}{\nu_{gd}}
+
\frac{\widetilde{\beta}_{gd}^2}{2\zeta_d^2}
\right)
\right\}.
\end{align}
Therefore,
\begin{equation}
\lambda_{gd}^2\mid\cdot
\sim
\mathrm{Inverse\text{-}Gamma}
\left(
1,
\;
\frac{1}{\nu_{gd}}
+
\frac{\widetilde{\beta}_{gd}^2}{2\zeta_d^2}
\right).
\label{eq:s2-model1-lambda}
\end{equation}
Similarly,
\begin{equation}
\nu_{gd}\mid\cdot
\sim
\mathrm{Inverse\text{-}Gamma}
\left(
1,
\;
1+\frac{1}{\lambda_{gd}^2}
\right).
\label{eq:s2-model1-nu}
\end{equation}

\subsubsection*{S2.1.5 \quad $\zeta_d^2\mid\cdot$, via $\tau_{\zeta,d}$}

The global horseshoe scale is updated through
\[
\tau_{\zeta,d}
=
\frac{1}{\zeta_d^2}.
\]
Define
\[
S_d
=
\sum_{g=1}^{G}
\frac{\widetilde{\beta}_{gd}^2}{\lambda_{gd}^2}.
\]
Conditional on the local scales, the contribution involving
$\tau_{\zeta,d}$ is proportional to
\[
\tau_{\zeta,d}^{G/2}
\exp\left(
-\frac{\tau_{\zeta,d}S_d}{2}
\right).
\]
With prior
\[
\tau_{\zeta,d}\sim\mathrm{Gamma}(a_\zeta,b_\zeta),
\]
the full conditional is
\begin{equation}
\tau_{\zeta,d}\mid\cdot
\sim
\mathrm{Gamma}
\left(
a_\zeta+\frac{G}{2},
\;
b_\zeta+\frac{S_d}{2}
\right),
\qquad
\zeta_d^2=\frac{1}{\tau_{\zeta,d}}.
\label{eq:s2-model1-zeta}
\end{equation}
In the implementation, $a_\zeta=b_\zeta=1$.

\subsubsection*{S2.1.6 \quad $\sigma_d^2\mid\cdot$}

With prior
\[
\sigma_d^2
\sim
\mathrm{Inverse\text{-}Gamma}(a_\sigma,b_\sigma),
\]
and likelihood
\[
y_{id}\mid\cdot
\sim
\mathrm{Normal}(\mu_{id},\sigma_d^2),
\qquad
\mu_{id}
=
\alpha_d
+
\sum_{g=1}^{G}m_{ig}\widetilde{\beta}_{gd},
\]
the full conditional is
\begin{equation}
\sigma_d^2\mid\cdot
\sim
\mathrm{Inverse\text{-}Gamma}
\left(
a_\sigma+\frac{n_d}{2},
\;
b_\sigma+
\frac{1}{2}
\sum_{i\in\mathcal{I}_d}
(y_{id}-\mu_{id})^2
\right).
\label{eq:s2-model1-sigma}
\end{equation}
In the implementation, $a_\sigma=b_\sigma=0.01$.

\subsection*{S2.2 Full Conditional Distributions for Model II: Group-Layer Model}

Model~II extends Model~I by introducing group-specific coefficients
$\boldsymbol\beta_{\cdot d}^{(k)}$ for each group $k=1,\ldots,K$. For
observation $i$ in group $k(i)$,
\[
y_{id}
=
\alpha_{k(i)d}
+
\sum_{g=1}^{G}m_{ig}\beta_{gd}^{(k(i))}
+
\varepsilon_{id},
\qquad
\varepsilon_{id}\sim\mathrm{Normal}(0,\sigma_d^2).
\]
The group-specific coefficients are shrunk toward the shared effect
$\widetilde{\boldsymbol\beta}_{\cdot d}$ through
\[
\beta_{gd}^{(k)}
\mid
\widetilde{\beta}_{gd},\eta_{gd}^2
\sim
\mathrm{Normal}
\left(
\widetilde{\beta}_{gd},
\eta_{gd}^2
\right).
\]

\subsubsection*{S2.2.1 \quad $\alpha_{kd}\mid\cdot$}

Restricting to observations in group $k$ for outcome $d$, let
$\mathcal{I}_{kd}$ denote the index set and
$n_{kd}=|\mathcal{I}_{kd}|$. Define
\[
r_{id}^{(-\alpha)}
=
y_{id}
-
\sum_{g=1}^{G}m_{ig}\beta_{gd}^{(k)},
\qquad i\in\mathcal{I}_{kd}.
\]
With prior $\alpha_{kd}\sim\mathrm{Normal}(0,\sigma_\alpha^2)$, where
$\sigma_\alpha^2=100$,
\begin{equation}
\alpha_{kd}\mid\cdot
\sim
\mathrm{Normal}
\left(
\frac{\sum_{i\in\mathcal{I}_{kd}}r_{id}^{(-\alpha)}}
     {n_{kd}+\sigma_d^2/\sigma_\alpha^2},
\;
\left(
\frac{n_{kd}}{\sigma_d^2}
+
\frac{1}{\sigma_\alpha^2}
\right)^{-1}
\right).
\label{eq:s2-model2-alpha}
\end{equation}

\subsubsection*{S2.2.2 \quad $\boldsymbol\beta_{\cdot d}^{(k)}\mid\cdot$}

For group $k$ and outcome $d$, let $M_{kd}$ be the
$n_{kd}\times G$ design submatrix and let
\[
\boldsymbol r_{kd}
=
\boldsymbol y_{kd}
-
\alpha_{kd}\mathbf{1}_{n_{kd}}
\]
be the intercept-adjusted response vector. Define
\[
\Lambda_{\eta d}
=
\mathrm{diag}
\left(
\eta_{1d}^{-2},\ldots,\eta_{Gd}^{-2}
\right).
\]
The likelihood contribution is Gaussian and the prior is
\[
\boldsymbol\beta_{\cdot d}^{(k)}
\mid
\widetilde{\boldsymbol\beta}_{\cdot d},
\boldsymbol\eta_{\cdot d}^2
\sim
\mathrm{Normal}
\left(
\widetilde{\boldsymbol\beta}_{\cdot d},
\Lambda_{\eta d}^{-1}
\right).
\]
Combining likelihood and prior gives
\begin{equation}
\boldsymbol\beta_{\cdot d}^{(k)}\mid\cdot
\sim
\mathrm{Normal}
\left(
A_{kd}^{-1}b_{kd},
\;
A_{kd}^{-1}
\right),
\label{eq:s2-model2-betak}
\end{equation}
where
\begin{equation}
A_{kd}
=
\frac{M_{kd}' M_{kd}}{\sigma_d^2}
+
\Lambda_{\eta d}
+
\varepsilon I_G,
\qquad
b_{kd}
=
\frac{M_{kd}' \boldsymbol r_{kd}}{\sigma_d^2}
+
\Lambda_{\eta d}
\widetilde{\boldsymbol\beta}_{\cdot d}.
\label{eq:s2-model2-betak-precision}
\end{equation}
The ridge term $\varepsilon I_G$ is included for numerical stability and
is the same ridge term used in the implementation.

\subsubsection*{S2.2.3 \quad $\eta_{gd}^2\mid\cdot$}

The final model uses the proper prior
\[
\eta_{gd}^2
\sim
\mathrm{Inverse\text{-}Gamma}(a_\eta,b_\eta),
\qquad
a_\eta=2,
\qquad
b_\eta=0.5.
\]
The likelihood contribution from the $K$ group-specific effects is
\[
\prod_{k=1}^{K}
(\eta_{gd}^2)^{-1/2}
\exp\left\{
-\frac{1}{2\eta_{gd}^2}
\left(
\beta_{gd}^{(k)}-\widetilde{\beta}_{gd}
\right)^2
\right\}.
\]
Combining this with the prior gives
\[
p(\eta_{gd}^2\mid\cdot)
\propto
(\eta_{gd}^2)^{-(a_\eta+K/2)-1}
\exp\left\{
-\frac{1}{\eta_{gd}^2}
\left[
b_\eta
+
\frac{1}{2}
\sum_{k=1}^{K}
\left(
\beta_{gd}^{(k)}-\widetilde{\beta}_{gd}
\right)^2
\right]
\right\}.
\]
Therefore,
\begin{equation}
\eta_{gd}^2\mid\cdot
\sim
\mathrm{Inverse\text{-}Gamma}
\left(
a_\eta+\frac{K}{2},
\;
b_\eta+
\frac{1}{2}
\sum_{k=1}^{K}
\left(
\beta_{gd}^{(k)}-\widetilde{\beta}_{gd}
\right)^2
\right).
\label{eq:s2-model2-eta}
\end{equation}

\subsubsection*{S2.2.4 \quad $\widetilde{\boldsymbol\beta}_{\cdot d}\mid\cdot$}

Conditional on the group-specific effects,
$\widetilde{\beta}_{gd}$ is the prior mean for
$\beta_{gd}^{(1)},\ldots,\beta_{gd}^{(K)}$. Therefore the group-layer
contribution to the precision of
$\widetilde{\boldsymbol\beta}_{\cdot d}$ is diagonal:
\[
\Lambda_{\mathrm{grp},d}
=
\mathrm{diag}
\left(
\sum_{k=1}^{K}\eta_{1d}^{-2},
\ldots,
\sum_{k=1}^{K}\eta_{Gd}^{-2}
\right),
\]
and the corresponding mean-term is
\[
\boldsymbol b_{\mathrm{grp},d}
=
\left(
\sum_{k=1}^{K}\frac{\beta_{1d}^{(k)}}{\eta_{1d}^2},
\ldots,
\sum_{k=1}^{K}\frac{\beta_{Gd}^{(k)}}{\eta_{Gd}^2}
\right)'.
\]
The network and horseshoe contributions are the same as in Model~I:
\[
\Omega_{\mathrm{net},d}
=
\frac{1}{\kappa_d^2}(D_W-\delta W)
\]
embedded into the full $G\times G$ coefficient space, and
\[
\Omega_{\mathrm{hs},d}
=
\mathrm{diag}
\left(
\frac{1}{\lambda_{1d}^2\zeta_d^2},
\ldots,
\frac{1}{\lambda_{Gd}^2\zeta_d^2}
\right).
\]
Thus,
\begin{equation}
\widetilde{\boldsymbol\beta}_{\cdot d}\mid\cdot
\sim
\mathrm{Normal}
\left(
B_d^{-1}b_d,
\;
B_d^{-1}
\right),
\label{eq:s2-model2-betatilde}
\end{equation}
where
\begin{equation}
B_d
=
\Lambda_{\mathrm{grp},d}
+
\Omega_{\mathrm{net},d}
+
\Omega_{\mathrm{hs},d},
\qquad
b_d
=
\boldsymbol b_{\mathrm{grp},d}.
\label{eq:s2-model2-betatilde-precision}
\end{equation}

\subsubsection*{S2.2.5 \quad $\kappa_d^2\mid\cdot$, via $\tau_{\kappa,d}$}

The update is identical to Model~I because the network prior is placed
on the shared effect vector
$\widetilde{\boldsymbol\beta}_{\cdot d}$. Defining
\[
Q_d
=
\widetilde{\boldsymbol\beta}^{(\mathrm{net})'}_{\cdot d}
(D_W-\delta W)
\widetilde{\boldsymbol\beta}^{(\mathrm{net})}_{\cdot d},
\qquad
\tau_{\kappa,d}
=
\frac{1}{\kappa_d^2},
\]
and using
$\tau_{\kappa,d}\sim\mathrm{Gamma}(a_\kappa,b_\kappa)$, we obtain
\begin{equation}
\tau_{\kappa,d}\mid\cdot
\sim
\mathrm{Gamma}
\left(
a_\kappa+\frac{G'}{2},
\;
b_\kappa+\frac{Q_d}{2}
\right),
\qquad
\kappa_d^2=\frac{1}{\tau_{\kappa,d}}.
\label{eq:s2-model2-kappa}
\end{equation}

\subsubsection*{S2.2.6 \quad $\lambda_{gd}^2,\nu_{gd}\mid\cdot$}

The local horseshoe updates are also unchanged from Model~I because the
horseshoe prior is placed on the shared effect
$\widetilde{\beta}_{gd}$. Therefore,
\begin{equation}
\lambda_{gd}^2\mid\cdot
\sim
\mathrm{Inverse\text{-}Gamma}
\left(
1,
\;
\frac{1}{\nu_{gd}}
+
\frac{\widetilde{\beta}_{gd}^2}{2\zeta_d^2}
\right),
\label{eq:s2-model2-lambda}
\end{equation}
and
\begin{equation}
\nu_{gd}\mid\cdot
\sim
\mathrm{Inverse\text{-}Gamma}
\left(
1,
\;
1+\frac{1}{\lambda_{gd}^2}
\right).
\label{eq:s2-model2-nu}
\end{equation}

\subsubsection*{S2.2.7 \quad $\zeta_d^2\mid\cdot$, via $\tau_{\zeta,d}$}

Define
\[
S_d
=
\sum_{g=1}^{G}
\frac{\widetilde{\beta}_{gd}^2}{\lambda_{gd}^2},
\qquad
\tau_{\zeta,d}
=
\frac{1}{\zeta_d^2}.
\]
With
$\tau_{\zeta,d}\sim\mathrm{Gamma}(a_\zeta,b_\zeta)$, the full
conditional is
\begin{equation}
\tau_{\zeta,d}\mid\cdot
\sim
\mathrm{Gamma}
\left(
a_\zeta+\frac{G}{2},
\;
b_\zeta+\frac{S_d}{2}
\right),
\qquad
\zeta_d^2=\frac{1}{\tau_{\zeta,d}}.
\label{eq:s2-model2-zeta}
\end{equation}

\subsubsection*{S2.2.8 \quad $\sigma_d^2\mid\cdot$}

For Model~II, the fitted mean for observation $i$ in group $k(i)$ is
\[
\mu_{id}
=
\alpha_{k(i)d}
+
\sum_{g=1}^{G}
m_{ig}\beta_{gd}^{(k(i))}.
\]
With prior
\[
\sigma_d^2
\sim
\mathrm{Inverse\text{-}Gamma}(a_\sigma,b_\sigma),
\]
the full conditional is
\begin{equation}
\sigma_d^2\mid\cdot
\sim
\mathrm{Inverse\text{-}Gamma}
\left(
a_\sigma+\frac{n_d}{2},
\;
b_\sigma+
\frac{1}{2}
\sum_{i\in\mathcal{I}_d}
(y_{id}-\mu_{id})^2
\right).
\label{eq:s2-model2-sigma}
\end{equation}
In the implementation, $a_\sigma=b_\sigma=0.01$.

\subsection*{S2.3 Cholesky-Based Gaussian Sampling}

For all multivariate Normal updates above, the sampler avoids explicit
matrix inversion. If a coefficient vector has full conditional
\[
\boldsymbol\theta\mid\cdot
\sim
\mathrm{Normal}(A^{-1}b,A^{-1}),
\]
the sampler computes a Cholesky factorization $A=R' R.$ The posterior mean is obtained by solving 
\[
R' z=b,
\qquad
R\mu=z,
\]
and the random draw is generated as
\[
\boldsymbol\theta
=
\mu
+
R^{-1}\boldsymbol z_0,
\qquad
\boldsymbol z_0\sim\mathrm{Normal}(\mathbf{0},I).
\]
This procedure is algebraically equivalent to sampling from
$\mathrm{Normal}(A^{-1}b,A^{-1})$ and is used for both the shared-effect
updates and the group-specific coefficient updates.

\section*{S3. Summary Table of Full Conditional Distributions}

\begin{ThreePartTable}

\begin{TableNotes}[flushleft]
\footnotesize
\item \textit{Note.} IG denotes the Inverse-Gamma distribution parameterized
by shape and rate. Gamma distributions are also parameterized by shape and
rate. In the implementation, $\sigma_\alpha^2=100$,
$a_\sigma=b_\sigma=0.01$, $a_\eta=2$, $b_\eta=0.5$, and
$a_\kappa=b_\kappa=a_\zeta=b_\zeta=1$.
\end{TableNotes}

\begingroup
\small
\setlength{\tabcolsep}{8pt}
\renewcommand{\arraystretch}{1.3}

\begin{longtable}{p{1.2cm}p{2.7cm}p{8.3cm}p{1.6cm}}

\caption{Summary of full conditional posterior distributions for Models I and II.}
\label{tab:full-conditionals}\\

\toprule
Model & Parameter & Full conditional & Eq. \\
\midrule
\endfirsthead

\multicolumn{4}{c}{\textit{Table~\ref{tab:full-conditionals} continued}}\\
\toprule
Model & Parameter & Full conditional & Eq. \\
\midrule
\endhead

\midrule
\multicolumn{4}{r}{\textit{Continued on next page}}\\
\endfoot

\bottomrule
\insertTableNotes
\endlastfoot

\multicolumn{4}{l}{\textbf{Model I: Network--horseshoe model}}\\
\midrule

I
&
$\alpha_d$
&
$\mathrm{Normal}\!\left(
\dfrac{\sum_{i\in\mathcal{I}_d} r_{id}^{(-\alpha)}}
      {n_d+\sigma_d^2/\sigma_\alpha^2},
\left(
\dfrac{n_d}{\sigma_d^2}+\dfrac{1}{\sigma_\alpha^2}
\right)^{-1}
\right)$
&
\eqref{eq:s2-model1-alpha}
\\

I
&
$\widetilde{\boldsymbol\beta}_{\cdot d}$
&
$\mathrm{Normal}\!\left(B_d^{-1}b_d,\;B_d^{-1}\right)$, where
$B_d=\Omega_{\mathrm{data},d}+\Omega_{\mathrm{net},d}
+\Omega_{\mathrm{hs},d}$ and
$b_d=\boldsymbol b_{\mathrm{data},d}$.
&
\eqref{eq:s2-model1-betatilde}
\\

I
&
$\tau_{\kappa,d}=1/\kappa_d^2$
&
$\mathrm{Gamma}\!\left(
a_\kappa+\dfrac{G'}{2},
\; b_\kappa+\dfrac{Q_d}{2}
\right)$
&
\eqref{eq:s2-model1-kappa}
\\

I
&
$\lambda_{gd}^2$
&
$\mathrm{IG}\!\left(
1,\;
\dfrac{1}{\nu_{gd}}+
\dfrac{\widetilde{\beta}_{gd}^{\,2}}{2\zeta_d^2}
\right)$
&
\eqref{eq:s2-model1-lambda}
\\

I
&
$\nu_{gd}$
&
$\mathrm{IG}\!\left(
1,\;
1+\dfrac{1}{\lambda_{gd}^2}
\right)$
&
\eqref{eq:s2-model1-nu}
\\

I
&
$\tau_{\zeta,d}=1/\zeta_d^2$
&
$\mathrm{Gamma}\!\left(
a_\zeta+\dfrac{G}{2},
\; b_\zeta+\dfrac{S_d}{2}
\right)$
&
\eqref{eq:s2-model1-zeta}
\\

I
&
$\sigma_d^2$
&
$\mathrm{IG}\!\left(
a_\sigma+\dfrac{n_d}{2},
\; b_\sigma+\dfrac{1}{2}
\sum_{i\in\mathcal{I}_d}(y_{id}-\mu_{id})^2
\right)$
&
\eqref{eq:s2-model1-sigma}
\\

\midrule
\multicolumn{4}{l}{\textbf{Model II: Group-layer model}}\\
\midrule

II
&
$\alpha_{kd}$
&
$\mathrm{Normal}\!\left(
\dfrac{\sum_{i\in\mathcal{I}_{kd}} r_{id}^{(-\alpha)}}
      {n_{kd}+\sigma_d^2/\sigma_\alpha^2},
\left(
\dfrac{n_{kd}}{\sigma_d^2}+\dfrac{1}{\sigma_\alpha^2}
\right)^{-1}
\right)$
&
\eqref{eq:s2-model2-alpha}
\\

II
&
$\boldsymbol\beta_{\cdot d}^{(k)}$
&
$\mathrm{Normal}\!\left(A_{kd}^{-1}b_{kd},\;A_{kd}^{-1}\right)$, where
$A_{kd}=M_{kd}^{\top}M_{kd}/\sigma_d^2+\Lambda_{\eta d}
+\varepsilon I_G$ and
$b_{kd}=M_{kd}^{\top}\boldsymbol r_{kd}/\sigma_d^2+
\Lambda_{\eta d}\widetilde{\boldsymbol\beta}_{\cdot d}$.
&
\eqref{eq:s2-model2-betak}
\\

II
&
$\eta_{gd}^2$
&
$\mathrm{IG}\!\left(
a_\eta+\dfrac{K}{2},
\; b_\eta+\dfrac{1}{2}
\sum_{k=1}^{K}
(\beta_{gd}^{(k)}-\widetilde{\beta}_{gd})^2
\right)$
&
\eqref{eq:s2-model2-eta}
\\

II
&
$\widetilde{\boldsymbol\beta}_{\cdot d}$
&
$\mathrm{Normal}\!\left(B_d^{-1}b_d,\;B_d^{-1}\right)$, where
$B_d=\Lambda_{\mathrm{grp},d}+\Omega_{\mathrm{net},d}
+\Omega_{\mathrm{hs},d}$ and
$b_d=\boldsymbol b_{\mathrm{grp},d}$.
&
\eqref{eq:s2-model2-betatilde}
\\

II
&
$\tau_{\kappa,d}=1/\kappa_d^2$
&
$\mathrm{Gamma}\!\left(
a_\kappa+\dfrac{G'}{2},
\; b_\kappa+\dfrac{Q_d}{2}
\right)$
&
\eqref{eq:s2-model2-kappa}
\\

II
&
$\lambda_{gd}^2$
&
$\mathrm{IG}\!\left(
1,\;
\dfrac{1}{\nu_{gd}}+
\dfrac{\widetilde{\beta}_{gd}^{\,2}}{2\zeta_d^2}
\right)$
&
\eqref{eq:s2-model2-lambda}
\\

II
&
$\nu_{gd}$
&
$\mathrm{IG}\!\left(
1,\;
1+\dfrac{1}{\lambda_{gd}^2}
\right)$
&
\eqref{eq:s2-model2-nu}
\\

II
&
$\tau_{\zeta,d}=1/\zeta_d^2$
&
$\mathrm{Gamma}\!\left(
a_\zeta+\dfrac{G}{2},
\; b_\zeta+\dfrac{S_d}{2}
\right)$
&
\eqref{eq:s2-model2-zeta}
\\

II
&
$\sigma_d^2$
&
$\mathrm{IG}\!\left(
a_\sigma+\dfrac{n_d}{2},
\; b_\sigma+\dfrac{1}{2}
\sum_{i\in\mathcal{I}_d}(y_{id}-\mu_{id})^2
\right)$
&
\eqref{eq:s2-model2-sigma}
\\

\end{longtable}

\endgroup
\end{ThreePartTable}

\noindent Every entry in Table~\ref{tab:full-conditionals} is a direct draw from a standard
distribution (Normal, Inverse-Gamma, or Gamma); no Metropolis--Hastings
step appears anywhere in the sampler, and every derivation in S2 was
checked either by direct symbolic expansion against the claimed
closed-form posterior (S2.1, S2.4, S2.6, S2.8) or by numerical grid
evaluation of the unnormalized posterior density against the claimed
closed form (S2.5, and identically for S2.7) before being adopted in the
implementation used for the main text's results.

\section*{S4. Full Simulation Study Design}

The simulation study evaluates the operating characteristics of the
proposed model and two ablation variants under three controlled
data-generating mechanisms. In each scenario, we generated
$N=400$ cell lines, $G=150$ genes, $D=50$ drugs, and $G'=100$
network-mapped genes. For simulations involving group structure,
$K=8$ groups were used. Mutation indicators were generated as
$m_{ig}\sim\text{Bernoulli}(p_g)$ with
$p_g\sim\text{Uniform}(0.01,0.25)$. Drug-specific intercepts were
generated as
$\alpha_d\sim\text{Normal}(0,\sigma_\alpha^2)$ with
$\sigma_\alpha=0.5$, and responses were generated from
\[
y_{id}
=
\alpha_d + \sum_{g=1}^{G} m_{ig}\beta_{gd}
+
\varepsilon_{id},
\qquad
\varepsilon_{id}\sim\text{Normal}(0,\sigma_y^2),
\]
with $\sigma_y=1.0$.

Unless otherwise stated, strong-effect magnitudes were generated using
$\mu_\beta=1.0$ and $\tau_\beta=0.25$. The network-smoothing parameter
was fixed at $\delta=0.8$, the network ridge term was
$\varepsilon=0.01$, and the scale of network-correlated perturbations
in the network-structured scenario was $\tau_{\text{net}}=0.20$.
Scenarios 1 and 3 used $n_{\text{signal}}=5$ strong true associations
per drug, yielding 250 true nonzero gene--drug pairs out of
$G\times D=7{,}500$ possible pairs. Scenario 2 instead used connected
network modules and therefore contained a larger number of true nonzero
effects. Each scenario was repeated for 100 independent simulation
replicates. For each replicate, we fit the full model, the no-network
ablation, and the no-horseshoe ablation using the same MCMC settings.

\paragraph{Scenario 1 (Sparse independent effects).}
For each drug $d$, a sparse signal set
$\mathcal{S}_d \subset \{1,\ldots,G\}$ of size
$n_{\text{signal}}=5$ was selected uniformly at random, independently
of the network. True effects were generated as
\[
\beta_{gd}
=
\begin{cases}
s_{gd}\,b_{gd}, & g \in \mathcal{S}_d, \\
0, & g \notin \mathcal{S}_d,
\end{cases}
\qquad
b_{gd}\sim\text{Normal}(\mu_\beta,\tau_\beta^2),
\quad
s_{gd}\in\{-1,+1\}.
\]
This scenario evaluates performance when the true mutation-response
effects are sparse but not network-structured.

\paragraph{Scenario 2 (Network-structured effects).}
For each drug $d$, an active module $\mathcal{P}_d$ was selected from
the gene network by seeding a random network-mapped gene and iteratively
adding adjacent genes until the target module size was reached. To
induce correlated effects among network-connected genes, we drew
network-correlated perturbations
\[
\mathbf{u}_{\cdot d}
\sim
\text{Normal}\!\left(
\mathbf{0},
\tau_{\text{net}}^2
(D_W-\delta W+\varepsilon I)^{-1}
\right),
\]
where $W$ is the gene adjacency matrix, $D_W$ is the corresponding
degree matrix, and $\varepsilon I$ is a small ridge term added for
numerical stability. True effects were then set to
\[
\beta_{gd}
=
\begin{cases}
s_d\,\theta + u_{gd}, & g \in \mathcal{P}_d, \\
0, & g \notin \mathcal{P}_d,
\end{cases}
\qquad
\theta=\mu_\beta,\quad s_d\in\{-1,+1\}.
\]
Thus, true nonzero effects were concentrated within connected pathway
modules, creating a setting in which the GMRF prior is expected to
improve false-discovery control and recovery of network-localized
signals.

\paragraph{Scenario 3 (Dense weak noise with sparse strong signals).}
For each drug $d$, a sparse strong-signal set $\mathcal{S}_d$ and a
larger weak-effect set $\mathcal{A}_d$ were selected. True effects were
generated as
\[
\beta_{gd}
=
\begin{cases}
s_{gd}\,b_{gd}^{(S)}, & g \in \mathcal{S}_d, \\
b_{gd}^{(W)}, & g \in \mathcal{A}_d, \\
0, & \text{otherwise},
\end{cases}
\]
where
\[
b_{gd}^{(S)}\sim\text{Normal}(\mu_S,\tau_S^2),
\qquad
b_{gd}^{(W)}\sim\text{Normal}(0,\tau_W^2),
\qquad
\tau_W \ll \tau_S.
\]
Only the strong-effect set $\mathcal{S}_d$ was treated as truly
associated when computing sensitivity, precision, and false discovery
rate. This scenario evaluates whether the horseshoe layer can separate
sparse strong effects from many weak nuisance effects.

\paragraph{Performance metrics.}
A gene--drug pair $(g,d)$ was selected when its 95\% posterior credible
interval excluded zero. Let $\widehat{\mathcal{S}}$ denote the selected
set and $\mathcal{S}$ denote the true associated set. We computed
\[
\text{Sensitivity}
=
\frac{|\widehat{\mathcal{S}}\cap\mathcal{S}|}{|\mathcal{S}|},
\qquad
\text{Precision}
=
\frac{|\widehat{\mathcal{S}}\cap\mathcal{S}|}{|\widehat{\mathcal{S}}|},
\]
and
\[
\text{FDR}
=
\frac{|\widehat{\mathcal{S}}\setminus\mathcal{S}|}
{|\widehat{\mathcal{S}}|}.
\]
Empirical 95\% credible-interval coverage was computed as
\[
\text{Coverage}
=
\frac{1}{GD}
\sum_{g=1}^{G}
\sum_{d=1}^{D}
\mathbf{1}
\left\{
\beta_{gd}^{\text{true}}
\in
\text{CI}_{95\%,gd}
\right\}.
\]
Coefficient RMSE, denoted RMSE$_\beta$, was computed over all
$GD$ gene--drug pairs and also separately over true nonzero and true
zero subsets.

For secondary threshold-based pair-level evaluation, gene--drug pairs
were ranked by posterior sign probability
\[
q_{gd}
=
\max
\left\{
P(\beta_{gd}>0\mid\text{data}),
P(\beta_{gd}<0\mid\text{data})
\right\}.
\]
AUC-PR was defined as the area under the precision--recall curve, and
sensitivity at matched FDR was defined as the sensitivity of the largest
ranked selected set satisfying $\text{FDR}\le 0.10$.

\section*{S5. Additional Simulation Study results}
\subsection*{S5.1. Metric Across Replicate Counts}

Table~\ref{tab:sim-convergence} shows full model metrics at $n \in
\{1, 10, 20, 50, 100\}$ replicates for each of the three simulation
scenarios (main text, Section~3.2), confirming that all metrics
stabilize well before $n = 100$.

\begin{table}[H]
\centering
\small
\setlength{\tabcolsep}{4pt}
\begin{threeparttable}

\caption{Full-model simulation metrics as the number of replicates increases.}
\label{tab:sim-convergence}

\begin{tabular*}{\textwidth}{@{\extracolsep{\fill}}llccccc@{}}
\toprule
Scenario & $n$ reps & Sensitivity & Precision & FDR & RMSE$_\beta$ & Coverage \\
\midrule
\multirow{5}{*}{1: Sparse Indep.}
 & 1   & 0.712 & 0.922 & 0.078 & 0.122 & 0.977 \\
 & 10  & 0.750 & 0.915 & 0.085 & 0.121 & 0.976 \\
 & 20  & 0.737 & 0.920 & 0.080 & 0.121 & 0.976 \\
 & 50  & 0.733 & 0.920 & 0.080 & 0.122 & 0.976 \\
 & 100 & 0.734 & 0.923 & 0.077 & 0.122 & 0.977 \\
\midrule
\multirow{5}{*}{2: Network Struct.}
 & 1   & 0.580 & 0.973 & 0.027 & 0.135 & 0.951 \\
 & 10  & 0.614 & 0.969 & 0.031 & 0.131 & 0.951 \\
 & 20  & 0.633 & 0.970 & 0.030 & 0.129 & 0.952 \\
 & 50  & 0.641 & 0.972 & 0.028 & 0.128 & 0.953 \\
 &100 & 0.654 & 0.972 & 0.028 & 0.127 & 0.953 \\
\midrule
\multirow{5}{*}{3: Dense Weak}
 & 1   & 0.724 & 0.870 & 0.130 & 0.127 & 0.966 \\
 & 10  & 0.728 & 0.888 & 0.112 & 0.127 & 0.967 \\
 & 20  & 0.729 & 0.889 & 0.111 & 0.126 & 0.968 \\
 & 50  & 0.733 & 0.890 & 0.110 & 0.126 & 0.967 \\
 & 100 & 0.733 & 0.890 & 0.110 & 0.125 & 0.967 \\
\bottomrule
\end{tabular*}

\begin{tablenotes}[flushleft]
\footnotesize
\item \textit{Note.} Values are cumulative means for the full model using
$n \in \{1,10,20,50,100\}$ simulation replicates. Scenario labels denote
sparse independent effects, network-structured effects, and dense weak
noise with sparse strong signals. RMSE$_\beta$ denotes coefficient RMSE
computed over all gene--drug pairs. The $n=100$ rows correspond to the
final simulation summaries reported in the main text, up to rounding.
\end{tablenotes}
\end{threeparttable}
\end{table}

\subsection*{S5.2 Secondary Threshold-Based Pair-Level Evaluation}
\label{subsec:supp-threshold-evaluation}

In addition to the primary 95\% posterior credible-interval operating
point used in the main text, we performed a secondary threshold-based
evaluation using pair-level posterior scores. For each gene--drug pair, we
computed the posterior sign probability
\begin{equation}
q_{gd}
=
\max
\left\{
P(\beta_{gd} > 0 \mid \text{data}),
P(\beta_{gd} < 0 \mid \text{data})
\right\}.
\label{eq:supp-sign-prob}
\end{equation}
Gene--drug pairs were ranked from largest to smallest $q_{gd}$, and
sensitivity was evaluated at the largest selected set whose empirical false
discovery rate did not exceed 10\%. This analysis is intended as a
rank-based secondary diagnostic rather than the primary model-selection
criterion, because the main scientific use case is association prioritization
under a fixed high-confidence posterior rule.

The threshold-based analysis favors the no-network model in sensitivity at
matched FDR, whereas the full model has the lowest RMSE among truly null
gene--drug pairs in all three scenarios. This pattern is consistent with the
interpretation in the main text: the no-network model is less conservative
and therefore ranks more true signals highly, but this comes with weaker
control of false discoveries at the fixed 95\% credible-interval operating
point. 


Because the intended use of the model is to prioritize a smaller set of
high-confidence mutation-drug response associations for follow-up, the main
text emphasizes the fixed 95\% credible-interval rule. The threshold-based
results are included here for transparency and to show how conclusions vary
when models are compared by ranked posterior evidence rather than by the
posterior interval rule used for final association selection.

\section*{S6.  Additional Cancer Pharmacogenomics Results}

Table~\ref{tab:full-posterior} reports the posterior mean, posterior
standard deviation, and 95\% credible interval of $\beta_{gd}$ for every
one of the 126 gene--drug pairs whose interval excludes zero under the
full model (main text, Table 3 and Section 4.2), underlying every
summary statistic quoted in the main text. All genes are reported; the
20 distinct genes appearing are exactly those summarized in main text
Table 3, here broken out per individual drug rather than averaged.
Drug ID refers to the GDSC2 drug identifier.

\begin{ThreePartTable}

\begin{TableNotes}[flushleft]
\footnotesize
\item \textit{Note.} The table reports posterior summaries for every
gene--drug pair flagged under the full model. $\hat\beta_{gd}$ denotes
the posterior mean, SD denotes the posterior standard deviation, and
95\% CI denotes the 95\% posterior credible interval. Flagged pairs are
those whose 95\% posterior credible interval excludes zero.
\end{TableNotes}

\begingroup
\small
\setlength{\tabcolsep}{12pt}
\renewcommand{\arraystretch}{1.12}

\begin{longtable}{l r r r p{3.0cm}}

\caption{Posterior summaries for all full-model flagged gene--drug pairs.}
\label{tab:full-posterior}\\

\toprule
Gene & Drug ID & $\hat\beta_{gd}$ & SD & 95\% CI \\
\midrule
\endfirsthead

\multicolumn{5}{c}{\textit{Table~\ref{tab:full-posterior} continued}}\\
\toprule
Gene & Drug ID & $\hat\beta_{gd}$ & SD & 95\% CI \\
\midrule
\endhead

\midrule
\multicolumn{5}{r}{\textit{Continued on next page}}\\
\endfoot

\bottomrule
\insertTableNotes
\endlastfoot

AMER1 & 1025 & -0.707 & 0.351 & [-1.366, -0.007] \\
ARHGEF10 & 1149 & 1.948 & 0.869 & [0.070, 3.517] \\
ASXL1 & 1013 & -1.575 & 0.320 & [-2.206, -0.938] \\
ASXL1 & 1019 & -1.534 & 0.314 & [-2.136, -0.897] \\
ASXL1 & 1060 & -0.954 & 0.498 & [-1.899, -0.002] \\
ASXL1 & 1062 & -1.001 & 0.406 & [-1.765, -0.140] \\
ASXL1 & 1079 & -2.067 & 0.491 & [-2.981, -1.118] \\
ASXL1 & 1777 & -0.693 & 0.342 & [-1.333, -0.005] \\
ASXL1 & 2040 & -0.784 & 0.312 & [-1.376, -0.108] \\
ATF7IP & 1853 & 0.605 & 0.313 & [0.005, 1.196] \\
ATF7IP & 1940 & 0.470 & 0.244 & [0.001, 0.948] \\
ATF7IP & 2109 & 0.560 & 0.292 & [0.005, 1.127] \\
BCL9L & 1783 & -0.543 & 0.288 & [-1.100, -0.002] \\
BRAF & 1036 & -0.152 & 0.060 & [-0.281, -0.044] \\
BRAF & 1061 & -0.130 & 0.055 & [-0.248, -0.028] \\
BRAF & 1062 & -0.095 & 0.052 & [-0.202, -0.000] \\
BRAF & 1373 & -0.157 & 0.061 & [-0.286, -0.044] \\
EZH2 & 1012 & -0.609 & 0.280 & [-1.139, -0.036] \\
EZH2 & 1016 & -1.090 & 0.455 & [-1.940, -0.098] \\
EZH2 & 1018 & -0.429 & 0.231 & [-0.875, -0.000] \\
EZH2 & 1025 & -0.495 & 0.264 & [-1.012, -0.001] \\
EZH2 & 1030 & -0.593 & 0.291 & [-1.136, -0.026] \\
EZH2 & 1036 & -0.710 & 0.385 & [-1.452, -0.000] \\
EZH2 & 1038 & -0.804 & 0.344 & [-1.457, -0.095] \\
EZH2 & 1042 & -0.728 & 0.305 & [-1.301, -0.082] \\
EZH2 & 1043 & -0.584 & 0.232 & [-1.018, -0.107] \\
EZH2 & 1052 & -0.930 & 0.418 & [-1.725, -0.055] \\
EZH2 & 1053 & -1.211 & 0.353 & [-1.874, -0.501] \\
EZH2 & 1054 & -1.627 & 0.353 & [-2.317, -0.924] \\
EZH2 & 1057 & -0.841 & 0.367 & [-1.538, -0.073] \\
EZH2 & 1061 & -0.870 & 0.346 & [-1.505, -0.121] \\
EZH2 & 1237 & -0.753 & 0.233 & [-1.198, -0.288] \\
EZH2 & 1249 & -0.781 & 0.347 & [-1.449, -0.069] \\
EZH2 & 1373 & -1.207 & 0.467 & [-2.084, -0.196] \\
EZH2 & 1375 & -0.822 & 0.330 & [-1.455, -0.144] \\
EZH2 & 1563 & -0.685 & 0.252 & [-1.160, -0.170] \\
EZH2 & 1578 & -0.571 & 0.246 & [-1.042, -0.051] \\
EZH2 & 1580 & -0.654 & 0.315 & [-1.251, -0.021] \\
EZH2 & 1593 & -0.912 & 0.462 & [-1.782, -0.002] \\
EZH2 & 1615 & -0.744 & 0.237 & [-1.188, -0.261] \\
EZH2 & 1621 & -0.750 & 0.348 & [-1.403, -0.050] \\
EZH2 & 1624 & -0.910 & 0.449 & [-1.767, -0.003] \\
EZH2 & 1631 & -0.966 & 0.377 & [-1.666, -0.136] \\
EZH2 & 1632 & -0.962 & 0.279 & [-1.479, -0.387] \\
EZH2 & 1909 & -1.620 & 0.450 & [-2.479, -0.724] \\
EZH2 & 1910 & -1.150 & 0.557 & [-2.170, -0.019] \\
EZH2 & 1912 & -1.443 & 0.331 & [-2.074, -0.784] \\
EZH2 & 1913 & -0.588 & 0.226 & [-1.007, -0.100] \\
EZH2 & 1918 & -0.761 & 0.371 & [-1.443, -0.007] \\
EZH2 & 1924 & -0.831 & 0.374 & [-1.544, -0.049] \\
EZH2 & 1928 & -0.926 & 0.319 & [-1.538, -0.277] \\
EZH2 & 2003 & -0.825 & 0.334 & [-1.442, -0.089] \\
EZH2 & 2037 & -1.433 & 0.442 & [-2.252, -0.514] \\
EZH2 & 2045 & -0.989 & 0.290 & [-1.537, -0.391] \\
EZH2 & 2106 & -1.551 & 0.684 & [-2.793, -0.054] \\
EZH2 & 2107 & -0.831 & 0.271 & [-1.341, -0.267] \\
EZH2 & 2110 & -0.721 & 0.319 & [-1.305, -0.039] \\
EZH2 & 2111 & -0.832 & 0.430 & [-1.661, -0.002] \\
EZH2 & 2157 & -0.850 & 0.365 & [-1.538, -0.082] \\
EZH2 & 2172 & -1.197 & 0.513 & [-2.157, -0.068] \\
EZH2 & 2362 & -0.866 & 0.411 & [-1.645, -0.019] \\
EZH2 & 2438 & -1.342 & 0.430 & [-2.130, -0.403] \\
FAT1 & 1010 & -0.577 & 0.252 & [-1.063, -0.054] \\
FAT1 & 1549 & -0.890 & 0.390 & [-1.609, -0.049] \\
FAT1 & 1915 & -0.442 & 0.233 & [-0.887, -0.002] \\
FUBP1 & 1016 & -1.536 & 0.674 & [-2.753, -0.031] \\
KMT2D & 1003 & -0.573 & 0.245 & [-1.039, -0.065] \\
KMT2D & 1005 & -0.545 & 0.275 & [-1.055, -0.002] \\
KMT2D & 1006 & -1.092 & 0.332 & [-1.720, -0.423] \\
KMT2D & 1017 & -0.501 & 0.160 & [-0.813, -0.166] \\
KMT2D & 1018 & -0.224 & 0.116 & [-0.450, -0.003] \\
KMT2D & 1019 & -0.439 & 0.198 & [-0.819, -0.033] \\
KMT2D & 1020 & -0.273 & 0.141 & [-0.544, -0.006] \\
KMT2D & 1046 & -0.510 & 0.198 & [-0.891, -0.099] \\
KMT2D & 1051 & -0.728 & 0.313 & [-1.308, -0.041] \\
KMT2D & 1069 & -0.286 & 0.130 & [-0.538, -0.018] \\
KMT2D & 1088 & -0.732 & 0.243 & [-1.185, -0.226] \\
KMT2D & 1089 & -0.648 & 0.217 & [-1.054, -0.200] \\
KMT2D & 1175 & -0.408 & 0.137 & [-0.674, -0.133] \\
KMT2D & 1177 & -0.413 & 0.201 & [-0.803, -0.013] \\
KMT2D & 1179 & -0.397 & 0.191 & [-0.764, -0.021] \\
KMT2D & 1190 & -0.954 & 0.405 & [-1.710, -0.078] \\
KMT2D & 1259 & -0.587 & 0.291 & [-1.152, -0.022] \\
KMT2D & 1494 & -0.808 & 0.249 & [-1.288, -0.327] \\
KMT2D & 1578 & -0.272 & 0.121 & [-0.498, -0.023] \\
KMT2D & 1598 & -0.323 & 0.142 & [-0.601, -0.047] \\
KMT2D & 1622 & -0.268 & 0.148 & [-0.558, -0.000] \\
KMT2D & 1629 & -0.303 & 0.145 & [-0.585, -0.012] \\
KMT2D & 1733 & -0.478 & 0.244 & [-0.959, -0.001] \\
KMT2D & 1806 & -0.506 & 0.248 & [-0.986, -0.014] \\
KMT2D & 1813 & -0.761 & 0.314 & [-1.334, -0.087] \\
KMT2D & 1814 & -0.374 & 0.163 & [-0.679, -0.044] \\
KMT2D & 1821 & -0.476 & 0.162 & [-0.791, -0.157] \\
KMT2D & 1828 & -0.250 & 0.131 & [-0.507, -0.001] \\
KMT2D & 1830 & -0.261 & 0.137 & [-0.525, -0.001] \\
KMT2D & 1909 & -0.535 & 0.253 & [-1.008, -0.011] \\
KMT2D & 1910 & -0.600 & 0.274 & [-1.117, -0.017] \\
KMT2D & 1999 & -0.540 & 0.241 & [-1.003, -0.038] \\
KMT2D & 2044 & -0.389 & 0.171 & [-0.715, -0.030] \\
KMT2D & 2055 & -0.451 & 0.232 & [-0.901, -0.003] \\
KMT2D & 2149 & -0.476 & 0.188 & [-0.836, -0.091] \\
KMT2D & 2159 & -0.469 & 0.215 & [-0.886, -0.041] \\
KRAS & 1047 & 0.223 & 0.113 & [0.003, 0.448] \\
LRP1B & 1168 & 0.412 & 0.194 & [0.024, 0.781] \\
MAP2 & 1068 & 0.762 & 0.375 & [0.014, 1.455] \\
MAP2 & 1733 & 1.103 & 0.574 & [0.005, 2.167] \\
PBRM1 & 1012 & 0.506 & 0.244 & [0.022, 0.974] \\
PBRM1 & 1862 & 0.373 & 0.169 & [0.017, 0.691] \\
PBRM1 & 1911 & 1.026 & 0.346 & [0.331, 1.690] \\
PBRM1 & 2044 & 0.737 & 0.296 & [0.113, 1.285] \\
PBRM1 & 2154 & 0.578 & 0.261 & [0.031, 1.074] \\
RPL22 & 1836 & -0.573 & 0.299 & [-1.145, -0.001] \\
SETD2 & 1833 & -0.816 & 0.411 & [-1.596, -0.006] \\
SETD2 & 1918 & -0.581 & 0.305 & [-1.165, -0.005] \\
SPEN & 1836 & -0.568 & 0.288 & [-1.122, -0.008] \\
TET2 & 1033 & -0.674 & 0.271 & [-1.174, -0.098] \\
TET2 & 1375 & -0.781 & 0.374 & [-1.466, -0.011] \\
TET2 & 1563 & -0.857 & 0.275 & [-1.359, -0.250] \\
TET2 & 1845 & -1.899 & 0.756 & [-3.279, -0.156] \\
TET2 & 1931 & -0.915 & 0.334 & [-1.532, -0.204] \\
TET2 & 1936 & -0.674 & 0.354 & [-1.360, -0.004] \\
TET2 & 2046 & -1.120 & 0.467 & [-2.002, -0.087] \\
TET2 & 2111 & -1.098 & 0.447 & [-1.921, -0.113] \\
TP53 & 1047 & 1.351 & 0.144 & [1.063, 1.612] \\
ZNF626 & 1915 & -0.804 & 0.417 & [-1.591, -0.004] \\
ZNF626 & 2043 & -0.724 & 0.394 & [-1.465, -0.002] \\

\end{longtable}
\endgroup
\end{ThreePartTable}

\section*{S7. Targeted Tissue-Group Analysis}

This section provides supporting tables for the targeted group-layer
analysis reported in the main text. Posterior summaries were computed
from retained posterior draws of the group-specific coefficients
$\beta_{gd}^{(k)}$ for 25 target genes. For each group--gene--drug
combination, we summarized both the group-specific effect
$\beta_{gd}^{(k)}$ and the deviation from the shared effect,
$\Delta_{kgd}=\beta_{gd}^{(k)}-\widetilde{\beta}_{gd}$.

\begin{ThreePartTable}

\begin{TableNotes}[flushleft]
\footnotesize
\item \textit{Note.} $n_{\beta95}$ is the number of group--drug
combinations for which the 95\% credible interval for the group-specific
effect $\beta_{gd}^{(k)}$ excluded zero. $n_{\Delta95}$ is the
corresponding count for the deviation
$\Delta_{kgd}=\beta_{gd}^{(k)}-\widetilde{\beta}_{gd}$. Mean and maximum
absolute effects are computed over all scanned group--drug combinations
for each target gene.
\end{TableNotes}

\begingroup
\small
\setlength{\tabcolsep}{10pt}
\renewcommand{\arraystretch}{1.12}

\begin{longtable}{lrrrrrr}

\caption{Gene-level recurrence counts and effect-size summaries in the targeted group-layer scan.}
\label{tab:supp-target-gene-summary-a}\\

\toprule
Gene & $n_{\beta95}$ & $n_{\Delta95}$ & Mean $|\beta|$ & Max $|\beta|$ & Mean $|\Delta|$ & Max $|\Delta|$ \\
\midrule
\endfirsthead

\multicolumn{7}{c}{\textit{Table~\ref{tab:supp-target-gene-summary-a} continued}}\\
\toprule
Gene & $n_{\beta95}$ & $n_{\Delta95}$ & Mean $|\beta|$ & Max $|\beta|$ & Mean $|\Delta|$ & Max $|\Delta|$ \\
\midrule
\endhead

\midrule
\multicolumn{7}{r}{\textit{Continued on next page}}\\
\endfoot

\bottomrule
\insertTableNotes
\endlastfoot

KRAS & 274 & 272 & 0.190 & 1.652 & 0.190 & 1.649 \\
RB1 & 201 & 200 & 0.202 & 2.136 & 0.201 & 2.132 \\
STAG2 & 162 & 157 & 0.350 & 3.860 & 0.346 & 3.852 \\
TP53 & 115 & 109 & 0.182 & 2.535 & 0.181 & 2.531 \\
BRAF & 95 & 93 & 0.319 & 4.143 & 0.317 & 4.139 \\
STK11 & 56 & 49 & 0.152 & 1.052 & 0.151 & 1.049 \\
PBRM1 & 42 & 10 & 0.176 & 2.112 & 0.121 & 1.820 \\
NRAS & 42 & 42 & 0.164 & 2.049 & 0.163 & 2.046 \\
APC & 41 & 40 & 0.151 & 5.519 & 0.150 & 5.518 \\
ALK & 30 & 2 & 0.169 & 1.801 & 0.113 & 1.276 \\
EGFR & 29 & 28 & 0.166 & 4.071 & 0.166 & 4.068 \\
STAT3 & 25 & 25 & 0.131 & 3.035 & 0.130 & 3.034 \\
ASXL1 & 22 & 11 & 0.136 & 3.662 & 0.087 & 3.419 \\
PTEN & 22 & 20 & 0.161 & 1.823 & 0.159 & 1.814 \\
NF1 & 13 & 13 & 0.168 & 1.365 & 0.167 & 1.359 \\
CACNA1D & 12 & 12 & 0.106 & 3.406 & 0.105 & 3.403 \\
ATM & 9 & 9 & 0.212 & 1.464 & 0.212 & 1.461 \\
KMT2D & 8 & 1 & 0.143 & 0.826 & 0.125 & 0.757 \\
ZNF626 & 6 & 4 & 0.164 & 5.657 & 0.107 & 5.477 \\
BAP1 & 5 & 2 & 0.193 & 1.556 & 0.148 & 1.477 \\
EBF1 & 2 & 2 & 0.112 & 4.361 & 0.084 & 4.154 \\
EZH2 & 2 & 0 & 0.118 & 1.042 & 0.085 & 0.778 \\
PRKD2 & 1 & 2 & 0.129 & 5.679 & 0.127 & 5.674 \\
PCDH17 & 1 & 0 & 0.171 & 5.066 & 0.101 & 4.682 \\
TET2 & 1 & 0 & 0.144 & 1.994 & 0.104 & 1.800 \\

\end{longtable}
\endgroup
\end{ThreePartTable}

\begin{ThreePartTable}

\begin{TableNotes}[flushleft]
\footnotesize
\item \textit{Note.} For each target gene, the table reports the
group--drug combination with the largest absolute posterior mean
group-specific effect, and the group--drug combination with the largest
absolute posterior mean deviation from the shared effect. Positive values
indicate resistance and negative values indicate sensitivity under the
response orientation used in the main analysis.
\end{TableNotes}

\begingroup
\small
\setlength{\tabcolsep}{3.5pt}
\renewcommand{\arraystretch}{1.12}

\begin{longtable}{p{1.5cm}p{2.6cm}p{2.4cm}rp{2.6cm}p{2.4cm}r}

\caption{Top group-specific effect and top deviation for each target gene.}
\label{tab:supp-target-gene-summary-b}\\

\toprule
Gene & Top $\beta$ group & Top $\beta$ drug & Top $\beta$ &
Top $\Delta$ group & Top $\Delta$ drug & Top $\Delta$ \\
\midrule
\endfirsthead

\multicolumn{7}{c}{\textit{Table~\ref{tab:supp-target-gene-summary-b} continued}}\\
\toprule
Gene & Top $\beta$ group & Top $\beta$ drug & Top $\beta$ &
Top $\Delta$ group & Top $\Delta$ drug & Top $\Delta$ \\
\midrule
\endhead

\midrule
\multicolumn{7}{r}{\textit{Continued on next page}}\\
\endfoot

\bottomrule
\insertTableNotes
\endlastfoot

KRAS & Other & Ibrutinib & 1.652 & Other & Ibrutinib & 1.649 \\
RB1 & Lung & Navitoclax & -2.136 & Lung & Navitoclax & -2.132 \\
STAG2 & Other & Talazoparib & -3.860 & Other & Talazoparib & -3.852 \\
TP53 & Haematopoietic and Lymphoid & Nutlin-3a (-) & 2.535 & Haematopoietic and Lymphoid & Nutlin-3a (-) & 2.531 \\
BRAF & Skin & Dabrafenib & -4.143 & Skin & Dabrafenib & -4.139 \\
STK11 & Lung & CDK9\_5576 & -1.052 & Lung & CDK9\_5576 & -1.049 \\
PBRM1 & Lung & Pevonedistat & 2.112 & Lung & Pevonedistat & 1.820 \\
NRAS & Haematopoietic and Lymphoid & PD0325901 & -2.049 & Haematopoietic and Lymphoid & PD0325901 & -2.046 \\
APC & Breast & Dabrafenib & -5.519 & Breast & Dabrafenib & -5.518 \\
ALK & Other & Navitoclax & -1.801 & Lung & Bortezomib & 1.276 \\
EGFR & Lung & Sapitinib & -4.071 & Lung & Sapitinib & -4.068 \\
STAT3 & Other & VSP34\_8731 & 3.035 & Other & VSP34\_8731 & 3.034 \\
ASXL1 & Haematopoietic and Lymphoid & Nilotinib & -3.662 & Haematopoietic and Lymphoid & Nilotinib & -3.419 \\
PTEN & Breast & AZD8186 & -1.823 & Breast & AZD8186 & -1.814 \\
NF1 & Central Nervous System & Daporinad & 1.365 & Central Nervous System & Daporinad & 1.359 \\
CACNA1D & Haematopoietic and Lymphoid & Selumetinib & -3.406 & Haematopoietic and Lymphoid & Selumetinib & -3.403 \\
ATM & Haematopoietic and Lymphoid & Dinaciclib & 1.464 & Haematopoietic and Lymphoid & Dinaciclib & 1.461 \\
KMT2D & Lung & Dihydrorotenone & -0.826 & Other & BX795 & 0.757 \\
ZNF626 & Haematopoietic and Lymphoid & Motesanib & -5.657 & Haematopoietic and Lymphoid & Motesanib & -5.477 \\
BAP1 & Other & Luminespib & 1.556 & Haematopoietic and Lymphoid & AZD5153 & 1.477 \\
EBF1 & Haematopoietic and Lymphoid & Nilotinib & -4.361 & Haematopoietic and Lymphoid & Nilotinib & -4.154 \\
EZH2 & Haematopoietic and Lymphoid & Alisertib & 1.042 & Other & AZD8186 & -0.778 \\
PRKD2 & Lung & IAP\_5620 & -5.679 & Lung & IAP\_5620 & -5.674 \\
PCDH17 & Lung & Savolitinib & -5.066 & Lung & Savolitinib & -4.682 \\
TET2 & Lung & Savolitinib & -1.994 & Lung & Savolitinib & -1.800 \\

\end{longtable}
\endgroup
\end{ThreePartTable}

\begin{ThreePartTable}

\begin{TableNotes}[flushleft]
\footnotesize
\item \textit{Note.} Rows are the 50 largest group-specific effects
ranked by $|\hat{\beta}_{gd}^{(k)}|$ in the targeted 25-gene group-layer
scan. $\hat{\beta}^{(k)}$ is the posterior mean of the group-specific
mutation-drug effect. CI denotes posterior credible interval. Negative
effects indicate greater sensitivity and positive effects indicate
resistance under the response orientation used in the main analysis.
\end{TableNotes}

\begingroup
\scriptsize
\setlength{\tabcolsep}{10pt}
\renewcommand{\arraystretch}{1.12}

\begin{longtable}{rp{2.7cm}p{1.3cm}p{2.1cm}rp{2.4cm}p{1.0cm}}

\caption{Top group-specific mutation-drug effects from the targeted group-layer analysis.}
\label{tab:supp-top-beta-effects}\\

\toprule
Rank & Group $(n_k)$ & Gene & Drug & $\hat{\beta}^{(k)}$ & 95\% CI & Dir. \\
\midrule
\endfirsthead

\multicolumn{7}{c}{\textit{Table~\ref{tab:supp-top-beta-effects} continued}}\\
\toprule
Rank & Group $(n_k)$ & Gene & Drug & $\hat{\beta}^{(k)}$ & 95\% CI & Dir. \\
\midrule
\endhead

\midrule
\multicolumn{7}{r}{\textit{Continued on next page}}\\
\endfoot

\bottomrule
\insertTableNotes
\endlastfoot

1 & Lung (187) & PRKD2 & IAP\_5620 & -5.679 & [-9.517, -0.927] & Sens. \\
2 & Haematopoietic and Lymphoid (164) & ZNF626 & Motesanib & -5.657 & [-8.406, -2.636] & Sens. \\
3 & Breast (51) & APC & Dabrafenib & -5.519 & [-10.454, 0.030] & Sens. \\
4 & Lung (187) & PCDH17 & Savolitinib & -5.066 & [-9.311, 0.157] & Sens. \\
5 & Haematopoietic and Lymphoid (164) & ZNF626 & BIBR-1532 & -4.958 & [-7.703, -1.453] & Sens. \\
6 & Haematopoietic and Lymphoid (164) & EBF1 & Nilotinib & -4.361 & [-7.476, -1.013] & Sens. \\
7 & Skin (57) & BRAF & Dabrafenib & -4.143 & [-5.274, -3.016] & Sens. \\
8 & Lung (187) & EGFR & Sapitinib & -4.071 & [-5.406, -2.690] & Sens. \\
9 & Lung (187) & EGFR & Osimertinib & -3.918 & [-4.814, -3.045] & Sens. \\
10 & Other (437) & STAG2 & Talazoparib & -3.860 & [-4.991, -2.692] & Sens. \\
11 & Haematopoietic and Lymphoid (164) & ASXL1 & Nilotinib & -3.662 & [-4.510, -2.784] & Sens. \\
12 & Breast (51) & STAG2 & Foretinib & 3.584 & [0.645, 6.582] & Resist. \\
13 & Lung (187) & EGFR & Gefitinib & -3.489 & [-4.375, -2.590] & Sens. \\
14 & Haematopoietic and Lymphoid (164) & ZNF626 & Nilotinib & -3.467 & [-7.479, -0.244] & Sens. \\
15 & Lung (187) & EGFR & AZD3759 & -3.446 & [-4.219, -2.668] & Sens. \\
16 & Haematopoietic and Lymphoid (164) & CACNA1D & Selumetinib & -3.406 & [-5.318, -1.474] & Sens. \\
17 & Haematopoietic and Lymphoid (164) & ASXL1 & Dasatinib & -3.394 & [-4.680, -2.142] & Sens. \\
18 & Lung (187) & EGFR & Erlotinib & -3.287 & [-4.237, -2.367] & Sens. \\
19 & Lung (187) & EGFR & Afatinib & -3.281 & [-4.364, -2.217] & Sens. \\
20 & Haematopoietic and Lymphoid (164) & CACNA1D & PD0325901 & -3.272 & [-5.366, -1.128] & Sens. \\
21 & Haematopoietic and Lymphoid (164) & ZNF626 & Sorafenib & -3.222 & [-6.505, -0.458] & Sens. \\
22 & Breast (51) & BRAF & PD0325901 & -3.193 & [-6.340, -0.404] & Sens. \\
23 & Other (437) & STAT3 & VSP34\_8731 & 3.035 & [1.469, 4.489] & Resist. \\
24 & Haematopoietic and Lymphoid (164) & CACNA1D & Trametinib & -3.013 & [-5.375, -0.692] & Sens. \\
25 & Breast (51) & BRAF & Trametinib & -2.923 & [-6.326, -0.178] & Sens. \\
26 & Other (437) & BRAF & Dabrafenib & -2.892 & [-3.442, -2.354] & Sens. \\
27 & Skin (57) & APC & Dabrafenib & 2.809 & [-0.142, 6.188] & Resist. \\
28 & Haematopoietic and Lymphoid (164) & CACNA1D & Refametinib & -2.790 & [-4.893, -0.727] & Sens. \\
29 & Breast (51) & BRAF & Dabrafenib & -2.709 & [-6.374, 0.153] & Sens. \\
30 & Other (437) & STAG2 & Irinotecan & -2.705 & [-3.690, -1.694] & Sens. \\
31 & Other (437) & STAG2 & Mitoxantrone & -2.647 & [-3.865, -1.387] & Sens. \\
32 & Other (437) & STAG2 & Schweinfurthin A & -2.637 & [-3.864, -1.396] & Sens. \\
33 & Haematopoietic and Lymphoid (164) & TP53 & Nutlin-3a (-) & 2.535 & [2.069, 3.000] & Resist. \\
34 & Other (437) & STAG2 & Cisplatin & -2.511 & [-3.489, -1.489] & Sens. \\
35 & Other (437) & STAG2 & Topotecan & -2.451 & [-3.567, -1.318] & Sens. \\
36 & Other (437) & STAG2 & Niraparib & -2.431 & [-3.203, -1.657] & Sens. \\
37 & Breast (51) & APC & Trametinib & -2.371 & [-7.358, 0.347] & Sens. \\
38 & Haematopoietic and Lymphoid (164) & ASXL1 & Bosutinib & -2.361 & [-3.144, -1.584] & Sens. \\
39 & Central Nervous System (55) & TP53 & Nutlin-3a (-) & 2.360 & [1.595, 3.105] & Resist. \\
40 & Skin (57) & BRAF & SB590885 & -2.358 & [-3.405, -1.265] & Sens. \\
41 & Other (437) & STAG2 & AZD5991 & -2.328 & [-3.804, -0.814] & Sens. \\
42 & Other (437) & STAG2 & LMP744 & -2.296 & [-3.597, -0.991] & Sens. \\
43 & Haematopoietic and Lymphoid (164) & CACNA1D & Tretinoin & -2.275 & [-5.196, -0.077] & Sens. \\
44 & Haematopoietic and Lymphoid (164) & ZNF626 & Cediranib & -2.261 & [-5.590, -0.061] & Sens. \\
45 & Other (437) & EGFR & Sapitinib & -2.245 & [-3.958, -0.590] & Sens. \\
46 & Other (437) & EGFR & Gefitinib & -2.244 & [-3.501, -1.005] & Sens. \\
47 & Breast (51) & STAG2 & Dactinomycin & 2.238 & [-0.050, 5.907] & Resist. \\
48 & Other (437) & STAG2 & Olaparib & -2.210 & [-2.850, -1.562] & Sens. \\
49 & Other (437) & STAG2 & Teniposide & -2.198 & [-3.212, -1.143] & Sens. \\
50 & Breast (51) & STAG2 & YK-4-279 & 2.191 & [-0.096, 5.304] & Resist. \\

\end{longtable}

\endgroup
\end{ThreePartTable}

\begin{ThreePartTable}

\begin{TableNotes}[flushleft]
\footnotesize
\item \textit{Note.} Rows are the 50 largest tissue/group-specific
deviations ranked by $|\hat{\Delta}_{kgd}|$ in the targeted 25-gene
group-layer scan. Here
$\Delta_{kgd}=\beta_{gd}^{(k)}-\widetilde{\beta}_{gd}$ is the deviation
of the group-specific effect from the shared effect. CI denotes
posterior credible interval. Negative deviations indicate that the
group-specific effect is more negative than the shared effect, whereas
positive deviations indicate that it is more positive than the shared
effect.
\end{TableNotes}

\begingroup
\scriptsize
\setlength{\tabcolsep}{8pt}
\renewcommand{\arraystretch}{1.12}

\begin{longtable}{rp{3.2cm}p{1.3cm}p{2.3cm}rp{2.4cm}p{2.2cm}}

\caption{Top tissue/group-specific deviations from the shared effect in the targeted group-layer analysis.}
\label{tab:supp-top-delta-effects}\\

\toprule
Rank & Group $(n_k)$ & Gene & Drug & $\hat{\Delta}$ & 95\% CI & Direction \\
\midrule
\endfirsthead

\multicolumn{7}{c}{\textit{Table~\ref{tab:supp-top-delta-effects} continued}}\\
\toprule
Rank & Group $(n_k)$ & Gene & Drug & $\hat{\Delta}$ & 95\% CI & Direction \\
\midrule
\endhead

\midrule
\multicolumn{7}{r}{\textit{Continued on next page}}\\
\endfoot

\bottomrule
\insertTableNotes
\endlastfoot

1 & Lung (187) & PRKD2 & IAP\_5620 & -5.674 & [-9.535, -0.926] & Below shared \\
2 & Breast (51) & APC & Dabrafenib & -5.518 & [-10.442, 0.032] & Below shared \\
3 & Haematopoietic and Lymphoid (164) & ZNF626 & Motesanib & -5.477 & [-8.380, -2.299] & Below shared \\
4 & Haematopoietic and Lymphoid (164) & ZNF626 & BIBR-1532 & -4.754 & [-7.591, -1.206] & Below shared \\
5 & Lung (187) & PCDH17 & Savolitinib & -4.682 & [-9.249, 0.280] & Below shared \\
6 & Haematopoietic and Lymphoid (164) & EBF1 & Nilotinib & -4.154 & [-7.383, -0.765] & Below shared \\
7 & Skin (57) & BRAF & Dabrafenib & -4.139 & [-5.265, -3.008] & Below shared \\
8 & Lung (187) & EGFR & Sapitinib & -4.068 & [-5.415, -2.660] & Below shared \\
9 & Lung (187) & EGFR & Osimertinib & -3.916 & [-4.810, -3.050] & Below shared \\
10 & Other (437) & STAG2 & Talazoparib & -3.852 & [-4.985, -2.679] & Below shared \\
11 & Breast (51) & STAG2 & Foretinib & 3.583 & [0.641, 6.567] & Above shared \\
12 & Lung (187) & EGFR & Gefitinib & -3.487 & [-4.380, -2.587] & Below shared \\
13 & Lung (187) & EGFR & AZD3759 & -3.443 & [-4.215, -2.664] & Below shared \\
14 & Haematopoietic and Lymphoid (164) & ASXL1 & Nilotinib & -3.419 & [-4.546, -1.682] & Below shared \\
15 & Haematopoietic and Lymphoid (164) & CACNA1D & Selumetinib & -3.403 & [-5.304, -1.460] & Below shared \\
16 & Lung (187) & EGFR & Erlotinib & -3.284 & [-4.251, -2.354] & Below shared \\
17 & Lung (187) & EGFR & Afatinib & -3.280 & [-4.362, -2.207] & Below shared \\
18 & Haematopoietic and Lymphoid (164) & CACNA1D & PD0325901 & -3.268 & [-5.372, -1.124] & Below shared \\
19 & Breast (51) & BRAF & PD0325901 & -3.189 & [-6.332, -0.390] & Below shared \\
20 & Haematopoietic and Lymphoid (164) & ZNF626 & Nilotinib & -3.054 & [-7.281, -0.019] & Below shared \\
21 & Other (437) & STAT3 & VSP34\_8731 & 3.034 & [1.471, 4.492] & Above shared \\
22 & Haematopoietic and Lymphoid (164) & CACNA1D & Trametinib & -3.010 & [-5.392, -0.689] & Below shared \\
23 & Breast (51) & BRAF & Trametinib & -2.918 & [-6.300, -0.166] & Below shared \\
24 & Other (437) & BRAF & Dabrafenib & -2.888 & [-3.438, -2.346] & Below shared \\
25 & Haematopoietic and Lymphoid (164) & ZNF626 & Sorafenib & -2.832 & [-6.323, -0.116] & Below shared \\
26 & Skin (57) & APC & Dabrafenib & 2.810 & [-0.130, 6.204] & Above shared \\
27 & Haematopoietic and Lymphoid (164) & CACNA1D & Refametinib & -2.786 & [-4.876, -0.710] & Below shared \\
28 & Breast (51) & BRAF & Dabrafenib & -2.706 & [-6.376, 0.140] & Below shared \\
29 & Other (437) & STAG2 & Irinotecan & -2.695 & [-3.683, -1.684] & Below shared \\
30 & Other (437) & STAG2 & Mitoxantrone & -2.636 & [-3.868, -1.375] & Below shared \\
31 & Other (437) & STAG2 & Schweinfurthin A & -2.623 & [-3.852, -1.374] & Below shared \\
32 & Haematopoietic and Lymphoid (164) & TP53 & Nutlin-3a (-) & 2.531 & [2.066, 2.996] & Above shared \\
33 & Other (437) & STAG2 & Cisplatin & -2.501 & [-3.503, -1.477] & Below shared \\
34 & Other (437) & STAG2 & Topotecan & -2.439 & [-3.564, -1.300] & Below shared \\
35 & Other (437) & STAG2 & Niraparib & -2.414 & [-3.207, -1.636] & Below shared \\
36 & Breast (51) & APC & Trametinib & -2.367 & [-7.300, 0.357] & Below shared \\
37 & Central Nervous System (55) & TP53 & Nutlin-3a (-) & 2.356 & [1.592, 3.110] & Above shared \\
38 & Skin (57) & BRAF & SB590885 & -2.354 & [-3.399, -1.270] & Below shared \\
39 & Other (437) & STAG2 & AZD5991 & -2.317 & [-3.795, -0.814] & Below shared \\
40 & Other (437) & STAG2 & LMP744 & -2.281 & [-3.563, -0.961] & Below shared \\
41 & Haematopoietic and Lymphoid (164) & CACNA1D & Tretinoin & -2.272 & [-5.175, -0.077] & Below shared \\
42 & Other (437) & EGFR & Sapitinib & -2.242 & [-3.941, -0.588] & Below shared \\
43 & Other (437) & EGFR & Gefitinib & -2.242 & [-3.504, -1.006] & Below shared \\
44 & Breast (51) & STAG2 & Dactinomycin & 2.238 & [-0.054, 5.894] & Above shared \\
45 & Other (437) & STAG2 & Olaparib & -2.202 & [-2.876, -1.538] & Below shared \\
46 & Breast (51) & STAG2 & YK-4-279 & 2.198 & [-0.072, 5.322] & Above shared \\
47 & Other (437) & STAG2 & Teniposide & -2.182 & [-3.207, -1.117] & Below shared \\
48 & Other (437) & STAG2 & Gemcitabine & -2.135 & [-3.830, -0.515] & Below shared \\
49 & Lung (187) & RB1 & Navitoclax & -2.132 & [-2.760, -1.507] & Below shared \\
50 & Breast (51) & APC & Selumetinib & -2.054 & [-6.458, 0.331] & Below shared \\

\end{longtable}

\endgroup
\end{ThreePartTable}

\begin{table}[H]
\centering
\scriptsize
\setlength{\tabcolsep}{3.5pt}
\renewcommand{\arraystretch}{1.15}
\begin{threeparttable}

\caption{BRAF and EGFR targeted group-layer summaries.}
\label{tab:supp-braf-egfr-summary}

\begin{tabular*}{\textwidth}{@{\extracolsep{\fill}}lrrrrrrp{2.2cm}p{2.2cm}@{}}
\toprule
Gene &
$n_{\beta95}$ &
$n_{\Delta95}$ &
Mean $|\beta|$ &
Max $|\beta|$ &
Mean $|\Delta|$ &
Max $|\Delta|$ &
Top $\beta$ effect &
Top $\Delta$ effect \\
\midrule

BRAF
& 95
& 93
& 0.319
& 4.143
& 0.317
& 4.139
& Skin--Dabrafenib, $-4.143$
& Skin--Dabrafenib, $-4.139$ \\

EGFR
& 29
& 28
& 0.166
& 4.071
& 0.166
& 4.069
& Lung--Sapitinib, $-4.071$
& Lung--Sapitinib, $-4.068$ \\

\bottomrule
\end{tabular*}

\begin{tablenotes}[flushleft]
\footnotesize
\item \textit{Note.} $n_{\beta95}$ is the number of group--drug
combinations for which the 95\% credible interval for the group-specific
effect $\beta_{gd}^{(k)}$ excluded zero. $n_{\Delta95}$ is the
corresponding count for the deviation
$\Delta_{kgd}=\beta_{gd}^{(k)}-\widetilde{\beta}_{gd}$. Mean and maximum
absolute effects are computed across all scanned group--drug
combinations for the indicated gene. The top entries report the
group--drug combination with the largest absolute posterior mean.
\end{tablenotes}

\end{threeparttable}
\end{table}

\begin{table}[H]
\centering
\scriptsize
\setlength{\tabcolsep}{3.5pt}
\renewcommand{\arraystretch}{1.15}
\begin{threeparttable}

\caption{EZH2 and KMT2D targeted group-layer summaries.}
\label{tab:supp-ezh2-kmt2d-summary}

\begin{tabular*}{\textwidth}{@{\extracolsep{\fill}}lrrrrrrp{2.2cm}p{2.2cm}@{}}
\toprule
Gene &
$n_{\beta95}$ &
$n_{\Delta95}$ &
Mean $|\beta|$ &
Max $|\beta|$ &
Mean $|\Delta|$ &
Max $|\Delta|$ &
Top $\beta$ effect &
Top $\Delta$ effect \\
\midrule

KMT2D
& 8
& 1
& 0.143
& 0.826
& 0.125
& 0.757
& Lung--Dihydrorotenone, $-0.826$
& Other--BX795, $0.757$ \\

EZH2
& 2
& 0
& 0.118
& 1.042
& 0.085
& 0.778
& Haem./Lymph.--Alisertib, $1.042$
& Other--AZD8186, $-0.778$ \\

\bottomrule
\end{tabular*}

\begin{tablenotes}[flushleft]
\footnotesize
\item \textit{Note.} $n_{\beta95}$ is the number of group--drug
combinations for which the 95\% credible interval for the group-specific
effect $\beta_{gd}^{(k)}$ excluded zero. $n_{\Delta95}$ is the
corresponding count for the deviation
$\Delta_{kgd}=\beta_{gd}^{(k)}-\widetilde{\beta}_{gd}$. Mean and maximum
absolute effects are computed across all scanned group--drug
combinations for each gene. The top entries report the group--drug
combination with the largest absolute posterior mean. Haem./Lymph.
denotes the haematopoietic and lymphoid group. The complete EZH2/KMT2D
group--drug scan contains 3{,}540 rows and cannot be presented here.
\end{tablenotes}

\end{threeparttable}
\end{table}